\documentclass[11pt,a4paper]{article}
\pdfoutput=1
\usepackage{amsmath}
\usepackage{color}
\usepackage{amsfonts}
\usepackage{amssymb}
\usepackage{graphicx}
\usepackage{geometry}
\usepackage{amssymb,epsfig,subfigure}
\usepackage{amssymb}
\usepackage{hyperref}
\usepackage{comment}
\usepackage[font=footnotesize]{caption}
\usepackage[T1]{fontenc}
\usepackage[utf8]{inputenc}
\usepackage{latexsym}
\usepackage{cancel}
\usepackage[english]{babel}
\usepackage{cite}

\usepackage{tikz}
\usetikzlibrary{positioning}

\makeatletter
\renewcommand\section{\@startsection {section}{1}{\z@}%
                                 {-3.5ex \@plus -1ex \@minus -.2ex}
                                   {2.3ex \@plus.2ex}%
                                   {\normalfont\large\bfseries}}
\renewcommand\subsection{\@startsection{subsection}{2}{\z@}%
                                   {-3.25ex\@plus -1ex \@minus -.2ex}%
                                     {1.5ex \@plus .2ex}%
                                     {\normalfont\bfseries}}
\renewcommand\subsubsection{\@startsection{subsubsection}{3}{\z@}%
                                   {-3.25ex\@plus -1ex \@minus -.2ex}%
                                     {1.5ex \@plus .2ex}%
                                     {\normalfont\itshape}}
\makeatother

\def\pplogo{\vbox{\kern-\headheight\kern -29pt
\halign{##&##\hfil\cr&{\ppnumber}\cr\rule{0pt}{2.5ex}&\ppdate\cr}}}
\makeatletter
\def\ps@firstpage{\ps@empty \def\@oddhead{\hss\pplogo}%
  \let\@evenhead\@oddhead 
}
\def\maketitle{\par
 \begingroup
 \def\thefootnote{\fnsymbol{footnote}}
 \def\@makefnmark{\hbox{$^{\@thefnmark}$\hss}}
 \if@twocolumn
 \twocolumn[\@maketitle]
 \else \newpage
 \global\@topnum\z@ \@maketitle \fi\thispagestyle{firstpage}\@thanks
 \endgroup
 \setcounter{footnote}{0}
 \let\maketitle\relax
 \let\@maketitle\relax
 \gdef\@thanks{}\gdef\@author{}\gdef\@title{}\let\thanks\relax}
\makeatother

\numberwithin{equation}{section}

\newcommand\eea{\end{eqnarray}}
\newcommand\bea{\begin{eqnarray}}

\def\beq{\begin{equation}}
\def\eeq{\end{equation}}

\newcommand{\be}{\begin{equation}}
\newcommand{\ee}{\end{equation}}
\newcommand{\ba}{\begin{align}}
\newcommand{\ea}{\end{align}}
\newcommand{\bg}{\begin{gather}}
\newcommand{\eg}{\end{gather}}
\newcommand{\bseq}{\begin{subequations}}
\newcommand{\eseq}{\end{subequations}}

\begin{document}
\setcounter{page}0
\def\ppnumber{\vbox{\baselineskip14pt
}}
\def\ppdate{
} \date{}

\author{ Horacio Casini\footnote{e-mail: casini@cab.cnea.gov.ar} and Marina Huerta
\footnote{e-mail: marina.huerta@cab.cnea.gov.ar}
\\
[7mm] \\
{\normalsize \it Centro At\'omico Bariloche and CONICET}\\
{\normalsize \it S.C. de Bariloche, R\'io Negro, R8402AGP, Argentina}
}

\bigskip
\title{\bf  Lectures on entanglement in quantum field theory\footnote{These notes grew from lectures given by the authors at Universidad Nacional de la Plata (2014), the Strings School at Bangalore (2015), It From Qubit School at Perimeter Institute (2016), Strings at Dunes IIP Natal (2016), CECS Valdivia (2016),  Universidad de Buenos Aires (2019),  It From Qubit School at YITP Kyoto (2019),   the ICTP schools (2018) and (2020), and TASI (2021). This manuscript was prepared to be published in the proceedings of TASI (2021).}  
\vskip 0.5cm}
\maketitle

\bigskip

\newpage

\tableofcontents

\vskip 1cm

\newpage


\section{Introduction}

Quantum theory gives a  simple synthesis for the description of all known physical systems. We have linear operators assigned to our questions about a system and states given a probabilistic answer to these questions. For an observable operator $O$ we can get $\langle O \rangle$ from experiments, and the same quantity is the focus of the theory. In this perspective, quantum mechanics looks like a sort of non-commutative information theory.     
To measure $O$, we have to be able to affect the state with such operation, and nothing prevents us then from repeating the operation to measure $O^n$. Linear combinations of $O^n$ form a closed (commutative) algebra, and the knowledge of the expectation values $\langle O^n\rangle$ in this algebra contains the information of the probabilities of all eigenvalues of $O$. More generally, if we have access to several operators $O_1, O_2, \cdots$ we can think that, in principle, our laboratory will be able to measure any product of these operators, and, trivially, linear combinations. Then, we can associate an algebra with a laboratory. 
 
Suppose we do not want to talk about different operators, but just about different laboratories. What number can we associate from the state to a given algebra or laboratory? If we have an operator, the expectation values $\langle O^n\rangle$ give statistical measures of the fluctuations of the state in these variables. In the same way for a whole algebra, if we do not want to distinguish particular operators in it, we are forced to describe its relation with the state by statistical measures of the state {\sl reduced} to the algebra. There are many possible statistical measures we can think about, the entropy is perhaps the most famous one. These quantities are in general not-operator expectation values (though we will see some of them, some Renyi entropies, which are related to expectation values); they are non-linear in the state and are less directly related to experiments. They are, in favourable situations, useful theoretical tools.             

Quantum field theory (QFT) is the quantum mechanics of systems that have a continuum of degrees of freedom distributed in space. It appears as low energy approximation in condensed matter systems and, for relativistic quantum theories, the description in terms of a QFT is a necessity. In these theories, without further information, we have a preferred set of algebras, labelled by the regions of space. Each region $R$ is attached to the algebra of the degrees of freedom localized in $R$. This is the setting of the algebraic approach to describe QFT which deals with the local algebras of operators, their mutual relations, and their relations with the state.  

In QFT different reasons may also drive us to study whole algebras instead of single operators. The operator content is usually described by field operators $\phi(x)$, and it is possible to describe a model by a Hamiltonian constructed in a simple way out of $\phi(x)$. But it may turn out, because of strong interactions, that the experimentally available variables are very different, encoded in other fields $\tilde{\phi}(x)$, not easily related to $\phi(x)$. Then, it may be the case that the theoretically preferable description is not clear from the experimentally available data. It also happens that we can have different descriptions of the same theory in terms of different fields. A simple example is a free field $\phi(x)$ and $:\phi(x)^3:$. These two fields give place to the same algebra of operators and the same theory. In the bootstrap approach to conformal field theory (CFT), the description is in terms of all possible local fields of a model (for example all powers and derivatives of a free field). This is conceptually similar to describing the model by the full algebra of operators. This is a ``coordinate-free'' description, in the sense that we do not require a particular set of fields generating the algebra.  
   
The vacuum state gives a preferred state, and natural theoretical objects in this context are the statistical measures of the vacuum reduced to different regions. The entropy in this context is usually called entanglement entropy (EE). The state of mind in this line of investigation is guided by an as yet unsolved question: can we uniquely describe a QFT in a universal way by its entanglement entropies?   
   
The main purpose of these lectures is to show some aspects of QFT in a different light using some tools of quantum information and think about them in different terms to most traditional approaches. There has been some progress in this line of research in recent years, especially in conceptual issues about renormalization group irreversibility, energy bounds, symmetries and order parameters. A major application of ideas of quantum information theory (QIT) to QFT is in the context of holographic theories. Most of the recent progress on understanding issues about quantum gravity relies on the fact that entanglement quantities are geometrized for the bulk dual description of holographic QFT. Progress on holographic theories will however not be covered by these lectures.    

We will assume some basic knowledge of QFT, and some QFT in curved space will be used eventually. We will introduce some notions of information theory in the first chapter, and then some more will be presented along the way as needed. We will not enter much into the interesting concepts of the theory and act only as users of some specific tools. The mathematics of the algebraic approach to quantum field theory is beyond the scope of these lectures. We will content ourselves with thinking in a more pedestrian way about a QFT as the continuum limit of a cut-off theory (a lattice theory for example).  
 The quantities that survive this limit and are independent of the way we arrive at it, are the ones that by definition belong to the QFT itself. 

\subsection{Some general bibliography}

For reviews on EE in QFT see for example \cite{Nishioka:2018khk,Calabrese:2009qy,Casini:2009sr}. 
For a review focused on holography see the book \cite{rangamani2017holographic}.
For an introduction to mathematical aspects of entanglement in QFT see \cite{Witten:2018lha, Hollands:2017dov}.

\newpage

\section{Algebras, states, and some basic tools of quantum information}
 
For simplicity, we will think mostly in terms of finite-dimensional Hilbert space in this chapter.

\subsection{Algebras}

 An algebra of operators is a set of operators (matrices) closed under linear combinations, products, and taking adjoints, and we will always include the multiples of the identity (the ``numbers'') in the algebra. In symbols
\be
1\in {\cal A}\,, \hspace{.5cm} a,b \in {\cal A}\implies  \hspace{.5cm} \alpha a +\beta b \in {\cal A}\,,\hspace{.5cm} a b\in {\cal A}\,,\hspace{.5cm} a^\dagger \in {\cal A}\,.   
\ee

von Neumann theorem gives an elegant characterization of the sets of operators that form an algebra of operators (any operator algebra with the unit in finite dimensions, a {\sl von Neumann algebra} in infinite dimensions). Let us define the {\sl commutant} of a set of operators as the sets of operators that commute with it
\be
{\cal A}' =\left\{b;\,\, [b,a]=0, \forall a\in {\cal A}   \right\}\,. 
\ee
von Newman theorem tells whatever ${\cal A}$ is, ${\cal A}'$ is an algebra, and that ${\cal A}$ is an algebra iff
\be
{\cal A}={\cal A}''\,. 
\ee
Then the algebra {\sl generated} by $\{a_1, a_2, \cdots \}$ is $\{a_1, a_2, \cdots \}''$. It is the smallest algebra containing $\{a_1, a_2, \cdots \}$. 

The general form of these algebras in finite dimensions is a sum over blocks of full matrix algebras   
\begin{eqnarray}
{\cal A} &=& \bigoplus_i \, M^{(i)}_{m_i\times m_i}\otimes 1_{n_i\times n_i}\,,\\
{\cal A}' &=& \bigoplus_i \,1_{m_i\times m_i} \otimes N^{(i)}_{n_i\times n_i} \,,
\end{eqnarray}
where $M^{(i)}$ is an arbitrary matrix of $m_i \times m_i$ ad $N^{(i)}$ is an arbitrary matrix of $n_i \times n_i$.

The {\sl centre} of both algebras ${\cal Z}={\cal A}\cap {\cal A}'$ is the set of operators in ${\cal A}$ that commutes with all other operators from ${\cal A}$ (and from ${\cal A}'$). ${\cal Z}$ is an Abelian (commutative) algebra, whose elements write in a sum over blocks form ${\cal Z}=\oplus_i \lambda_i 1_{m_i n_i\times m_i n_i}$. The language of algebras then allows us to consider also commutative algebras, which include the case of classical physics, where all operators commute with each other.

More explicitly, in  a basis which diagonalizes the center we have
\begin{equation}
\small \small\small
({\cal A}\cup {\cal A}')''={\cal Z}'=
\left(\begin{array}{ccc}
  M^{(1)}\otimes N^{(1)} & 0 & 0\\
  0 & M^{(2)}\otimes N^{(2)} & 0 \\
  0 & 0 & \ddots
\end{array}\right), \,  {\cal Z}=
\left(\begin{array}{ccc}
  \lambda_1\,1_{m_1 n_1\times m_1 n_1} & 0 & 0 \\
  0 & \lambda_2\, 1_{m_2 n_2\times m_2 n_2} &  0 \\
  0 & 0 & \ddots
\end{array}\right)\,.
\end{equation} 
The algebra ${\cal A}$ is then selected by choosing all the $N^{(i)}=1$, and is isomorphic (as an algebra) to its {\sl standard representation}
\begin{equation}\small
 \left(\begin{array}{ccc}
  M^{(1)} & 0 & 0 \\
  0 & M^{(2)} & 0\\
  0 & 0 & \ddots
\end{array}\right)\,.
\end{equation}
Technically, in an abstract way, this representation is defined such that the minimal projectors of the algebra (projectors such that there are no other projectors onto smaller subspaces) are of dimension $1$.  

\subsection{States and density matrices}

What is a state? It is something that takes an operator and gives a number, its expectation value, in a linear way. Then it is a linear function $\omega: {\cal A}\rightarrow {\mathbf{C}}$. A state has to give positive expectation values to hermitian operators with positive spectrum, and has to be normalized, such that the total probability of orthogonal projectors that add up to the identity is one. Then a state is defined such 
\be
\omega(\alpha a+ \beta b)=\alpha \,\omega(a)+\beta\, \omega(b)\,, \hspace{.5cm} \omega(a a^\dagger)\ge 0\,, \hspace{0.5cm} \omega(1)=1\,. 
\ee 
 
To any state for a finite dimensional ${\cal A}$ corresponds a unique {\sl density matrix} $\rho_\omega$, which is defined as the unique element of the algebra $ \rho_\omega\in  {\cal A}$ such that it reproduces the expectation values through  
\begin{equation}
\omega (a)=\textrm{tr} \rho_\omega a\,, \,\,\,\forall a\in {\cal A}\,,
 \end{equation} 
where the trace on the right hand side is evaluated in the standard representation. 

Then we have the interesting fact that a state in an algebra selects an operator in the algebra itself. This is lost in some infinite-dimensional algebras but is replaced by something else we will see later.  
The general form of a density matrix in the standard representation is 
\begin{equation}
\rho= \left(\begin{array}{ccc}
  p_1\, \rho^{(1)} & 0 & 0 \\
  0 & p_2 \, \rho^{(2)} & 0\\
  0 & 0 & \ddots
\end{array}\right)\,,\label{matrtr}
\end{equation}
where $p_i\ge 0$, $\sum_i p_i=1$, are probabilities, and $\rho^{(i)}$ are $m_i\times m_i$ non negative hermitian matrices of trace $1$. 

\subsection{Entanglement}

If we have two independent systems the algebra is equivalent to the tensor product of algebras ${\cal A}_{12}={\cal A}_1\otimes {\cal A}_2$. The operators in the two algebras commute to each other. This is equivalent to the impossibility of a causal connection between the laboratories. Acting on the state with operations on ${\cal A}_1$ cannot alter the state in a form that can be detected by ${\cal A}_2$. However, the state may have correlations between the two algebras. If these correlations are purely classical, in the sense that we can create the state using uncorrelated states and classical communication (a telephone), the state is said to have no entanglement. These states are of the form
\be
\omega_{12} = \sum_\lambda p_\lambda\, \omega_1^\lambda \otimes \omega_2^\lambda\,.
\ee
A state that is not of this form is said to have entanglement. It can be shown that any entangled state can be used to do some task that is not possible to do in classical physics (violation of Bell's inequalities, teleportation, etc). 

\subsection{Entropy}
Once we have a state in an algebra represented by an operator (the density matrix) we could compute a functional to get a number out of it. This will be an intrinsic property of the state and the algebra and will depend on nothing else. One of the most interesting functionals is the entropy. For a density matrix in the form (\ref{matrtr}) this is   
\begin{equation}
S(\rho)=-\textrm{tr}(\rho\log \rho)=H(\{p_k\})+\sum_k p_k S(\rho^{(k)})\,,
\end{equation}
where $S(\rho^{(k)})=-\textrm{tr}(\rho^{(k)} \, \log(\rho^{(k)})$, and
\begin{equation}
H(\{p_k\})=-\sum_k p_k \log(p_k)\,,
\end{equation}
is the classical Shannon entropy of a probability distribution. 

Why the entropy is particularly interesting depends on its interpretations and properties. 

First, for a thermal state in the canonical ensemble 
\be
\rho=Z^{-1}\, e^{-\beta H}\,,\hspace{.7cm} Z=\textrm{tr} \, e^{-\beta H}\,,
\ee
it gives precisely the thermodynamic entropy
\be 
S= \beta (E-F) \,,\hspace{.6cm} E= \langle H\rangle\,,\hspace{.6cm} F=-T \log Z\,.
\ee
For the microcanonical ensemble, the state has equal probabilities for all vector states lying in some energy shell. If this is a space of dimension $N$, we get
\be
\rho= N^{-1} {\bf 1}\,,\hspace{1cm} S=\log N\,,
\ee
which is Boltzmann's formula. 

The entropy vanishes if and only if the state is pure in the algebra, that is, if it cannot be written as non trivial mixing of other states $\omega= p \omega_1+(1-p) \omega_2$, $p\in (0,1)$. For full matrix algebras this is the case of vector states $\rho=|\psi\rangle \langle\psi|$. The entropy always increases under mixing $S(\omega)\ge  p S(\omega_1)+(1-p) S(\omega_2)$. 

When the state $\omega$ in the Hilbert space is a vector state we have the equality for the entropies of commutant algebras 
\begin{equation}
S_{{\cal A}}(\omega)=S_{{\cal A}'}(\omega)\,.
\end{equation}
In this case all the non zero eigenvalues of $\rho_{\cal A}$ and $\rho_{{\cal A}'}$ coincide.

The entropy has several different information theory ``operational'' interpretations. In the pure classical case, the algebra is diagonal and the density matrix is a collection of probabilities $\{p_i\}$, $i=1\, \cdots, N$. The entropy $S=-\sum_1^N p_i \log p_i$ then has the interpretation of the average amount of information per letter that can be conveyed with an alphabet of $N$ letters appearing with probabilities $p_i$ in a message. This is Shannon's theory. For example, if a certain alphabet has entropy $S(p)$, and another (that can have a different number of letters) has entropy $S(q)$, we can convey the same information with $m$ letters of the first one and $n=m \frac{S(p)}{S(q)}$ of the second one. This is however an asymptotic statement, it applies to the limit of long messages. This is obviously used for compressing messages. For example, a message using an alphabet where the probabilities of some letters are small and for other letters is large (then small entropy per letter) can be equally represented by a shorter message with maximal entropy per letter, where all letters have the same probability of appearing in the message. This compression is entirely analogous to the one between the canonical and microcanonical ensembles in the thermodynamical limit: a huge reduction of space but keeping essentially the same information and the same total entropy.  

In the opposite limit of a pure quantum mechanical situation, consider a full matrix algebra ${\cal A}$ (trivial centre) and a pure global state $\omega$.  In this case, the entropy $S_{\cal A}(\omega)=S_{{\cal A}'}(\omega)$ is called {\sl entanglement entropy} and is a measure of the amount of entanglement between ${\cal A}$ and ${\cal A}'$. Its operational interpretation is that given another pure state $\sigma$ we can convert $n$ copies of $\omega$ to $n \frac{S(\omega)}{S({\sigma})}$ of the second one and back (in the limit of large $n$), by using only local operations in the two laboratories and classical communication.  In particular, we can transform reversibly the state  $\omega$ into $S(\omega)/\log(2)$ pair of maximally entangled qubits, for a large number of copies.   

The entropy is monotonically increasing under the inclusion of algebras in the classical discrete case. In the quantum case, this is not true because of entanglement. For example, for a global pure state, if we enlarge ${\cal A}$ we make ${\cal A}'$ smaller, but the entropies of the complementary algebras remain equal. Or more simply, the global pure state has zero entropy, while there can be non zero entanglement entropy in a subalgebra. 
As Sch\"odringer put this perplexing nature of entanglement, we can have complete knowledge of a system (a pure state in the global Hilbert space) and know nothing about a subsystem (for example the reduced state be maximally mixed, proportional to the identity, with maximal entropy).

\subsection{Modular flow and modular Hamiltonian}

A density matrix is a positive operator and we can write it (if it does not have zero eigenvalues) as
\be
\rho= \frac{e^{-K}}{\textrm{tr}\,e^{-K}}\,. \label{termal}
\ee
$K$ is called the {\sl modular Hamiltonian}. In this way, we can think of entropy of a density matrix as the canonical entropy for an equilibrium state at temperature $1$ for a different (dimensionless) Hamiltonian $K$. The ``time'' notion associated with the state through the modular Hamiltonian is implemented by the unitary evolution in the algebra
\be
U(\tau)=\rho^{i \tau}\sim e^{-i \tau K}\,. \label{modflow}
\ee      
The evolution of operators $O(\tau)= U(\tau) O U(-\tau)$ is called the modular flow. Notice the modular flow is a one parameter group that leaves invariant operators expectation values
\be
\textrm{tr}\,\rho \, O(\tau)=\textrm{tr}\,\rho \, O\,.\label{ppp}
\ee
It is also a purely quantum mechanical object and becomes trivial in the classical case where density matrices and operators are diagonal matrices. 

There is a way to say that a state has the thermal like form (\ref{termal}) with respect to $K$ just looking at the behaviour of correlations under time evolution $U(\tau)$ dictated by $K$, and going to imaginary $\tau$. This is useful in some infinite dimensional cases where the density matrix is ill defined but the modular flow exists. We have for such state 
\be
\langle O_1(i)\,O_2(0)\rangle= \textrm{tr}\,(\rho\, O_1(i)\,O_2(0)) =\textrm{tr}\,(\rho\, (\rho^{-1}\,O_1(0)\,\rho)\,O_2(0)) =\langle O_2(0)\,O_1(0)\rangle\,. \label{kms}
\ee
This is called the KMS condition (Kubo, Martin, Schwinger). Notice it gives a sort of periodicity in imaginary time $\tau$, except for the order in the operators. 
 This is related to the thermal partition function being given by a path integral with a periodic imaginary time in QFT. The modular flow can be characterized as the unique one-parameter group of automorphisms of the algebra keeping the state invariant (\ref{ppp}) and satisfying the KMS condition.  

\subsection{Relative entropy}
Relative entropy is defined for two states and the same algebra. It is given by
\begin{equation}
S(\rho|\sigma) = \textrm{tr}( \rho \log (\rho)-\rho \log(\sigma))\,.
\end{equation}
It is a central quantity in QIT, more important than the entropy itself. The entropy is derived from the relative entropy of a state and the totally mixed state proportional to the identity
\be
S(\rho|{\bf 1}/d)=\log(d)-S(\rho)\,. 
\ee
But more generally, as we will see, it is a well-defined quantity for the infinite-dimensional algebras of the type that appear in QFT while the entropy is not.  

It has a thermodynamics interpretation too. If we take the second state $\sigma$ as the canonical thermal state at temperature $T$ we get
\be
S(\rho|\sigma)= \beta (F(\rho) - F(\sigma))\,,\hspace{.7 cm} F(\rho)= \langle H\rangle_\rho - T S(\rho)\,.
\ee
In other words, it is proportional to the difference of free energies between the states. Since the relative entropy is always positive
\be
S(\rho|\sigma)\ge 0\,,\label{fff}
\ee
where the equality holds only for identical states, the thermodynamic interpretation is that the free energy, at a certain fixed temperature, is minimal for the thermal state. 

We can write the formula for the relative entropy in the general case in a manner resembling this thermodynamical formula in terms of the free energy using the modular Hamiltonian for $\sigma$:
\be
S(\rho|\sigma)=(\langle K_\sigma \rangle_\rho - \langle K_\sigma  \rangle_\sigma)-   (S(\rho)-S(\sigma))=\Delta \langle K_\sigma\rangle- \Delta S\,. 
\ee

The relative entropy is monotonically increasing with the algebra, a quite important and deep property, 
\be
S_{\cal A}(\rho|\sigma)\le S_{\cal B}(\rho|\sigma) \,,\hspace{.7cm} {\cal A}\subseteq {\cal B}\,.\label{mon}
\ee
$\rho$ and $\sigma$ are thought here as global states reduced to ${\cal A}$ and ${\cal B}$.
This property and (\ref{fff}) suggest the relative entropy is a measure of distinguishability between two states. As we look at the states in smaller algebras we can distinguish them less. In fact, there is a precise result giving an operational interpretation to this idea. Suppose we have a state $\sigma$ and make $N$ measurements of it, and want to ascertain the probability $p$ that the outcome of these measurements come close (say in a neighbourhood $\epsilon$) of the expectation values predicted by the state $\rho$, which is to be interpreted as a  theoretical model. If the states are different, as we make more measurements (e.g. the best measurements designed to test the difference between the states),  this probability will decay exponentially with $N$ as
\be
p\sim e^{-N S(\rho|\sigma)}\,.    
\ee
This interpretation accounts for the relative entropy being non-symmetric between the two states. Suppose we have two probability distributions for flipping a coin. Take $\rho=(1,0)$ and $\sigma=(1/2,1/2)$. If we flip $\sigma$ there will be a probability $2^{-N}=e^{-N \log 2}$ to get the distribution of outcomes that $\rho$ produces, which is all on the same side of the coin. This matches the relative entropy $S(\rho|\sigma)=\log(2)$. If we were to flip $\rho$ instead, we will never get close to the distribution $(1/2,1/2)$, and this is consistent with $S(\sigma|\rho)=\infty$.    

\subsection{Mutual information and strong subadditivity}
\label{mumu}

The relative entropy requires two different states for the same algebra. With only one state we can however produce two states if our algebra has the structure of a tensor product ${\cal A}_{12}={\cal A}_1\otimes {\cal A}_2$. We have the two states $\omega_{12}$ and $\omega_1\otimes \omega_2$ in   ${\cal A}_{12}$. Their relative entropy is called mutual information 
\be
I({\cal A}_1, {\cal A}_2)= S(\omega_{12}|\omega_1\otimes \omega_2)=S({\cal A}_1)+S({\cal A}_2)-S({\cal A}_{12})\,,
\label{sss} \ee
and nicely can be written in terms of a combination of entropies. 

The mutual information is positive and the positivity of the right-hand side of (\ref{sss}) tell us the subadditivity property of the entropy. Monotonicity of relative entropy gives the monotonicity of the mutual information under increasing any of the two algebras. If we take a product  ${\cal A}_{123}={\cal A}_1\otimes {\cal A}_2\otimes {\cal A}_3$ from $I(1,2)\le I(1,23)$ we get 
\be
S({\cal A}_{12})+S({\cal A}_{23})\ge S({\cal A}_{123})+S({\cal A}_{2})\,.
\ee
This is called the strong subadditive property of the entropy. We will use it a good deal in QFT. 

Mutual information is a measure of correlations (both classical and quantum mechanical) that the state produces between the two algebras. It is zero only for product states. The interesting point is that any kind of correlation is detected by mutual information. We have the bound
\be  
I(1,2)\ge \frac{1}{2}\frac{|\langle O_1 O_2\rangle -\langle O_1 \rangle \langle O_2 \rangle|^2}{\|O_1\|^2 \|O_2\|^2 }\,,
\ee
where $O_1$, $O_2$ are operators in the two algebras and $\|O\|$ is the norm of the operator, the size of its largest eigenvalue. 

\subsection{Evolution and the second law}

Time evolution is somehow not as well adapted as we would wish to a general view of algebras and states as isolated worlds. 
To be sure, the modular evolution is a natural evolution arising from the state and the algebra but it keeps the state and the expectation values invariant. For a system interacting with an environment, the problem is that in general there is no mapping between states for different times, no evolution law. The reason is that the global evolution corresponding to different initial global states giving place to the same initial state in the subsystem will give in general different final states in the subsystem. Then, there is no function taking an initial to a final state in the subsystem itself.   

Anyway, if such a mapping between states can be defined (and this is a big if), it has to be what is called a completely positive trace-preserving map (CPTP). These are
linear maps of density matrices in one algebra into density matrices in another
one, which are physical in the sense they are combinations of operations such
as unitary evolution, reduction to a subsystem and enlarging the system with a new
independent subsystem. The general expression of a CPTP map is 
\begin{equation}
\rho^\prime=\sum_i M_i \rho M_i^\dagger\,,\hspace{2cm} \sum_i M_i^\dagger M_i=I\,,
\end{equation}
for matrices $M_i$ with arbitrary dimension, i.e., not necessarily square
matrices.

The monotonicity of relative entropy can be generalized to CPTP maps. We have
\begin{equation}
S(\rho_1|\rho_0)\ge S(\rho_1^\prime|\rho_0^\prime)\,.
\end{equation}
The monotonicity property (\ref{mon}) is a particular case of monotonicity under CPTP maps.
 Such CPTP maps then generally entail the loss of distinguishability
between states and therefore are typically irreversible. 

The second law states that the entropy of an isolated system cannot decrease.
Of course, a completely isolated system in quantum mechanics evolves unitarily
and the entropy does not change. We have to soften the condition of being
completely isolated to allow for some interchange of information with
the ambient space, and we have to assume this evolution law actually exists. 
As a model for this evolution let us consider the case of a fixed quantum
system with state $\rho(t)$ evolving under  CPTP maps. Assume, following the idea of an ``isolated'' system, that the total energy $E$ is
conserved. Also assume that time evolution preserves a thermal equilibrium
state $\rho_T=e^{-H/T}/\textrm{tr}(e^{-H/T})$ at
some temperature $T$. 
Then the relative entropy $S(\rho(t),\rho_T)$ is decreased by the CPTP
evolution, and we have for $t_1<t_2$,
\begin{equation}
F(\rho(t_2))-F(\rho_T)<F(\rho(t_1))-F(\rho_T)\,,
\end{equation}
where we used that the thermal state is invariant under time evolution.
Expressing this relation in terms of entropy and energy, and considering all
the involved energies are the same by assumption, we have
\begin{equation}
S(t_2)>S(t_1)\,,
\end{equation}
as required by the second law of thermodynamics. 

Another ``microcanonical'' version of this is
when the totally random state $\rho_0=I/n$, where $n$ is the dimension of the
Hilbert space is preserved under a CPTP evolution. 
 The second law follows from the fact that
the relative entropy is in this case
\begin{equation}
S(\rho(t)|\rho_0)=\log(n)-S(\rho(t))\,.
\end{equation}
Then, the increase in entropy follows again by the decrease of relative entropy.

A CPTP map can be thought of as a unitary evolution in the full system starting from a decoupled initial state of the system and the environment (the state of the environment being fixed for different states of the system, otherwise we would not have a map of states for the subsystem itself). This is where the irreversibility leaks in. So these proofs of the second law secretly contain the ``Stosszahlansatz'' (molecular chaos) assumption of Boltzmann's H theorem.

\subsection{Purification and Tomita-Takesaki theory}
\label{tt}

According to the context, the idea of purification is called with different names, such as GNS (Gelfand-Naimark-Segal) construction, or thermofield double. The question is the following. Given a system with an impure state can we always think of this system as a subsystem of a larger one, where the state is pure? The answer is yes. 

Let us see this in the simplest case of a full matrix algebra ${\cal A}$, which is the algebra of operators on a Hilbert space ${\cal H}_1$. Let a general density matrix be 
\be
\rho=\sum_i p_i |i\rangle \langle i |\,,\label{dm}
\ee
 where $p_i$, are the eigenvalues of $\rho$ and $|i\rangle$ the corresponding eigenvectors. We can take a Hilbert space ${\cal H}={\cal H}_1\otimes {\cal H}_2$ with ${\cal H}_2$ a copy of ${\cal H}_1$, and define the vector state 
\begin{equation}
\vert \Omega \rangle = \sum_i \sqrt{p_i}  \vert i \,\tilde{i}\rangle\,.\label{laba}
\end{equation}
We have $\rho=\textrm{tr}_{{\cal H}_2}\vert \Omega\rangle \langle \Omega \vert$, which is the reduced density matrix to the ${\cal H}_1$ factor. This proves that purification can be achieved. 
The orthonormal base $\{\vert\tilde{i}\rangle\}$ for ${\cal H}_2$ in (\ref{laba}) is arbitrary, and different basis correspond to different purifications of $\vert \Omega \rangle$ in the same space. These different basis correspond to unitary transformations in ${\cal A}'$. A vector $\vert \Omega\rangle$ in a tensor product space can always be written in the form (\ref{laba}). This is called the Schmidt decomposition of $\vert \Omega \rangle$ in ${\cal H}_1\otimes {\cal H}_2$.

A less explicit but more pretty way to arrive at this same purification is called the GNS construction. It is very general and mathematically natural. One labels the vectors of the bigger Hilbert space with the names of the elements of the algebra: $a \in {\cal A}\rightarrow |a\rangle$, and defines the scalar product through the state $\omega$ as $\langle a | b\rangle = \omega(a^\dagger b)$. The representation of the algebra on this Hilbert space is $a|b\rangle=|ab\rangle$, and the state $\omega$ is represented simply by the vector $|1\rangle$, $\langle 1| a|1\rangle=\omega(a)$. For a matrix algebra of dimension $n\times n$ the Hilbert space of the GNS purification has in general dimension $n^2$.

To construct a density matrix written like (\ref{dm}) one can produce an ensemble of the different pure states according to the probabilities drawn from a dice. The fact that this same state in ${\cal A}$ comes from a reduced pure state in a bigger space reminds us that this noisy ensemble is indistinguishable from the purely quantum noise produced by entanglement with other systems. Purification also tells us that entropy can always be interpreted as entanglement entropy with another part of the universe.

If the eigenvalues $p_i$ are all different from zero, the purification of the state $\omega$ on ${\cal A}$ comes together with the modular operators $\Delta$ and $J$. These are defined as
\begin{eqnarray}
\Delta&=&\sum_{i,j} \frac{p_i}{p_j}\vert i \,\tilde{j}\rangle\langle i\, \tilde{j}\vert\,,\\
J&=&\sum_{i j} \vert i \,\tilde{j}\rangle \langle j\,\tilde{i}\vert  *\,.\label{jota}
\end{eqnarray}
The modular reflection $J$ is an antiunitary operator product of a transposition of the basis of ${\cal H}_1$ and ${\cal H}_2$, with the operator $*$ of complex conjugation of the vector components written in the basis $\{\vert i \tilde{j} \rangle\}$. It is independent of the actual eigenvalues $p_i$. The modular operator $\Delta$ is positive and can be written $\Delta=\rho\otimes \rho^{-1}$. There are several magic rules for these operators that can be checked from the definition
\be
\Delta \vert \Omega \rangle=J \vert \Omega \rangle=\vert \Omega \rangle\,,\hspace{.5cm} J \Delta = \Delta^{-1} J\,, \hspace{.5cm} J^\dagger=J=J^{-1}\,. 
\ee 
 Defining the operator $S=J \Delta^{\frac{1}{2}}=\Delta^{-\frac{1}{2}} J$ we have 
\be
S a |\Omega \rangle = a^\dagger |\Omega \rangle\,, \hspace{.5cm} a \in {\cal A}\,.
\ee
Interestingly, the modular reflection maps the algebra in its commutant (and vice versa) 
\be
J {\cal A} J= {\cal A}'\,.
\ee
This allows us to define for any operator of $a\in {\cal A}$ a reflected copy $\bar{a}= J a J \in {\cal A}'$ which satisfies a ``reflection positivity'' relation
\begin{equation}
 \langle \Omega\vert a \bar{a} \vert \Omega \rangle=\langle \Omega\vert a J a \vert \Omega \rangle=\langle \Omega\vert a \Delta^{\frac{1}{2}} S a \vert \Omega \rangle=\langle \Omega\vert a \Delta^{\frac{1}{2}} a^\dagger \vert \Omega \rangle\geq 0\,.\label{rere}
 \end{equation}

These operators and relations are what is called Tomita-Takesaki theory. One interest of this theory is that the scope extends to infinite dimensional algebras of the type associated to regions in QFT. So these operators are well defined in QFT even if density matrices will not be well defined. The operator $J$ is related to CPT symmetry in QFT, but as we see can be defined for much more general quantum systems. The condition we imposed that the density matrix should have no zeros is described in a more general context saying that the vector $|\Omega\rangle$ is 
 cyclic and separating for ${\cal A}$: a vector such that ${\cal A} \vert \Omega \rangle$ span the whole Hilbert space and such no $a\in {\cal A}$, $a\neq 0$,  can annihilate the state, $a\,|\Omega \rangle \neq 0$. The modular flow intrinsic to the algebra described in (\ref{modflow}) can be implemented in the space of purification with the operator  
\be 
 U(\tau)=\Delta^{i \tau}\,,\hspace{.5cm}
 U(\tau)\,{\cal A}\, U(-\tau)= {\cal A}\,,\hspace{.5cm}
 U(\tau)\,{\cal A}'\, U(-\tau)= {\cal A}'\,,\hspace{.5cm} J U(\tau) J=U(\tau)\,,\hspace{0.5cm} U(\tau)|\Omega \rangle=|\Omega \rangle\,.
 \ee   

\subsection{Exercises}

\begin{itemize}

\item[1.-] Jaynes describes the canonical ensemble as a state of which our only knowledge is the expectation value of the energy. Our uncertainty about other characteristics of the state should force us to choose the maximal entropy state with $E$ fixed. Show that if $\langle O_1\rangle, \langle O_2\rangle,\cdots, \langle O_n\rangle$ are fixed expectation values for Hermitian operators, the maximal entropy state is of the form
\be
\rho=c\, e^{-\sum_{i=1}^n \lambda_i \,O_i } 
\ee  
where the ``chemical potentials'' $\lambda_i$ have to be fixed such as to reproduce the expectation values $\langle O_i\rangle$. How does this formula change if the operators are not Hermitian?

\item[2.-] Consider the algebra ${\cal J}$ generated by the operators of angular momentum $J_i$, $i=1,2,3$ (it contains all polynomials in these operators). We know the algebra acts on a space where there are only representations of angular momentum $j=0,1,2$. Given the expectation values 
\begin{equation}
\langle \vec{J}^2\rangle= \frac{8}{3}\,,\hspace{.6cm}
\langle (\vec{J}^2)^2 \rangle = \frac{40}{3}\,,
\end{equation}   
compute the entropy of ${\cal J}$ if: a) the state is rotational invariant; b) $\langle J_z\rangle= 1$. 
 
 \noindent (Answer: a) $4/3 \log(3)+ 1/3 \log(5)$;   b) $\log(3)$.)

\item[3.-] Consider the euclidean path integral for a scalar field $\phi$ with a source and think the averages on the path integral as averages over classical probability distributions in the space of fields. We can define two probability distributions, with and without source
\be
P[\phi]= Z[0]^{-1} e^{-S[\phi]}\,, \hspace{.7cm} P_J[\phi]= Z[J]^{-1} \, e^{-S[\phi]+(J\cdot \phi)} \,.
\ee
Assume the average $\langle \phi\rangle_{J=0}=0$. Show that the relative entropies 
\bea
S(P|P_J)=W[J]\,, \hspace{1cm}
S(P_J|P)=\Gamma[\phi_{\textrm{cl}}]\,,
\eea
where $W[J]=\log(Z[J]/Z[0])$ is the euclidean free energy (normalized such that $W[0]=0$) and $\Gamma[\phi_{\textrm{cl}}]=(J\cdot \phi_{\textrm{cl}})-W[J]$ is the euclidean effective action, where $\phi_{\textrm{cl}}(x)=\frac{\delta W[J]}{\delta J(x)}=\langle \phi(x)\rangle|_J$. The QFT functionals are indeed relative entropies too!  

\end{itemize}

\subsection{Notes and references}
 von Neumann algebras, density matrices, and entropic quantities, as well as their properties and applications, are described in the books \cite{ohya2004quantum,petz2007quantum}. The older review \cite{wehrl1978general} is very useful. For a QIT perspective see for example \cite{nielsen2002quantum,vedral}. Several proofs of the second law are reviewed in \cite{sagawa2013second}. General von Neumann algebras, including the Tomita Takesaki theory, and applications to QFT, are introduced in \cite{Haag:1992hx}. Jaynes paper mentioned in exercise $1$ is \cite{jaynes1957information}.

\newpage
 
\section{Entanglement entropy in QFT}
\label{EEQFT}

Quantum field theory has different formulations, and all of them are useful for some purposes. The basic one is to define it by a collection of field operators $\phi(x)$ with certain properties.  The operators of the quantum theory, however, are not the $\phi(x)$, which are too singular. For example, if $\phi(x)$ were an operator, $|\psi\rangle=\phi(x)|0\rangle$ would be a vector, and we know that $\langle \psi|\psi \rangle= \langle 0|\phi^\dagger(x)\phi(x)|0\rangle=\infty$ because correlations diverge at the coincidence points. Real operators (as opposed to the operator values distributions $\phi(x)$) are defined by smearing the field 
\be
\phi_\alpha=\int d^dx\, \alpha(x)\, \phi(x)\,,
\ee
where the smearing function $\alpha(x)$ has compact support and is smooth enough. In general, we have to smear in a $d$ dimensional region ($d$ being the space-time dimension), and this is always enough to produce an operator, but for some fields, such as a free scalar, smearing in a spatial $d-1$ dimensional surface suffices.

It is well known that the quantum fields and the Hilbert space can be reconstructed from the vacuum expectation values of products of fields, this is, the correlation functions. This is Wightman's reconstruction theorem,  
\be
\{\phi(x)\,, \, {\cal H}\} \leftrightarrow \langle 0| \phi(x_1)\cdots \phi(x_n)|0\rangle\,.  \label{recon}
\ee
This tells us the full information of the theory is contained in the statistics of vacuum fluctuations. 

\subsection{Formulation in terms of local algebras}
Let us take a region $W$ of spacetime and consider the algebra ${\cal A}(W)$ of operators localized in $W$. We can build it from the smeared operators $\phi_\alpha$ with the support of $\alpha$ included in $W$. There is a minor technical point here. In general $\phi_\alpha$, even if it is an operator, it is not a bounded one. That means that its spectrum is not bounded, and there are some vectors in the Hilbert space which are not in the domain of $\phi_\alpha$. Hence, issues of domain interpose to multiply operators and form an algebra. But this is not a problem: it is enough to take the projectors in a spectral decomposition of $\phi_\alpha$ or to consider $e^{i \phi_\alpha}$ for hermitian $\phi_\alpha$, which is unitary and bounded.

\begin{figure}[t]
\begin{center}  
\includegraphics[width=0.50\textwidth]{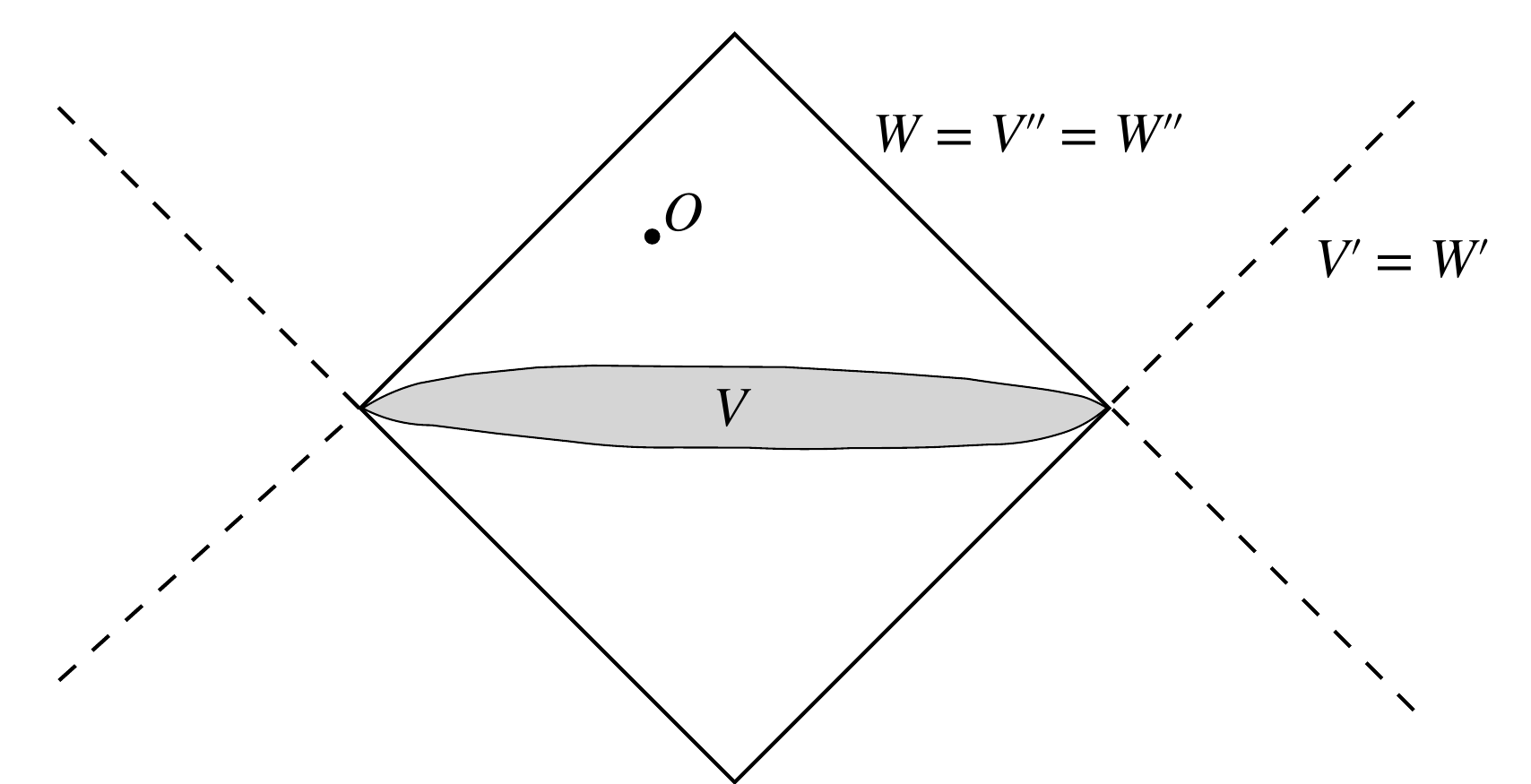}
\captionsetup{width=0.9\textwidth}
\caption{The causal complement $V'$ of the region $V$ (shaded), and the causal completion $V''=W$. }
\label{causal}
\end{center}  
\end{figure}

Haag-Kastler description of a QFT takes as fundamental objects the algebras ${\cal A}(W)$ instead of fields. This is the natural setting of investigations about EE. Two absolutely minimal conditions have to hold for these algebras. The first is that operators localized in a region have to be so in a larger one   
\be
V\subseteq W\rightarrow  {\cal A}(V) \subseteq {\cal A}(W)\,.
\ee
The second is that operators localized in spatially separated regions should commute to each other. Otherwise, we could be able to send superluminal information (see exercise 1). To write this with a nice notation let us define the causal complement of a region $V$ as (see figure \ref{causal})
\be
V'=\{x | x \,\,\textrm{spacelike to }y\,, \,\,\, \forall y \in V\}\,.  
\ee
Then the causality axiom writes 
\be
W\subseteq V'\rightarrow {\cal A}(W)\subseteq ({\cal A}(V))'\,.
\ee
This does not hold as such for fermions, but the usual extension replacing commutativity by anticommutativity covers the case of fermion operators.

In general, not all pairs of different regions have different algebras. Looking at figure \ref{causal}, we can think that an operator $O$  localized in the diamond-shaped region $W$ would belong to the algebra of operators generated by the thin region $V$ around the surface $t=0$. Heuristically, this is because Heisenberg operators should obey some causal equations of motion that would allow us to determine them at $t>0$ from the initial data at $t=0$. If our theory has a stress tensor, we could construct an operator, integrating $T_{00}$ at $t=0$ in the region $V$, that acts locally as the Hamiltonian inside $V$. Then, evolving with it, we could push the operator to larger times. We will assume that this causal evolution holds, and consequently only be interested in regions with a diamond-shaped form.\footnote{There are however counterexamples to this assumption. For example, a generalized conformal free field $\phi_\Delta$ of conformal dimension $\Delta$, as the ones that are dual to free bulk massive fields in the large $N$ limit of holographic theories. These do not have a stress tensor.} We will call these regions causal regions.  Their technical definition is regions such that $W=W''$. They are the domain of dependence of pieces of space-like surfaces. 

\begin{figure}[t]
\begin{center}  
\includegraphics[width=0.29\textwidth]{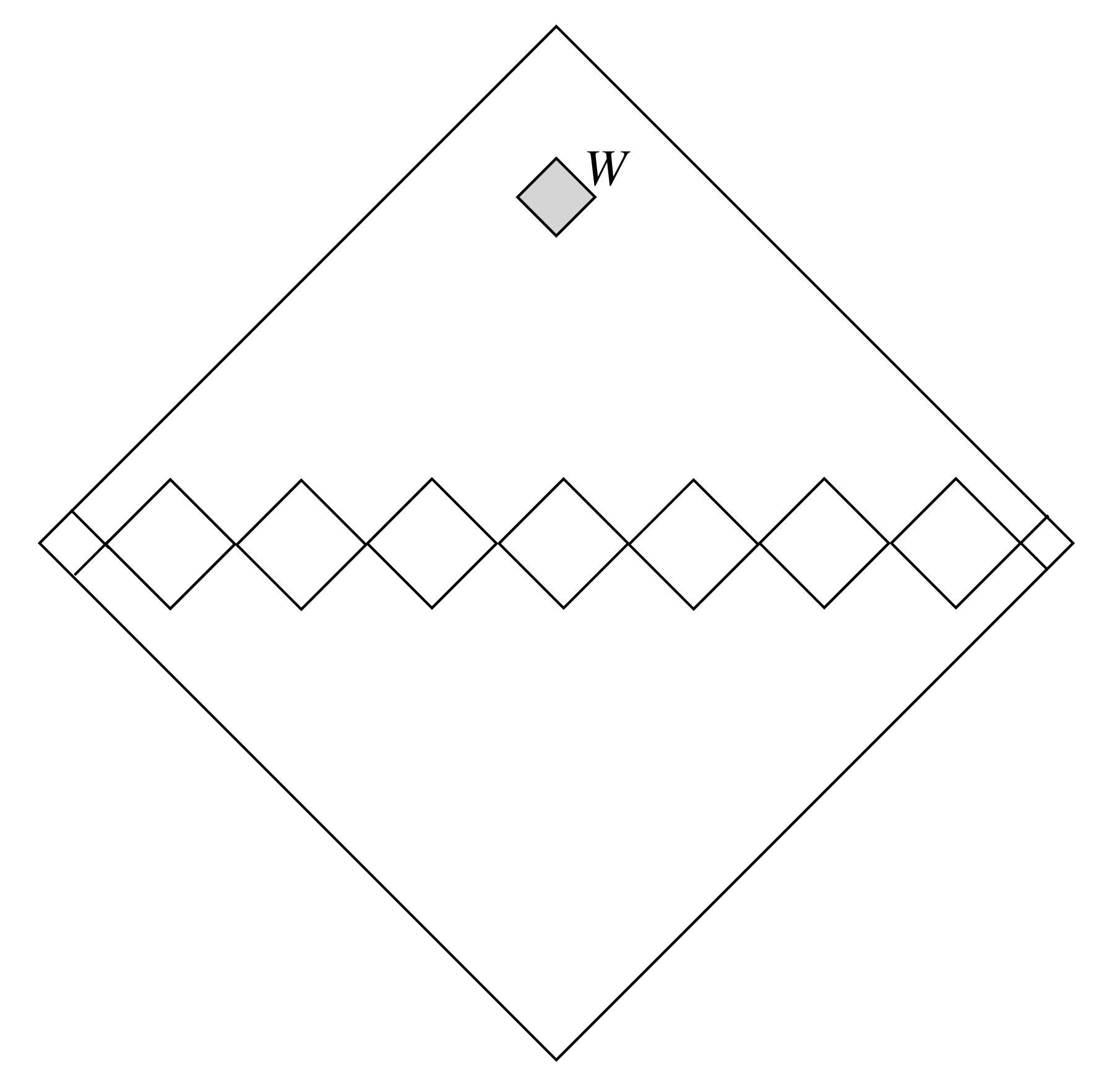}
\captionsetup{width=0.9\textwidth}
\caption{Algebras of small regions near $t=0$ generating the algebra containing $W$.}
\label{causal1}
\end{center}  
\end{figure}

Where is the information of the particular theory in this perspective? It turns out that is not in the algebras themselves. All local algebras for all theories in all dimensions are supposed to be isomorphic mathematical objects, von Neumann algebras of type III$_1$. This should not surprise us much; after all, the harmonic oscillator, the hydrogen atom, and a scalar field have isomorphic Hilbert spaces and global algebras (all bounded operators of these Hilbert spaces). The information of the QFT is encoded in the relations of the different algebras with each other. That is, in the net of algebras, the way they intersect each other and share operators. In figure \ref{causal1},  ${\cal A}_W$ is supposed to be included in the algebra generated by the ones of all the small diamonds around $t=0$. But the way it is generated by the algebras at $t=0$ should depend on the dynamics.

A sharper way of differentiating models is to evaluate ``correlations'' between the algebras in the vacuum. This is precisely what the mutual information $I(V,W)$ between two spatially separated regions does. A natural unsolved question is the analogue to the reconstruction (\ref{recon}),
\be
\{{\cal A}(W), {\cal H}\}\leftrightarrow I(V,W) \hspace{1cm}?  
\ee

The definition of a causal region $W=W''$ fits nicely with the one of von Neumann algebras ${\cal A}''={\cal A}$.  One could wonder about a pre-established relation between space-time causal structure and operator theory. About the important theme of the possible relations between algebras and regions, the very few things that are actually rigorously known will not satisfy our wishes. However, it seems that symmetries are encoded in simple degradations of the most perfect possible relations between algebras and regions. We will come back to this issue in the last chapter.    

\subsection{Entropy and the continuum limit: a free field example}

Another way to think about QFT that is common in the mind of most physicists is to think in a discrete model, such as a lattice, and then take the continuum limit, putting more points in the lattice and at the same time, doing, in general, judicious changes in the lattice model,  aiming to arrive at some specific QFT for the long-distance physics. Alternatively, we can think of a theory with a different sort of cutoff, and take the limit by removing the cutoff. There may be many ways to cut off a theory, but all of them should arrive at the same QFT, say, to the same correlation functions. In this process, there may be quantities that make sense for the discrete model but that diverge as the cutoff is removed. Or they may not diverge but depend on the regularization that was used. Only the quantities that are well defined in the limit belonging to the continuum theory. For example, the question of how many degrees of freedom there are in a given volume does not make sense for any QFT. With the EE something similar happens: it is always divergent as we remove the cutoff. However, some interesting pieces can be extracted that make perfect sense in the continuum.    

To make this more concrete and visualize the continuum limit we can consider some models where the entropy is easy to evaluate in a computer. This is the case with free fields. If we have a free scalar in a lattice, the operator content is described by the field and momentum ${\phi_a,\pi_a}$ at different points of the lattice, labelled by $a$. The algebra is defined by the relation $[\phi_a,\pi_b]=i \delta_{ab}$. For a quadratic Hamiltonian, the fundamental state is {\sl Gaussian}. A Gaussian state is one where the two-point correlation function has all the information on the state. The multi-point correlators are given by combinations of products of the two-point functions (as happens for Gaussian probability distributions). In QFT this is usually called Wick's theorem for vacuum correlators.

In this situation, it is quite easy to evaluate the entropy from the two-point function. For a free scalar, we have for a region $W$ in the lattice
\begin{equation}
S(W)=\textrm{tr}\left(( C+1/2)\log (C+1/2)-(C-1/2)\log (C-1/2)\right)\,,\hspace{.3cm} C=\sqrt{X P}\,,  \label{for}
\end{equation}
where $X$, $P$, are the matrices of correlators in the region $W$,
\begin{equation}
X_{ab}=\left\langle \phi _{a}\phi _{b}\right\rangle|_{a,b \in W}  \,, \hspace{.6cm}  P_{ab}=\left\langle \pi _{a}\pi _{b}\right\rangle|_{a,b \in W}  \,. 
\end{equation}

Analogous formulas hold for free fermion Gaussian states. Notice the huge reduction in computation for Gaussian states. For $N$ lattice points we have to evaluate eigenvalues of a $N\times N$ matrix, while the density matrix is $2^N\times 2^N$ for fermions and infinite-dimensional for bosons. Non-Gaussian states, even for free theories, are not simple to treat numerically.  

To use this formula we need vacuum correlators. These we obtain from the Hamiltonian. For a free scalar field, setting the lattice spacing to $1$, we have a discrete Hamiltonian
\be
H=\frac{1}{2}\sum \pi _{a}^{2}+\frac{1}{2}\sum_{a}m^2\, \phi_a^2+\frac{1}{2}\sum_{a<b, |a-b|=1}    (\phi_a -\phi_{b})^2=\frac{1}{2}\sum \pi _{a}^{2}+\frac{1}{2}\sum_{a,b}\phi _{a}K_{ab}\phi
_{b}\,.
\ee
The vacuum (fundamental state) correlators in the full lattice are 
\begin{equation}
\left\langle \phi _{a}\phi _{b}\right\rangle =\frac{1}{2}(K^{-
\frac{1}{2}})_{ab}\,, \hspace{.5cm}
\left\langle \pi _{a}\pi _{b}\right\rangle =\frac{1}{2}(K^{\frac{1
}{2}})_{ab}\,, \label{p}
\end{equation}
which follows by diagonalizing the matrix $K$ (in the full lattice!). In particular, for a square lattice, the correlators are given by integrals in momentum space. For example, for a two dimensional infinite lattice (space-time dimension $d=3$) we have, writing $a=(i,j)$, $i,j \in Z$, for the lattice points,  
\begin{eqnarray}
\small
\hspace{-.7cm}\langle \phi_{(0,0)}\phi_{(i,j)}\rangle &=&\frac{1}{8\pi^2}\int_{-\pi}^{\pi}dk_1 \int_{-\pi}^{\pi}dk_2\frac{\cos(i k_1)\cos(j k_2)}{\sqrt{2
(1-\cos(k_1))+2
(1-\cos(k_2)+m^2)}}\,, \\
\hspace{-.7cm}\langle \pi_{(0,0)}\pi_{(i,j)}\rangle &=& \frac{1}{8\pi^2}\int_{-\pi}^{\pi} dk_1\int_{-\pi}^{\pi}dk_2 \cos(i k_1)\cos(j k_2)}{\sqrt{2
(1-\cos(k_1))+2
(1-\cos(k_2))+m^2}\,.
\end{eqnarray}

We need the correlators only in the region $W$ of the lattice if we are interested in computing the entropy in $W$. If we are interested in a region $W$ in the continuum limit, we have to take a region of the shape of $W$ in the lattice and make it grow by scaling its size. This is equivalent to making the lattice denser. If the field is massless and free this is all. If there is a mass, however, we have to aim for a certain fixed mass in the continuum. If the size of $W$ is $L$, then we have to take for example $L m$ fixed as we grow $W$ in the lattice, changing the mass of the lattice model to smaller values in lattice units as we scale the region. An analogous (but more complicated) renormalization of Hamiltonian coupling constants is necessary for interacting models.

\begin{figure}[t]  
\begin{subfigure}
\centering
\includegraphics[width=0.22\textwidth]{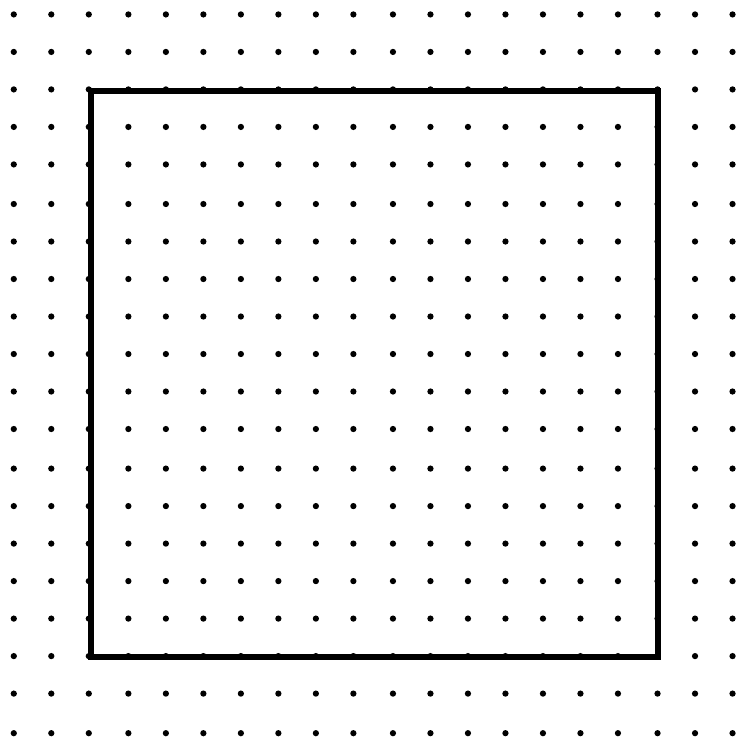}
\end{subfigure}
\begin{subfigure}
\centering
\includegraphics[width=0.34\textwidth]{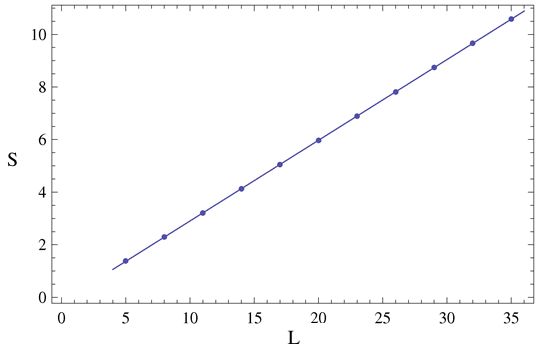}
\end{subfigure}
\begin{subfigure}
\centering
\includegraphics[width=0.4\textwidth]{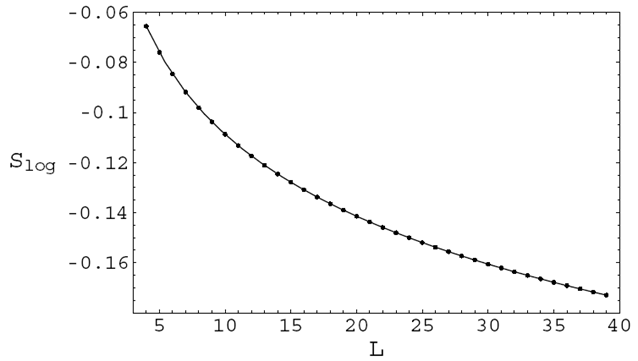}
\end{subfigure}
\captionsetup{width=0.9\textwidth}
\caption{A square on the lattice. Area law and logarithmic term.}
\label{square}
\end{figure}

Let us see what we get for a massless field. Figure \ref{square} shows the entropy of a square as we make the side $L$ larger in the lattice, or, equivalently,  the cutoff smaller. The entropy is shown in the middle panel and is clearly growing linearly with $L$. As the boundary of the square is proportional to $L$ this is called an {\sl area law} (as opposed to a volume law for the entropy of thermal states. The terminology is adapted to three-dimensional space). But this is not the end of the story. Hardly visible in the entropy linear behaviour is a sub-leading logarithmic term that can be extracted by fitting the data and subtracting the linear term. This is shown in the right panel of figure \ref{square}. Plugging back the lattice spacing, that we call $\epsilon$, we then get
\be
S= 0.75\, \frac{\textrm{4 L}}{\epsilon}-0.047\, \log(L/\epsilon)+\textrm{cons}=0.75\, \frac{\textrm{perimeter}}{\epsilon}-0.047\, \log(L/\epsilon)+\textrm{cons}\,.
\ee
This shows the continuum limit $\epsilon\rightarrow 0$ is divergent as $\epsilon^{-1}$ in this $d=3$ example. 

\begin{figure}[t]  
\begin{subfigure}
\centering
\hspace{3.3cm}\includegraphics[width=0.25\textwidth]{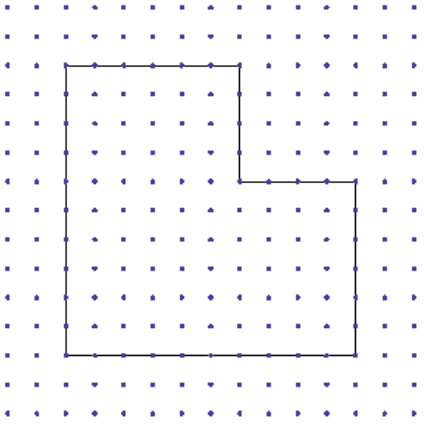}
\end{subfigure}
\hspace{2cm}
\begin{subfigure}
\centering
\includegraphics[width=0.25\textwidth]{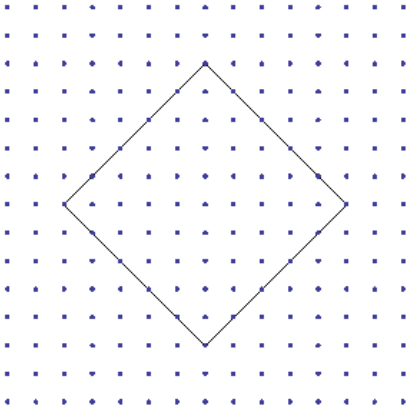}
\end{subfigure}
\captionsetup{width=0.9\textwidth}
\caption{Left: a shape with $6$ right angle corners. Right: a square rotated with respect to the lattice.   }
\label{muchos}
\end{figure}

Now we can look at other shapes. For example, the left panel of figure \ref{muchos} gives 
\be
S=  0.75\, \frac{\textrm{perimeter}}{\epsilon}-6\frac{0.047}{4}\, \log(L/\epsilon)+\textrm{cons}\,.
\ee
Notice the area term has the same coefficient, but the logarithmic term is increased by $6/4$. We interpret this as arising from the fact that the new shape has $6$ corners instead of the $4$ the square had. One can check drawing shapes with more corners that the logarithmic coefficient scales with the number of corners.

Now we check if this result depends on the way the regions are placed concerning the lattice. The right panel of figure \ref{muchos} gives
\be
S=  0.85\, \frac{\textrm{perimeter}}{\epsilon}-0.047\, \log(L/\epsilon)+\textrm{cons}\,.
\ee 
This tells something bad about the area term (perimeter in $d=3$). It does not have the rotational symmetry that the continuum theory has. It never forgets the lattice. These terms are called non-universal because they depend on the way (the kind of lattice for example) we arrive at the continuum. On the other hand, the logarithmic coefficient seems to be independent of the regularization. This is called a {\sl universal} number. It belongs to the continuum theory.  

\subsection{General structure of the EE}
\label{ge}
Let us pause for a moment and think about the physical origin of these results. It is evident that the area term depends on the boundary of the region. It has to be generated by the entanglement of modes very near the boundary, across the two sides of the boundary. As we make the cutoff smaller, more and more of these modes appear, and this accounts for their contribution being divergent with $\epsilon\rightarrow 0$. They are short distance modes, very correlated with other short wavelength UV modes nearby, but practically uncorrelated with distant modes. Then, it is natural that their contribution is independently added along the boundary and gives an area increasing term. The divergent logarithmic term should have a similar ultraviolet origin, but due to features of UV entanglement near the vertices. 
This explains why it scales with the number of vertices. In fact, we can expect (and check) that its general contribution would be of the form
\be
S_{\log}=\sum_{v} s(\theta_v) \log(\epsilon)\,,
\ee
where the sum is over the different vertices, $\theta_v$ is the vertex angle, and $s(\theta)$ is a dimensionless universal function. Vertices do not have associated any dimensionful geometrical quantity to compensate for a power of the cutoff.  The $\log(\epsilon)$ behaviour comes from the scale invariance of the model at short distances and occurs even if we take a massive scalar. The reason is that as we zoom in the vertex region by a fixed amount, we find almost independent modes of the scaled size that contribute the same to the entanglement associated with the vertex feature. The number of contributions grows with the number of zooms that fit in the geometry. This grows with the logarithm of the cutoff that marks the shortest wavelength available.  

 But why is this coefficient universal (independent of the choice of cutoff) if of ultraviolet origin? On one side, it is very difficult to modify the coefficient of a logarithmic law by changing the definition of $\epsilon$. But more importantly, to compensate for the log dimensions, $\log(\epsilon)$ should be accompanied by a $\log(L)$ term,  with opposite coefficient. This is now a finite term measuring entanglement in large scale physics, and dependent on the correlations of the continuum QFT itself. Therefore, it must correspond to a universal term. Notice we write $\log(L)$ for some scale $L$ of the region. This accounts for how it changes under scaling the region,  but we do not know how this term precisely depends on its shape. This may be quite complicated. Any change in the shape changes the constant term. We have said nothing about the constant term, but it is the one containing most of the universal information.       

With these heuristic ideas in mind, we can try to imagine what we can expect in more general cases of QFT at different dimensions, conformal or with mass scales,  for different region shapes, etc. Let us call $\Sigma$ to the boundary of the region $W$. The divergent terms should come as a series in inverse powers of $\epsilon$ (as $\epsilon$ is the cutoff, we always expand for $\epsilon\rightarrow 0$ and eliminate all positive powers) accompanied with coefficients that depend on the boundary of the region,
\be 
S(W)= \sum_i C_i(\Sigma) \times \epsilon^{-\lambda_i}+S_0(W)\,. 
 \ee 
The coefficients  $C_i(\Sigma)$ should be: a) local and extensive on the boundary, or in other words, they should be expressed as integrals on the boundary, b) depend only on UV physics and the geometry, c) independent of the state because all reasonable states look like the vacuum at very short distances, and d) have dimension $\lambda_i$. The finite term $S_0$ does depend on the state and is non-local.     
 
Suppose further that we have a nice ``geometric'' cutoff to the continuum model itself, that does not introduce new geometrical tensors, as the lattice directions would introduce (of course we could as well consider these tensors introduced by the lattice in the analysis below). Take the UV theory that describes the short scale physics to be a CFT perturbed by a relevant operator $\phi_{UV}$ of dimension $\frac{d-2}{2}\le \Delta_{UV} < d$.\footnote{The first inequality is called the unitarity bound: no field can have less dimension than the free massless scalar. The second is the condition for the operator to be relevant and is needed for the perturbed theory to converge to the CFT at short scales.} That is,  the action is perturbed as
\be
S_{QFT_{UV}}=S_{CFT_{UV}}+g_{UV} \int d^dx\, \phi^{UV}_\Delta\,, 
\ee
with coupling constant $g_{UV}$ of dimension $ [g_{UV}]=d-\Delta_{UV}>0$. We also take the surface $\Sigma$ smooth. Then the structure of the divergent terms is perturbative in $g_{UV}$ in the UV. The possible divergent terms are of the form 
\be
 \sum \epsilon^{-\lambda} g_{UV}^n \times \int_\Sigma d\sigma\, (\textrm{curvatures})\,, \label{ssaa}
\ee
where $n\ge 0$ is an integer, and the total dimension of the term is $0$. The curvatures can be polynomials in the spacetime curvature tensor at the point (which are zero in flat space) or in the extrinsic or intrinsic curvatures of the surface. Note not many of these types of terms are expected in general, because the more curvatures and the more powers of $g_{UV}$ the dimensions increase, making $\lambda$ smaller. The highest possible $\lambda$ is for $n=0$ and no curvatures, which just gives an area term, $\epsilon^{-(d-2)}\,\int_\Sigma d\sigma$. This is the reason the area term is always the most divergent possible term. Another simplification arises in most cases from the idea that the entropy in a pure state is the same for complementary regions (we will come back to this point in the last lecture). This implies the curvature terms in (\ref{ssaa}) have to have even dimensions, to be invariant under changes of orientation on the surface.   

Let us see this more concretely in some examples. Let us take a spherical ball in $d=4$ and analyse different regimes of the vacuum EE.

\subsubsection*{CFT}

For a CFT we have no scales except the radius $R$ of the sphere and the cutoff.
According to the above, we have
\be
S=c_2 \frac{\int_\Sigma d\sigma}{\epsilon^2}+ 2 A\, \left(\int_\Sigma d\sigma\, \hat{R}\right) \log(\epsilon) - 4 A\log(R) +c_0=c_2 \frac{4 \pi R^2}{\epsilon^2}+ 4 A \, \log(\epsilon) - 4 A\log(R) +c_0\,.
\ee 
The logarithmic term is allowed because we have the intrinsic curvature $\hat{R}$ of the sphere, with dimension $2$, whose integral over the sphere is just $2$ (the Euler number). We will show the coefficient $A$ is the so-called ``trace anomaly'', more correctly the coefficient of the Euler term in the trace anomaly, and is a universal quantity of the CFT.  For a scalar field, for example, the logarithmic coefficient results $4A=\frac{1}{90}$. The constants $c_2$ and $c_0$ are non-universal, this last one changes under redefinition of $\epsilon$ due to the logarithmic term.  

More generally, for any even-dimensional CFT, there is a logarithmic term with a universal coefficient in the entanglement entropy of a smooth surface. This is allowed by dimensional analysis: we need to integrate curvature polynomial with dimension $(d-2)$ over the $(d-2)$-dimensional boundary surface.  On the other hand, smooth surfaces in odd-dimensional CFT's have no logarithmic contribution because integrating over the boundary a polynomial of the curvature with even dimensions will not give a dimensionless quantity. In these odd-dimensional cases, the universal coefficient in the expansion of the entropy of a sphere is the constant term. This is universal because it cannot be changed by integrals of local terms in the boundary, representing contributions from the cutoff. 

 For regions different from a sphere in $d=4$, there is another independent logarithmic term  proportional to the integral of extrinsic curvature terms on the surface (more terms appear in higher dimensions). The coefficient of $\log(\epsilon)$ of the new contribution is proportional to another universal trace anomaly coefficient $C$. Solodukhin's formula for the logarithmic coefficient is
\begin{equation}
c_{\textrm{log}}=2A \,\chi(\partial V)+\frac{C}{2 \pi}   \int_{\partial V}(k_i^{\mu \nu}k^i_{\nu \mu} -\frac{1}{2} k_i^{\mu \mu}k^i_{\mu \mu})\,.\label{general}
\end{equation}
Here $\chi(\partial V)$ is the Euler number of the surface,  $k^i_{\mu\nu}=-\gamma^\alpha_\mu \gamma^\beta_\nu \partial_\alpha n^i_\beta$ is 
the second fundamental form, $n^\mu_i$ with $i=1,2$ are a pair of unit vectors orthogonal to $\partial V$, and $\gamma_{\mu\nu}=\delta_{\mu\nu}-n^i_\mu n^i_\nu$ is 
the induced metric on the surface. 
For example, for a cylinder of length $L$ and radius $R$, $L\gg R$, the coefficient is $c_{\textrm{log}}=\frac{C}{2}\frac{L}{R}$, which for a free massless scalar field gives $c_{\textrm{log}}=\frac{L}{240 R}$.

\subsubsection*{QFT-UV}
Consider the UV regime of a $d=4$ QFT, that is, we select a sphere with a short radius $R$ in a theory with mass scales, but which becomes asymptotically conformal for short scales.  We have the previous result but corrected by terms perturbative in the coupling constant $g_{UV}$. This coupling constant with positive dimensions, determine the departure of the theory from a CFT as we move to larger scales. The perturbative series usually starts with $g_{UV}^2$. We have 
\be
S=(4 \pi R^2)\left(c_2 \,\epsilon^{-2}+ c_{2}'\,g_{UV}^2 \,\epsilon^{2 \left(\Delta_{UV} -(d+2)/2\right)} +\cdots\right)- 4 A_{UV} \, \log(R/\epsilon) +c_0 +c_0'\, g_{UV}^2 \, R^{2(d-\Delta_{UV})}+\cdots\,.\label{uvi}
\ee
The area term can receive new divergent contributions only if $\Delta_{UV}\ge 3$ (or $\Delta_{UV}\ge (d+2)/2$ in general dimensions)\footnote{As we will explain later, the corrections to the area term are controlled by correlators of traces of the stress tensor. The definition of the dimension $\Delta_{UV}$ in this formula changes for free scalars where a non-improved stress tensor must be used.}. We have renamed $A$ to $A_{UV}$ to have in mind this is the anomaly coefficient of the CFT to which the QFT tends for short scales. We have included the leading correction to the finite term (the last term), with a form that is determined by dimensional analysis,  assuming a $g_{UV}^2$ dependence\footnote{This is known to change for $\Delta_{UV}<d/2$.} This term goes to zero as $R\rightarrow 0$ and we approach more the CFT. This is a finite universal non-local contribution.  

\subsubsection*{QFT-IR}
In the opposite limit of large spheres, we expect the theory to approach a new CFT. Perturbative deviations from this CFT$_{IR}$ are controlled by an irrelevant operator $\phi_\Delta^{IR}$, with $\Delta_{IR}> d$, with action
\be
S=S_{CFT_{IR}}+g_{IR} \int d^dx\, \phi^{IR}_\Delta\,. 
\ee

With a bit of rethinking, our understanding of the ultraviolet contributions to the divergent terms applies equally well to other contributions in the IR regime. In fact, one can think that for regions very large in comparison with all physical scales $\Lambda$ in the model,  the contributions of the physics due to these scales has to be produced by correlations across the boundary at distances $\sim\Lambda$. Then, they can be considered local and extensive contributions on the boundary (assuming the curvatures of this boundary are also of IR scales). The divergent terms cannot be changed, and have to be the same as they were for small regions, but there will be finite renormalizations to these divergent local terms. In other words, for very large and smooth regions, the terms that come from the scales $\Lambda$ of the theory have the same geometric form as the divergent terms and follow by replacing the cutoff with finite physical scales.  

Therefore, for the sphere in $d=4$ at the IR (large radius) we expect  
\bea
&&S=(4 \pi R^2)\left(c_2 \,\epsilon^{-2}+ c_{2}'\,g_{UV}^2 \,\epsilon^{2 \left(\Delta_{UV} -(d+2)/2\right)}+\cdots -m^2 \right)\nonumber 
\\
&&\hspace{1.5cm}+ 4 A_{UV} \, \log(\epsilon)+4 X \log(m) -4 (A_{UV}-X) \log(R) +\tilde{c}_0 +c_0'\, g_{IR}^2 \, R^{2(d-\Delta_{IR})}+\cdots\,.
\eea
The coefficient of the local area term has received a universal finite contribution $-m^2$. In general, this is not perturbatively calculable and depends on the whole QFT between the fixpoints. We will show a compact formula for it later. The local logarithmic term multiplying the integral of the curvature over the surface of the sphere (which we have replaced by its value $2$) has also received a correction written here $4 X \log(m)$. As a consequence, the IR fixpoint has a new anomaly coefficient $A_{IR}=A_{UV}-X$ multiplying the non-local term $\log(R)$. We have also included the non-local perturbative term $g_{IR}^2 \, R^{2(d-\Delta_{IR})}$ that has the same form as the one in (\ref{uvi}) but now it decreases for large spheres because $\Delta_{IR}>d$.     

\subsubsection*{Singular boundaries}
We have explained the form of the divergent terms for smooth boundaries. 
The expansion is in terms of polynomials of the curvature of the surface, in addition to possible powers of the UV coupling constants. 
 The combination $(\hat{R}\epsilon^2)$ can be considered as a small perturbative parameter for $\epsilon$ small enough, and this is why only positive powers of the curvature appear in the expansion.  This is so unless the curvature of the boundary diverges. In that case, the above expansion does not apply around the singularity, and new divergent features appear.
 We have encountered this in the case of squares in $d=3$. There is a logarithmic term attached to the corners in this case. Analogous divergent terms appear in different dimensions and singular features. Generally, a feature with dimension $k$ will lead to a leading $\epsilon^{-k}$ divergence. But more complex behaviour is known for certain geometric features, including fractional powers.  

\subsubsection*{Finite terms}

The finite terms have a lot of information. They depend on the state. The usual volume increasing entropy for thermal states shows up in the finite term for regions large compared to the thermal length
\be
S_0\sim s\,\textrm{vol}\,,  
\ee
with $s$ the usual thermal entropy density. This would be larger than an area increasing term if the area term were not divergent. Super-area finite terms also appear for pure states different from the vacuum. For fermions systems with a Fermi surface, we have logarithmic corrections to the area 
\be
S_0\sim (R\mu)^{d-2} \log(R \mu)\,,  
\ee
with $\mu$ the chemical potential. Even in the vacuum, most of the information in the EE is in the dependency of the finite term on the shape of the region.   
We turn to this term in the next section. 

\subsection{Entropy: what does survive the continuum limit?}
\label{regent}

The EE in QFT is divergent and this cannot be avoided. We will see later that this divergence is implied by strong subadditivity and Lorentz invariance. Then, the question is what does survive the continuum limit. It was clear from the preceding discussion that certain coefficients in the entropy expansion are well-defined quantities. Now, we want to know how to extract all the useful information about the QFT from the EE.
  
The structure of the divergent terms gives us the clue to this question because they are always local and extensive along the boundary. Then, we have to make linear combinations of entropies for different regions such that each boundary appears with a total coefficient $0$ in the combination. The natural quantity with this property is the mutual information of two disconnected regions $A,B$, 
\be
I(A,B)=S(A)+S(B)-S(AB)\,.
\ee  
This is a measure of correlations that is positive and monotonic.

  \begin{figure}[t]  
\centering
\includegraphics[width=1\textwidth]{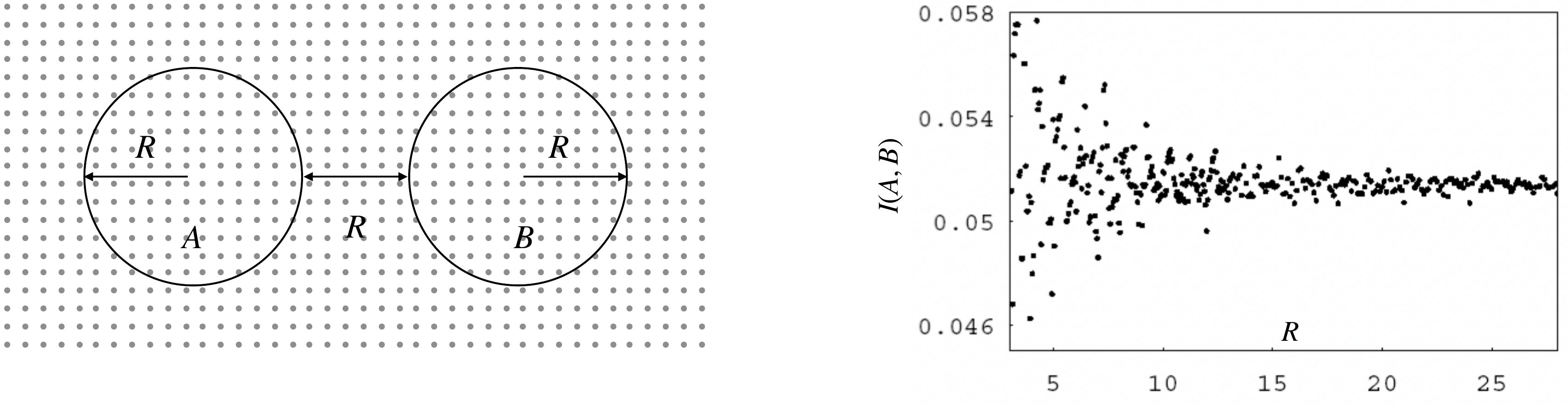}
\captionsetup{width=0.9\textwidth}
\caption{Mutual information of two circles of radius $R$ separated by $R$ in a square lattice.}
\label{mutuall}
\end{figure}  
 
 In figure (\ref{mutuall}) we see the calculation of the mutual information of two circles of radius $R$ separated by the same distance. This is in $d=3$ and for a free massless scalar in a square lattice. The figure shows the behaviour as we go to the continuum limit. A square lattice is badly adapted to circles, but nevertheless, the mutual information converges pretty well to a finite number. 
 
 Notice the value of this number. It is very small, $I(A,B)\sim 0.052$, a fraction of a bit for an infinite number of d.o.f. This speaks about the locality of the theory. To get large amounts of mutual information we have to take the circles closer. When they touch each other the mutual information diverges.

\begin{figure}[t]
\begin{center}  
\includegraphics[width=0.39\textwidth]{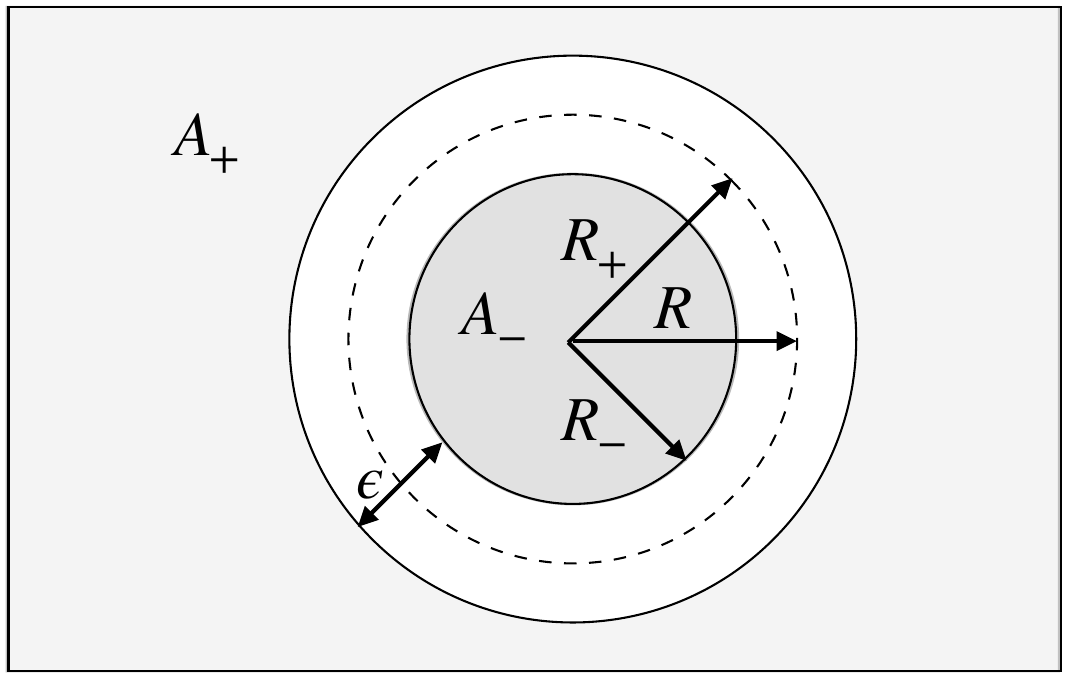}
\captionsetup{width=0.9\textwidth}
\caption{Entropy regularization with mutual information.}
\label{regu}
\end{center}  
\end{figure}

The mutual information should contain all universal information of the entropy. This is because it can be used to provide a standard regularization of the entropy. Instead of the entropy of $A$ with some cut-off we can take the mutual information between region $A^-$ with boundary contracted a distance $\epsilon/2$ from the boundary of $A$ and another $A^+$ which is the complement of $A$ contracted the same distance $\epsilon/2$, see figure \ref{regu},
\be
S_{\textrm{reg}}=\frac{1}{2}\,I(A^+,A^-)\,.  \label{reg}
\ee
If we make $\epsilon\rightarrow 0$ this has divergences in $\epsilon$ which are of the same form as the ones in the entropy, but now the coefficients in this expansion are universal. $\epsilon$ is now a short physical distance in the continuum model. When $\epsilon\rightarrow 0$, we can think heuristically $S(A^+),S(A^-)\rightarrow S(A)$, and $S(A^+\cup A^-)\rightarrow 0$, because it is the entropy of the total space in a pure state. Of course, this last statement is not strictly correct, this entropy is equivalent to the one of the thin shell in between $A^+$ and $A^-$, and is always divergent. But this is expected to be a purely UV geometric local additive divergence, independent of the details of the theory at larger scales. In this way, we expect $I(A^+,A^-)\sim 2 \, S(A)$ in what respect to the universal terms.  This is the reason for the factor $1/2$ in (\ref{reg}).

Another quantity that we can expect to be convergent is the combination in the strong subadditivity (SSA) relation for two intersecting regions,
\be
F(A,B)=S(A)+S(B)-S(A\cap B)-S(A\cup B)\,.
\ee
We have only to be careful that the regions cut each other smoothly such that the intersection and union are smooth too and do not contain new geometric features which carry new divergences not appearing already in $A,B$. This quantity is in fact not independent from the mutual information. It is a limit of a difference of mutual information as discussed in section \ref{mumu}. Any quantity formed out of linear combinations of entropies, and without divergences, can in fact be written in terms of the mutual information.

For two states and only one region, the relative entropy is a quantity of interest. This again is expected to be finite in most cases because the two states will not differ much from each other at very high energies, where they both should approach the vacuum.\footnote{In fact, the statement that the divergences of the entropy are independent of the state is better expressed in the finiteness of the relative entropy. The difference of EE between two states may contain state-dependent boundary terms depending on the way they are regularized.} Relative entropy can be defined using modular theory directly in the continuum without the need of a cut-off. This is called the Araki formula. We will not use it here and continue to think in terms of density matrices in cut-off theories. 

In summary, we can think of the EE in similar terms as we used to think about other QFT quantities such as the effective action. This is not well defined by itself, there are divergences that require a cut-off and the related ambiguities. However, the divergences are local and this much restricts the possible ambiguities. To get continuum quantities, we have to compare partition functions for different geometries or sources. This similarity with the effective action is not, as it might seem, simply a fortuity. We will see the relation when we explain the replica trick.  

\subsection{Algebras: what does survive the continuum limit?}
In the lattice, we may take our algebras for a region with a bit of discretion. We can add or subtract one point from the boundary and it will go to the same region in the continuum. We can also take for example a field operator and not the momentum operator for such a point in the boundary. This will be an algebra with a centre. All these possible choices are to be thought of as different regularizations. But how are algebras with centre and without centre both tend to the same object in the continuum? The reason is that in the continuum the two algebras will differ by operators localized at the $(d-2)$-dimensional boundary.  There is no true operator in the continuum QFT that can be localized in a $(d-2)$-dimensional surface. The smearing function would be too singular, and self correlations diverge (see exercise 2). A field is an operator-valued distribution that can be localized in a $(d-2)$-dimensional surface, but not an operator. The lattice operators in the surface will have too large self-correlations in the continuum and will decouple from the rest of the operators. Acting on the vacuum, they will give vectors with energy tending to infinity in the continuum, and this limit vector will not form part of the Hilbert space of the QFT. Therefore, in general, the algebras of QFT do not have any centre. Algebras without centres are called factors. 

What about regions of dimension $d-1$? Well, these in general are not causal regions. But a $(d-1)$-dimensional piece of a null surface can be a causal region. It turns out that, again, in most cases, smearing in a null surface is not enough to produce an operator, and these algebras are empty. For $d\ge 3$, the exception is the case of free fields or theories with a free UV fixpoint. For $d=2$, all theories have non-trivial null interval algebras: for example, the smeared stress tensor always give some null localized operators.  

This shows how to distinguish QFT's with free or interacting UV fixpoint, using the entropy. If the mutual information does not vanish when one of the regions is null it must be the first case. Mutual information with a $(d-2)$-dimensional surface always vanishes. 

\subsection{Density matrix: what does survive the continuum limit?}
\label{density}
In the continuum limit not only the entropy gets divergent (which can also be the case of some states for the harmonic oscillator) but the density matrix itself does not survive without a cutoff. The reason is simple to understand first in another context, the case of the large volume limit of ordinary thermodynamics. 

Suppose we have an infinite series of single-qubit states and assume the states of these qubits is decoupled. For still more simplicity take all the single-qubit states to be the same $\rho$. The density matrix of $n$ qubits is just the tensor product
\be
\rho^{(n)}=\rho^{\otimes n}\,. 
\ee
The eigenvalues are products of $n$ eigenvalues, and will all tend to zero as $n$ increases (unless $\rho$ is pure and has only one eigenvalue $1$ and the rest $0$). This means the limit of $n\rightarrow \infty$ cannot be a density matrix, which has to have some non zero eigenvalues. The problem is the number of degrees of freedom, the large tensor product, rather than the dimension of the Hilbert space. 

The same happens in the large volume limit of thermal systems and also for the algebras of finite regions in QFT. We will see later that, in certain cases, one can even map the situation of a finite region in vacuum QFT to the infinite volume thermal state in another space. 

In QFT, as in the case of the product of an infinite number of qubits, the state is perfectly well defined. We can evaluate the expectation value of operators without problem. But there is no density matrix.     
What replaces the density matrix is the modular flow, which is the continuum version of the unitary flow induced by $\rho^{i\tau}$ as a transformation on the operators and keeping the state invariant. It satisfies the KMS condition and this defines it uniquely.  

If we have the algebra of a region $W$ in QFT, the vacuum state in this algebra automatically comes together with a purification, given by the vacuum state in the full space. This gives us all the elements of the Tomita-Takesaki theory for any region. The modular operator $\Delta$ is the continuum version of $\rho_W\otimes \rho_{W'}^{-1}$. Note there is no need to normalize the density matrix in the expression of this operator because it comes in a product with the inverse of another matrix with the same eigenvalues. One can think this is the reason $\Delta$ has a good continuum limit even if $\rho$ does not. The full modular Hamiltonian is $\hat{K}=-\log \Delta$, which in terms of the cutoff version has the form
\be 
\hat{K}= K_W \otimes 1 - 1\otimes K_{W'}\,,\label{full}
\ee
whereas usual $K_{W}=-\log(\rho_W)$. This now has two pieces, one acting in $W$ and the other acting in $W'$ but making ``time'' go backwards in that region (because of the minus sign). It turns out, that this combination is much better behaved than each of the parts, and defines an operator in the continuum, while each of the parts does not.  We will come back to this point later within an explicit example. We have
\be
\Delta |0\rangle=e^{-\hat{K}} |0\rangle=|0\rangle\,,\hspace{.6cm}   \hat{K} |0\rangle=0\,. 
\ee
\subsection{Replica trick}
\begin{figure}[t]
\begin{center}  
\includegraphics[width=0.7\textwidth]{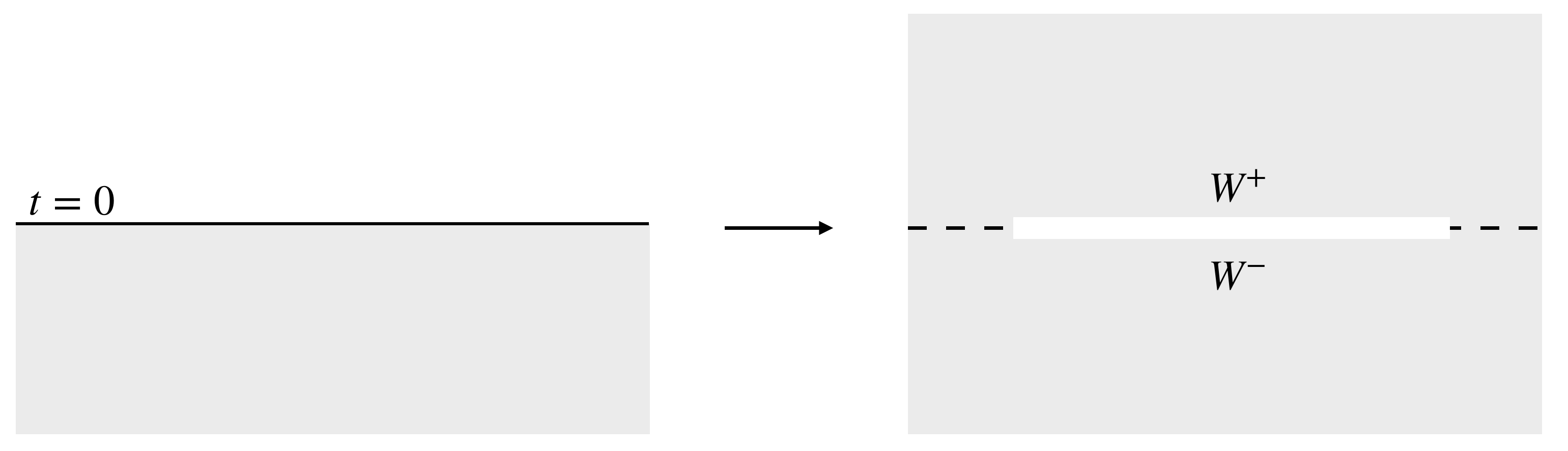}
\captionsetup{width=0.9\textwidth}
\caption{Path integral representation of the vacuum wave function and the reduced density matrix of a region. }
\label{replica}
\end{center}  
\end{figure} 

To compute the entropy one can start from the representation of the vacuum state in terms of a path integral in Euclidean space.  For definiteness, consider a scalar field $\hat{\phi}(\vec{x},t)$, and take the basis formed by eigenvectors of this field operator at time $t=0$ as $\hat{\phi} (\vec{x},0) \left|  \alpha\right>= \alpha(\vec{x})\left|  \alpha\right>$, where $\alpha$ is any real function on the space. The vacuum wave functional writes   
\begin{equation}
\Phi(\alpha)=\left<0|\alpha\right>=N^{-1/2}\int_{\phi(\vec{x},-\infty)}^{\phi(\vec{x},0)=\alpha(\vec{x})} D\phi\,\, e^{-S_E(\phi)}\,. \label{qs}
\end{equation} 
In order to select the vacuum state, the path integral is over the lower half space and with Euclidean time. This can be thought as Euclidean time evolution with an exponential damping factor for non zero energy states.  $S_E(\phi)$ is the Euclidean action and $N^{-1/2}$ is a normalization factor. The vacuum density matrix in this basis is
$\rho(\alpha^+,\alpha^-)=\left<\alpha^+ | 0\right> \left< 0| \alpha^-\right>=\Phi(\alpha^+)^*\Phi(\alpha^-)$. In order to trace over degrees of freedom in $W'$, the set complementary to $W$, one considers functions $\alpha^+=\beta + \alpha_{W}^+$, $\alpha^-=\beta+ \alpha^-_{W}$, which are equal to $\beta$ on $W'$, and sum over all possible functions $\beta$. Here $\beta=0$ on $W$ and $\alpha^\pm_W$ are zero on $W'$.  Using the representation (\ref{qs}), this construction amounts to take two copies of the half space,\footnote{The representation of the complex conjugate vacuum wave functional is the path integral on the upper half plane. In cases where the Euclidean action is complex, it must have the property $S_E(\phi)=S_E^*(T\phi)$, where $T\phi$ is the Euclidean time inversion of $\phi$.} glue them on $W'$ (see figure \ref{replica}), and take the path integral in this space,
\begin{equation}
\rho_W(\alpha_{W}^+,\alpha^-_{W})=\int D\beta\,\, \Phi(\beta + \alpha_{W}^+)^* \Phi(\beta+ \alpha^-_{W})= N^{-1}\int_{\phi(\vec{x},0^-)=\alpha^-_W(\vec{x}),\, x\in W^- }^{\phi(\vec{x},0^+)=\alpha_W(\vec{x}),\,x\in W^+} D\phi\,\, e^{-S_E(\phi)}\,. \label{opre}
\end{equation}
The arguments of the density matrix are the boundary conditions of the path integral on both sides of the cut on $W$.

\begin{figure}[t]
\begin{center}  
\includegraphics[width=0.8\textwidth]{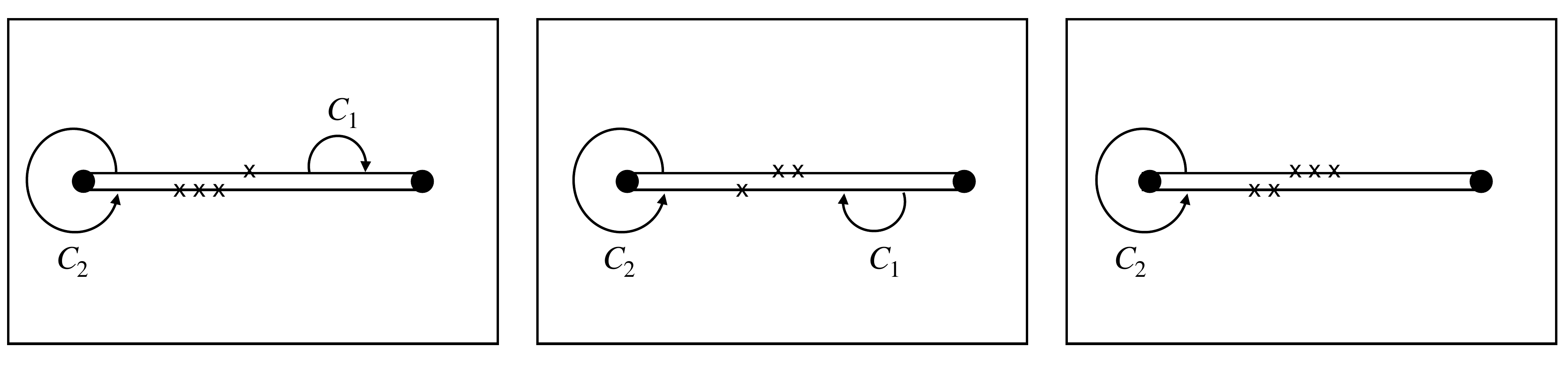}
\captionsetup{width=0.9\textwidth}
\caption{Path integral representation of $\textrm{tr} \rho^3$. Boundaries with the same number of crosses are identified.}
\label{replica1}
\end{center}  
\end{figure}

Now that we have a path integral representation of the density matrix we want to compute the entropy. This cannot be done directly. Instead, the replica trick follows by computing the traces $\textrm{tr}\rho_W^n$. A representation of this calculation with the path integral is realized by replicating the density matrix $n$ times. The product of matrices corresponds to sewing the upper boundary of one matrix with the lower one of the other. The trace finally corresponds to sewing the last and the first boundaries. See figure \ref{replica1}, where boundaries labelled with the same number of crosses are identified.  
 The resulting space is a $n$-sheeted $d$ dimensional Euclidean space with conical singularities of angle $2\pi n$ located at the boundary $\partial W$ of the region $W$. You can convince yourself that this manifold is flat everywhere and there is no singularity at all near the original cut, which has disappeared. Going around a circle $C_1$ in figure \ref{replica1} you come back to the original point after a $2\pi$ rotation, as usual. You will not notice any problems locally there. The only singular points are around the boundary of $W$. Following the path $C_2$ you have to turn an angle $2 \pi n$ before returning to the original point.

Finally we have  
\begin{equation}
\textrm{tr}\rho_W^{n}= \frac{Z(n)}{Z(1)^{n}}\,,
\label{dd}
\end{equation}
where $Z(n)$ is the functional integral on the $n$-sheeted manifold, and we have used the normalization factor $N=Z(1)$ in (\ref{opre}) to have $\textrm{tr} \rho_W=1$. 
From the knowledge of this trace, we can construct an entropic-like quantity by taking the logarithm. The Renyi entropies are defined as
\be
S_n=(1-n)^{-1} \log \textrm{tr}\rho^n \label{cheto}\,.
\ee
In particular the entropy follows by taking $n$ to $1$
\be
S=\lim_{n\rightarrow 1} S_n\,.
\ee
The replica method gives the Renyi entropies as
\be
S_n(W)=\frac{\log Z(n)-n \log Z(1)}{1-n}\,.
\ee
Eq. (\ref{cheto}) gives a representation of the Renyi entropies for integer $n$. The entanglement entropy follows by analytic continuation of $S_n$ down to $n=1$. Notice that there is in general no path integral representation for non-integer $n$. Hence, we have to compute the function of $n$ and do analytic continuation to get the entropy. 

Most of the explicit results about EE in QFT were obtained using the replica trick. The replica trick has a very appealing form because it is a quite geometric prescription, we have to do the same geometric manipulation for any theory. This geometric representation is connected with a geometric manifestation of the entropy in holographic theories, which are the ablest ones to profit from the power of the replica trick.  

There is not much to say about the Renyi entropies as information quantities. They do not enjoy many of the nice properties or operational interpretations the entropy does. The Renyi entropies are certainly additive for decoupled systems and measure the mixedness of the density matrix since they vanish for pure states too. From the point of view of the quantum theory, they have however an additional nice feature for integer $n$ that we explain in what follows.   

\subsection{Renyi twist operators}
One can interpret the calculation in figure \ref{replica1} in a different light. We can say that we have a new theory, which is just $n$ decoupled copies of the original one, with something inserted at $W$ at time $t=0$. In this new theory that consists of $n$ independent copies of the original one, we then evaluate the vacuum expectation value of what we have inserted. This that we have inserted just takes copy $i$ to copy $i+1$, including copy $n$ to copy $1$. This is a cyclic permutation of the theories but restricted to $W$. The original replicated theory has in fact an obvious global permutation symmetry between the copies. This symmetry is clearly not spontaneously broken. What we have inserted at $W$ is then an operator that implements the cyclic symmetry only at $W$. 

Thinking more generally, we can ask if given a global symmetry of a theory we can always find an operator that implements this global symmetry in a finite region $W$. This we can do exponentiating the charge density integrated over $W$ if we have a charge density and a Noether current. But, more generally, it can be done also for discrete symmetries as the present one. These local symmetry operators are called twist operators. In the present case, this is a twist operator in the $n$-replicated theory, that we can call Renyi twist operator $\tau^{(n)}_W$. We then have
\be
\textrm{tr} \rho_W^n=\langle 0^{(n)}|\tau^{(n)}_W|0^{(n)}\rangle=e^{-(n-1) S_n(W)}\,,
\ee
wherein the expectation value both, operators and states, are in the replicated theory. 

This shows something very interesting, which has had many applications. The Renyi entropies with integer $n$ are given by the logarithm of expectation values of operators. This establishes an important connection between difficult to grasp entropic quantities and normal operator expectation values. This relation is of course behind the simple representation of these quantities in the path integral. Almost anything we can insert in a path integral represents an operator insertion.  

But this is not a peculiar feature of QFT. The same holds for any  quantum system.  
 Let us consider a system
with global state $|0\rangle$, Hilbert space ${\cal H} = {\cal H}_1 \otimes {\cal H}_2$. In the replicated
model we take $n$ identical copies of this system. The global state is $|0^{(n)}\rangle=\otimes_{i=1}^n |0_i\rangle$, and let the basis of the $i^{th}$ Hilbert space be $\{|e^i_a\rangle\otimes |f^i_ b\rangle\}$ where  $\{|e^i_a\rangle\}$ is a basis of ${\cal H}_1$ and $\{ |f^i_ b\rangle\}$ a basis of ${\cal H}_2$. The twist operator is 
\be
\tau^{(n)}_1=\left(\otimes_{i=1}^n \sum_a |e^{i+1}_a \rangle \langle e^i_a |\right)\otimes {\bf 1}_2\,,
\ee
and we have
\be
\langle 0^{(n)}|\tau^{(n)}_1 |0^{(n)}\rangle= \textrm{tr}\rho_1^n\,.
\ee
The twist is a unitary operator, independent of the chosen basis. For two different Hilbert space factors we also have 
\be
\tau_{12}^{(n)}=\tau_{1}^{(n)}\tau_{2}^{(n)}. 
\ee

As operators, the Renyi operators may be measured, in contrast to the entropy. By knowing all the Renyi twists expectation values we could compute the entropy. But it is true that knowing enough expectation values of other operators we can also do it, at least in the finite-dimensional case.  

\subsection{The Rindler wedge}
\label{rw}
\begin{figure}[t]
\begin{center}  
\includegraphics[width=0.65\textwidth]{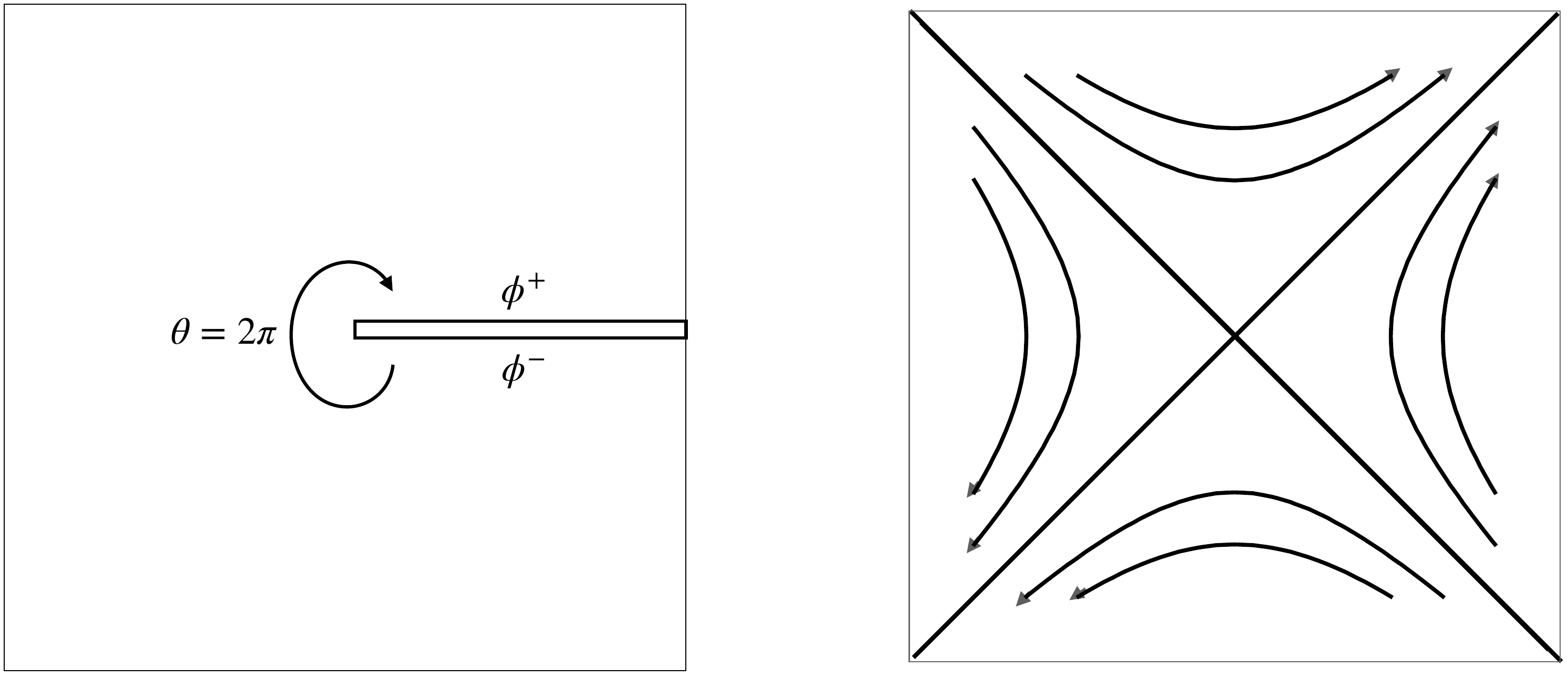}
\captionsetup{width=0.9\textwidth}
\caption{Left: Path integral representation of the density matrix of the right half space. Right: The modular flow of the Rindler wedge is given by a one parameter group of boosts.}
\label{rindler}
\end{center}  
\end{figure}

The Rindler wedge is the region $x^1>|t|$ in Minkowski space. It is the causal region corresponding to a half spatial plane $x^1>0$. The path integral representation of the density matrix is shown in figure \ref{rindler}. In this particular case, there is a rotational symmetry of the euclidean theory that allows us to express the density matrix as  
\be
\rho[\phi^+,\phi^-]=\langle \phi^+| R(2\pi) |\phi^- \rangle\,,
\ee
where $R(\theta)$ is the Euclidean rotation operator in the plane $x^0,x^1$. 
This follows by thinking of the path integral as an evolution in the angle variable from $0$ to $2\pi$, whose generator is the angular momentum.  
Once we go to real-time, this rotation between spatial and temporal coordinates is converted into a boost. This gives
\be
\rho\sim e^{- K}\label{pico}
\ee
where
\be
K=2 \pi \int_{x^1>0} d^{d-1}x\,  x^1\, T_{00}(x)\,.\label{pico1}
\ee
This is $2 \pi$ times the boost operator, restricted to act on the right Rindler wedge only. The $2\pi$ reminds us of the full Euclidean rotation.  

This result is rather interesting. First, it shows the modular Hamiltonian in this case is an integral of a local operator, the energy density operator. We say the modular Hamiltonian is local in this case. For other non-vacuum states or other regions, in general, the modular Hamiltonian can be rather non-local, containing products of local operators at different positions. Second, it shows there is a pre-established connection between entanglement in vacuum and energy that will give us much to talk about in another chapter. In fact, this density matrix is entirely produced by vacuum entanglement between the right and left wedges. As we mentioned before, a relativistic QFT is completely determined by the vacuum fluctuations manifested on the field expectation values. Here we strikingly see this: by reducing the vacuum state to the half-space we can learn about the energy density operator of the theory and hence about the Hamiltonian.      

The modular flow $\rho^{i\tau}$ in the Rindler wedge moves operators locally because it is just given by the orbits of a one-parameter group of boost transformations. Sometimes it is better to think of this flow not just on the algebra of the Rindler wedge but on the full space which is a purification of the wedge. There, the full modular Hamiltonian (\ref{full}) is proportional to the complete boost operator. 
The modular flow is shown in figure \ref{rindler}. In the language of Tomita-Takesaki theory, we have $\Delta=e^{-\hat{K}}$, where $\hat{K}$ is $2 \pi$ times the complete boost operator. This is in fact a well-defined operator in any QFT. On the other hand, while we can compute expectation values of the half modular Hamiltonian $K$, it is too singular to be an operator, in the sense that $K|0\rangle$ is not normalizable. 

What plays the role of the modular reflection $J$ in the Rindler wedge? It has to change the sign of the $\hat{K}$ and map operators on the right wedge to the left wedge and vice versa. The two wedges represent commutant algebras. It turns out that $J$ is the CRT operator (charge conjugation times reflection on $x^1$ times time reflection). It is an antiunitary operator mapping (for scalars)
\be
J \phi(x^0,x^1,x^2,\cdots) J= \phi^\dagger (-x^0,-x^1,x^2,\cdots)\,.  
\ee
This is just CPT times a rotation in $d=4$ and exists in any dimensions. In fact, Bisognano and Whichmann were the first to compute the modular operators of half-space, and they did it using the CPT theorem. One can show that all the expected properties of $\Delta$ and $J$ described in section \ref{tt} hold on polynomials of field operators, including the KMS condition of periodicity in imaginary modular parameter, using the analyticity properties of correlation functions.     

 Equations (\ref{pico}) and (\ref{pico1}) give the vacuum state in half-space as a thermal state of inverse temperature $2\pi$ for the boost operator. For an observer following a trajectory given by a boost orbit, the state should look like an ordinary thermal state concerning his notion of proper time $\tilde{\tau}$, because for these trajectories the proper time and the boost parameter $s$ are proportional $s=a \tilde{\tau}$. The proportionality constant is the proper acceleration $a$ of the observer, which is constant along boost orbits. The acceleration $a$ is also the inverse of the distance between the points in the hyperbolic orbit to the origin. This gives the relation $K= \tilde{H}/a$, locally along the trajectory, between the boost operator and the proper time Hamiltonian $\tilde{H}$ of the accelerated observer.
 Then, from the inverse temperature $2 \pi$ of the vacuum with respect to boosts, for such an observer, there is a thermal bath at (proper time) temperature 
\be
T=\frac{a}{2\pi}\,.
\ee
This is Unruh's effect: accelerated observers see the vacuum as a thermally excited state. It is a completely general phenomenon. We will see Hawking radiation of a black hole is intimately related to this effect. Note this temperature diverges as we approach the boundary $a\rightarrow \infty$, where entanglement with the complementary Rindler space is very high. 

\subsection{Conformal transformations and spheres}
For a CFT, Poincare symmetries are enlarged to the conformal group. These theories are characterized by having a traceless stress tensor $T_\mu^\mu=0$, on top of being symmetric $T_{\mu\nu}=T_{\nu\mu}$, and conserved  $\partial_\mu T^{\mu\nu}=0$. This allows to enlarge the number of conserved currents related to space-time symmetries. These currents write
\be  
j_\mu= a^\nu \,T_{\nu\mu} +  b^{\alpha \nu}\,x_\alpha\, T_{\nu\mu} + c\, x^\nu\, T_{\nu\mu} + d_\alpha\, (x^2 g^{\alpha\nu}-2\, x^\alpha x^\nu)\, T_{\nu\mu}\,.\label{das} 
\ee
The corresponding conserved charges depend on parameters $a^\mu$ determining translations, the antisymmetric $b^{\mu\nu}$ giving Lorentz transformations, $c$ dilatations, and $d^\mu$ the so-called special conformal transformations.   

\begin{figure}[t]
\begin{center}  
\includegraphics[width=0.30\textwidth]{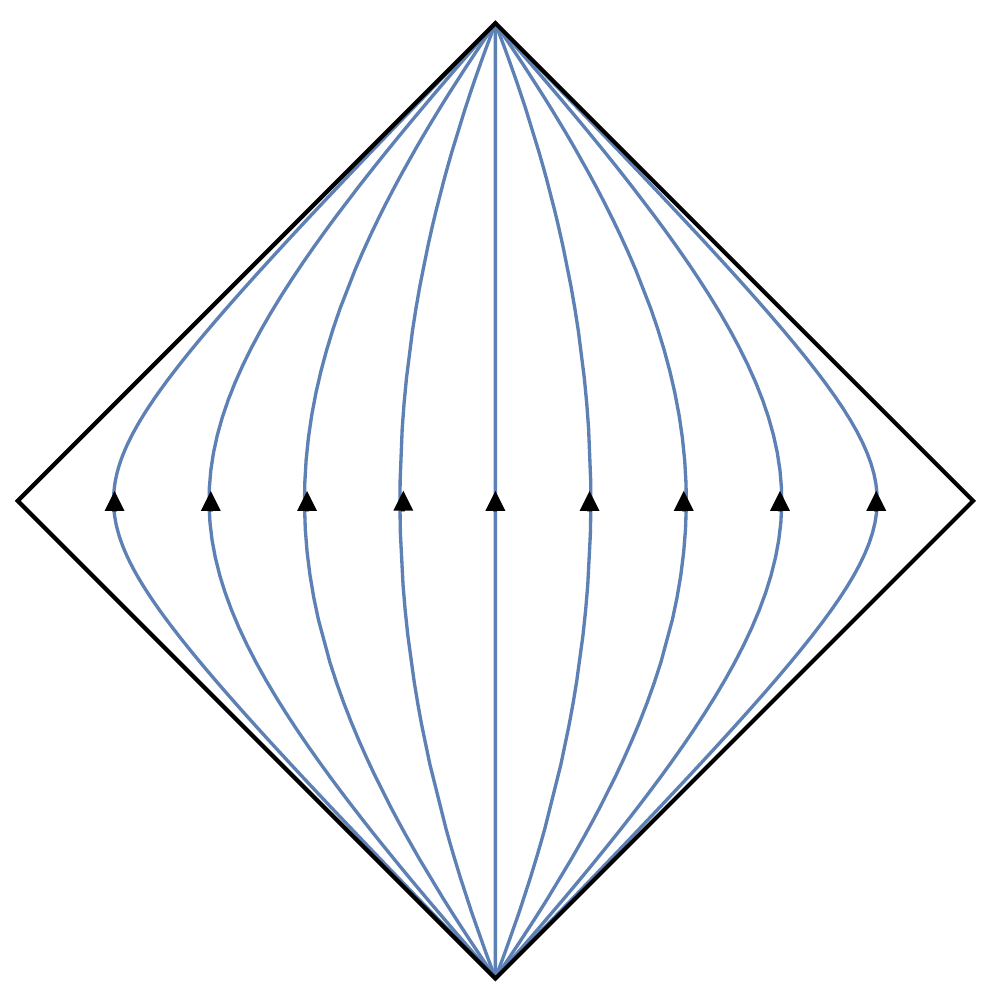}
\captionsetup{width=0.9\textwidth}
\caption{Sphere modular flow for a CFT.}
\label{smf}
\end{center}  
\end{figure}

Geometrically the conformal symmetries of Minkowski space correspond to the group of point transformations that keeps the angles invariant but can deform the distances. It keeps the null lines into null lines, maintaining the causal structure. It also sends planes to planes or spheres, and spheres to spheres or planes.  

Therefore, it is possible to transform conformally the Rindler wedge to the causal region with a sphere as a boundary, and with the same transformation, the algebras are mapped. As the vacuum is invariant under this transformation for a CFT, the modular Hamiltonian is the transformed modular Hamiltonian. We could work out the expression transforming coordinates and the stress tensor. But is simpler to see that as $K$ is a generator of the group for the Rindler wedge it must be a generator for the sphere. Then, it is an integral of some $j_0$ of the form given by (\ref{das}) on the interior of the sphere. The modular flow of the wedge is in the time direction at $t=0$, and the same has to be true for the modular flow of a sphere at $t=0$. This implies the generator we are looking for must be expressed entirely in terms of $T_{00}$ at $t=0$. It must also be invariant under rotations. We have to calibrate the overall constant to have the same form of Rindler's $K$ near the boundary of the sphere. It is easy to  get
\be
K=2\pi \int_{|\vec{x}|<R} d^{d-1}x\, \frac{R^2-r^2}{2R}\, T_{00}(\vec{x})\,.\label{modesf}
\ee    
The modular flow is given by the one-parameter conformal transformations generated by $K$. Conjugation with $\rho^{-i s K}$ gives a conformal transformation on operators that write
\be
r^{\pm}(s)=\frac{R(R+r^{\pm}-e^{\pm 2\pi s}(R-r^{\pm}))}{(R+r^{\pm}+e^{\pm 2\pi s}(R-r^{\pm}))}\,,
\ee
where $r^{\pm}=r\pm x^0$ are the mull radial coordinates. See figure \ref{smf}.
The modular flow freezes on the spatial boundary but still moves the null boundaries.

\subsection{Transformations to curved spaces and the EE of a sphere} 
\label{tow}

We have seen above, conformal transformations from Minkowski to Minkowski space. But more generally one can conformally transform a space $M$ of metric $g$ into another one $\tilde{M}$ of metric $\tilde{g}$. The rule is that the point transformation between these spaces $x'=f(x)$ should be such that
\be
g_{\mu\nu}'=\frac{\partial x^\alpha}{\partial x^{\mu\, '} }\, \frac{\partial x^\beta}{\partial x^{\nu\,'}} g_{\alpha\beta} =\Omega^{-2} \, \tilde{g}_{\mu\nu}\,, 
\ee
so that the metric induced by the transformation is proportional to the pre-existent metric. This rescaling of the metric does not change angles. 

Operators of a CFT can be transformed to the new space. Primary fields transform homogeneously. We have 
\be
\phi_1'(x'_1)=\Omega(x)^{\Delta} \,U\,\phi(x)\,U^{-1}\,,
\ee
for some unitary $U$. Taking the new state in the new space as the unitarily transformed state (the transformed vacuum is what is called the conformal vacuum) we get for the correlation functions
\be
\langle\phi_1'(x'_1)\cdots\phi_n'(x'_n)\rangle= \Omega(x_1)^{\Delta_1}\cdots \Omega(x_n)^{\Delta_n}\, \langle\phi(x_1)\cdots \phi(x_n)\rangle\,,
\ee
where the $\Delta_i$ are the field dimensions.

This transformation allows us to define the theory and the state in the new space given the knowledge of the theory in the original one. Notice that the operators are transformed locally and the factors in the transformation will not prevent the algebras from being likewise transformed to each other. From the algebraic point of view, the new theory is somehow the old one in disguise, it has the same algebras and states transformed unitarily,  only the label of the causal regions has changed. And the net of causal regions is identical too since conformal transformations preserve the causal structure. 

With conformal transformation, we can understand the structure of the reduced density matrix in other spaces. Local modular flows are clearly mapped to local flows. It is possible to map Minkowski space to a cylinder space-time $S^{d-1}\times R$, where $R$ is time. Then, we can compute the vacuum density matrix for spheres inside a cylinder (in vacuum) by mapping spheres in Minkowski space.
 
The causal development of a sphere in Minkowski can be mapped to a $H_{d-1}\times R$, where the $R$ is again time and $H_{d-1}$ is hyperbolic space. The metric is 
\be
ds^2= -d\tau^2+{\cal R}^2\,(du^2+\sinh^2(u) \, d\Omega_{d-2}^2)\,.
\ee
  The hyperbolic space is a maximally symmetric space, like the sphere, but with negative curvature. It is non-compact, with infinite volume. The vacuum state in the original sphere is mapped to an ordinary thermal state for the time $\tau$ at inverse temperature $T^{-1}=(2\pi {\cal R})$, where ${\cal R}$ is the curvature scale of the hyperbolic space. Therefore, we see in this new representation, the reduced density matrix on a ball looks like the infinite volume limit of a thermal state. It is a peculiar infinite volume though, in a space where the curvature is of the same size as the thermal length, and the geometry is curved in such a way that area and volume increase in the same way towards infinity, $\sinh^{d-2}(u)\sim \int du\, \sinh^{d-2}(u)\sim e^{(d-2) u}$.   

Let us consider the mapping of a ball to de Sitter space. We do this more carefully because it will allow us to compute the EE of a sphere in a CFT. 
 
 We start
 with the flat space metric for $d$-dimensional Minkowski space in
polar coordinates
 \beq
ds^2=-dt^2+dr^2+r^2\,d\Omega^2_{d-2}\,.
 \label{flat1}
 \eeq
Now with the coordinate transformation
 \be
t=R\,\frac{\cos\theta\,\sinh(\tau/R)}{1+\cos\theta\,\cosh(\tau/R)}\,,\hspace{.6cm}
 \label{trn2}
r=R\,\frac{\sin\theta}{1+\cos\theta\,\cosh(\tau/R)}\,,
 \ee
 the flat space metric  becomes
 \be
ds^2=\Omega^2\, \left[-\cos^2\!\theta\, d\tau^2+
R^2\left(d\theta^2+\sin^2\!\theta\,d\Omega^2_{d-2}\right)\right]\,,\hspace{.4cm}
 \label{round0}
\qquad\Omega=(1+\cos\theta\,\cosh(\tau/R))^{-1}\,.
 \ee
After eliminating the conformal factor $\Omega^2$, the remaining metric
corresponds to the static patch of $d$-dimensional de Sitter space with
curvature scale $R$. The latter identification may be clearer if we
transform to $\tilde{r}= R\sin\theta$, which puts the above metric in the
form
 \beq
ds^2=-\left(
1-\frac{\tilde{r}^2}{R^2}\right)\,d\tau^2+\frac{d \tilde{r}^2}{1-\frac{\tilde{r}^2}{R^2}}
+\tilde{r}^2\,d\Omega^2_{d-2}\,.
 \label{twor}
 \eeq
\begin{figure}[t]
\begin{center}  
\includegraphics[width=0.5\textwidth]{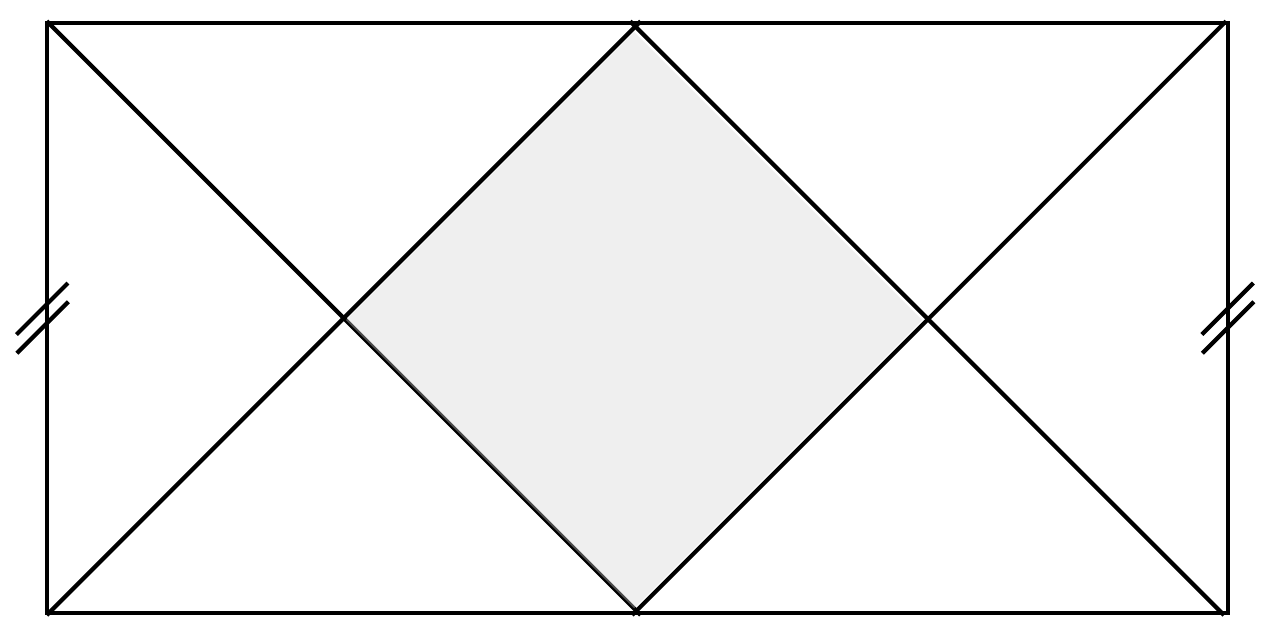}
\captionsetup{width=0.9\textwidth}
\caption{Static patch in de Sitter space. }
\label{ds}
\end{center}  
\end{figure}

We observe that
 \bea
\tau\rightarrow\pm\infty\,:&&(t,r)\rightarrow(\pm R,0)\label{junker9}\\
\theta\rightarrow\frac{\pi}{2}\,:&&(t,r)\rightarrow(0,R)\nonumber
 \eea

Note that $\theta=\pi/2$ corresponds to the cosmological horizon at the
boundary of the static patch. The static patch in de Sitter space is shown in the Penrose diagram of figure \ref{ds}. We
see that the new coordinates cover precisely the causal development of the ball $r\le R$ on the surface $t=0$, and this is mapped to the static patch.

Recall that the modular transformations acts geometrically inside the 
diamond corresponding to the ball, with modular Hamiltonian (\ref{modesf}) . Infinitesimal modular flow (see figure \ref{smf}) will move the time coordinate for points with $t=0$ as 
$\delta t= 2\pi(R^2-r^2)/(2R) \,\delta s$, with $\delta s$ the variation of the modular parameter. 
This corresponds
through the conformal mapping (\ref{trn2}) to the time translation
$\tau\rightarrow \tau+ 2 \pi R\,\delta s$ in de Sitter space. 
Therefore the modular
transformations act geometrically as time translations in the static
patch and the state in de Sitter geometry is thermal at temperature
$T=1/(2\pi R)$ with respect to the Hamiltonian $H_\tau$ generating
$\tau$ translations. Then, the density matrix is given by $\rho\sim
\exp\left[-2 \pi R\, H_{\tau}\right]$.

Therefore, the entanglement entropy for the sphere of
radius $R$ in flat space is equivalent to the thermodynamic entropy of
the thermal state in de Sitter space. We are going to use standard
thermodynamics to compute this entropy. Normalizing the
thermal density matrix $\rho=e^{-\beta H_{\tau}}/\textrm{tr}(e^{-\beta
H_{\tau}})$, we calculate the von Neumann entropy
 \bea
S=-\textrm{tr}(\rho\, \log \rho)&=&\beta \, \textrm{tr} (\rho H_\tau )+
\log \textrm{tr} (e^{-\beta H_\tau})\,,
 \nonumber\\
&=&\beta E\,-\,F \label{freeenergy}
 \eea
where $F=-\log Z$ denotes the `free energy' of the partition function
$Z=\textrm{tr} \left(\exp[-2 \pi R\,H_\tau]\right)$.

The energy term in (\ref{freeenergy}) is the expectation
value of the operator which generates $\tau$ translations. Since the
latter translations correspond to a Killing symmetry of the static
patch (\ref{round0}), $E$ is just the Killing energy which can be
expressed as
 \beq
E=\int_V d^{d-1}x \sqrt{h}\, \langle\, T_{\mu \nu}\,\rangle\, \xi^\mu\,
n^\nu= - \int_Vd^{d-1}x  \sqrt{-g}\,  \langle\,
T^\tau{}_\tau\,\rangle\,,
 \label{energy}
 \eeq
where the integral runs over $V$, a constant $\tau$ slice out to
$\theta=\pi/2$. Further $n^\mu \partial_\mu \equiv
\sqrt{|g^{\tau\tau}|}\,\partial_\tau$ is the unit vector normal to $V$
and $\xi^\mu\partial_\mu\equiv
\partial_\tau$ is the time translation Killing vector.

Now, for a CFT, after transforming the Minkowski vacuum to another space, the expectation value 
of the trace of the stress tensor needs not to be zero anymore. It has to be
 produced locally, however, out of some combination of curvatures with dimensionless coefficients, since there are no other scales in the theory. This is called the conformal anomaly. We have a general expression 
 \be  
\langle T_{\mu}^\mu\rangle = -2 (-)^{d/2} \, A\, E_d+\sum B_n\, I_n\,,\label{polo}
\ee
where $A$, $B_n$, are dimensionless coefficients (partially) characterizing the theory, $E_d$ is the Euler density, and $I_n$ are called Weyl invariants. Both $E_d$ and $I_n$ are constructed out of polynomials on the curvature tensor and have dimension $d$. As the curvature tensor has dimension $2$ the conformal anomalies are absent in odd dimensions. The Euler density is singled out because its integral is the Euler number, a topologically invariant of the space. For spaces that are conformal transformations of flat space, all the Weyl invariants vanish.

 The state in de Sitter, coming from a conformal transformation
of the Minkowski vacuum, which has maximal symmetry, is invariant under de Sitter symmetry group, and we have
 \beq
\langle\, T^\mu{}_\nu\,\rangle=\kappa \,\delta^\mu{}_\nu\,,
 \label{stud9x}
 \eeq
where $\kappa$ is some constant. Hence the expectation value of the
stress tensor is completely determined by the conformal anomaly. 
It turns out that as there are no divergences in this energy density, and the volume of the static patch is finite,  the energy term in (\ref{energy}) is finite for even $d$ and zero (no anomaly) for odd $d$.  

It remains to compute the contribution $F=-\log \textrm{tr}
\left(\exp[-2 \pi R\,H_\tau]\right)$. This can
be done, as usual, passing to imaginary time $\tau_{E}$ and
compactifying the Euclidean time with a period $\beta=2\pi R$. The
metric becomes
 \beq
ds^2=\cos^2\!\theta\, d\tau^2+
R^2\left(d\theta^2+\sin^2\!\theta\,d\Omega^2_{d-2}\right) \,.
 \label{none1}
 \eeq
This Euclidean manifold is precisely a $d$-dimensional sphere with
the radius of curvature $R$ written in unfamiliar coordinates.
The fact that this metric
 corresponds to the sphere may be more evident after the
coordinate transformation: $\sin \theta=\sin \theta_1
\sin \theta_2$ and $\tan (\tau/R) = \cos \theta_2 \tan \theta_1$, which
transforms the metric to
 \beq
ds^2=R^2\left(d\theta_1^2+\sin^2 \theta_1^2 d\theta_2^2 + \sin^2
\theta_1 \sin^2 \theta_2 d\Omega^2_{d-2}\right)\,.\nonumber
 \eeq  
 
 Note that the periodicity $\Delta
\tau=2\pi R$ is precisely that required to avoid a conical singularity
at $\theta=\pi/2$. Thus, one ends up with the Euclidean path integral
on $S^d$.

For even $d$ we only need to determine the coefficient of the
logarithmic term in the entropy, which is the only universal coefficient in this case. Now the free energy has a
general expansion
\begin{equation}
F=-\log Z=(\textrm{non-universal terms}) +a_{d+1}
\log{\delta}+(\textrm{finite terms})\,,
 \label{polo1}
\end{equation}
where $\delta$ is our short distance cut-off and the non-universal
terms diverge as inverse powers of $\delta$.  The coefficient $a_{d+1}$
for a conformal field theory is determined by the integrated conformal
anomaly. In order to see this, consider an
infinitesimal rescaling of the metric $g^{\mu\nu}\rightarrow(1-2 \delta
\lambda)g^{\mu\nu}$. 
\begin{equation}
\frac{2}{\sqrt{g}}\frac{\delta F}{\delta g^{\mu\nu}}=\langle\,
T_{\mu\nu}\,\rangle+(\textrm{divergent terms})\,.
 \label{none2x}
\end{equation}
In terms of the renormalized stress tensor $\langle\, T_{\mu\nu}
\,\rangle $, we have
\begin{equation}
\frac{\delta F}{\delta \lambda}=-\int d^dx\, \sqrt{g}\,
\langle\, T^\mu{}_\mu\,\rangle+(\textrm{divergent terms})\,,
\label{none3}
\end{equation}
which is the integrated trace anomaly. On the other hand, due to the
conformal invariance of the action, scaling the metric as above must
give the same result as keeping the metric constant but shifting the UV
regulator:  $\delta\rightarrow (1-\delta \lambda)\delta$. Combining
these expressions with eq.~(\ref{polo}), one finds
\begin{equation}
a_{d+1}= \int d^dx\, \sqrt{g}\,\langle\,T^\mu{}_\mu\,\rangle\,.
 \label{titres}
\end{equation}

Hence we are left to substitute eq.~(\ref{polo}) for the trace anomaly
and integrate over the $S^d$. Here we also need to observe that since
the sphere is conformally flat, all of the Weyl invariants $I_n$ vanish
for the sphere, while  the
integral of the Euler density on $S^d$ yields $2$. Hence for any CFT in
even dimensions, the universal contribution to the entanglement entropy
becomes
 \beq
S_{\textrm{log}}= (-1)^{\frac{d}2-1} 4\, A\, \log (R/\delta)\,,
 \label{none4}
 \eeq
 where we have completed the logarithmic divergence with the radius of the sphere. 
 In particular, we see that the coefficient of the
universal term in the entanglement entropy is proportional to the
central charge $A$. For $d=2$ the sphere is a single interval and there are no higher divergences than the logarithmic one. In $d=2$ the coefficient $A=c/12$, where $c$ is the Virasoro central charge of the CFT. We have
\be
S(R)=\frac{c}{3}\,\log(R/\epsilon)\,.
\ee

In the odd-dimensional case, we have seen $E=0$ and so
the entropy reduces to 
\be
S=\log Z=-F\,.
\ee
That is, the entanglement
entropy is simply minus the free energy on a sphere. The only universal part in this expression is the constant term of the entropy, which is then the constant term of $-F$
\be
S_{\textrm{cons}}=-F_{\textrm{cons}}\,.
\ee

\subsection{Entanglement at long distances}
\label{long}
Now we present an application of the fact that Renyi entropies come from expectation values of operators. We want to find how the mutual information behaves for two regions $A,B$ which are far away from each other. We know that the correlations vanish in this limit and the mutual information should vanish too, but we want to find the way it tends to zero. 
We consider a CFT.

Let the twist Renyi operator for $AB$ be
\be
\tau_{AB}^n=\tau_{A}^n \, \tau_B^n\,.
\ee
These operators act in the $n$-replicated QFT. From the perspective of long distances, these operators look like operators concentrated on two points. Then, they can be expanded into a basis of local operators at these two points. This operator expansion is very much like a multipolar expansion of classical fields but applied to operators. For our purposes here we do not need to know much about the details of this expansion, just that we should include all possibilities
\be
 \tau_{A}^n= c^0(A)+c^1_{\alpha}(A)\,\phi^{\alpha,i}(x)+ c^2_{(\alpha,i),(\beta,j)}(A)\, \phi^{\alpha,i}(x)\phi^{\beta,j}(x)+\cdots\,.  \label{expa}
\ee
The operators are evaluated at a point $x\in A$. The indices $\alpha$ denote the type
 of field in the original theory, and $i=1,\cdots,n$ the copy. Note that $c^0(A)$ and $c^1_{\alpha}(A)$ cannot depend on the copy.  
The Renyi mutual information is
\be
I_n(A,B)=S_n(A)+S_n(B)-S_n(AB)=(1-n)^{-1} \,\log\left(\frac{\langle \tau_{A}^n \, \tau_B^n\rangle}{\langle\tau_{A}^n \rangle \langle\tau_{B}^n \rangle}\right)\,.\label{rm}
\ee
In doing the expansion for long distances we have to order the operators in terms of increasing dimension. The lowest scaling dimension will give the largest long-distance correlator. Then, we start with the single copy operators of the lowest dimensions in (\ref{rm}). However, single copy operators cannot contribute to the mutual information. The reason is that in (\ref{expa}) both $c^0(A)$ and $c^1_{\alpha}(A)$ are functions of $n$, and we must have $\lim_{n\rightarrow 1} c^0(A)=1$ and $\lim_{n\rightarrow 1} c^1_{\alpha}(A)=0$ in the single copy limit where $\tau_A^1=1$. As we need two coefficients  $c^1_{\alpha}(A)$, $c^1_{\alpha}(B)$, for producing a correlator that affects the mutual information, the vanishing of the two coefficients should overcome the $(1-n)^{-1}$ factor in (\ref{rm}) and also make the contribution vanishes. They could contribute to the Renyi entropies for $n\neq 1$ however. 

This does not happen for the terms with operators in two copies in (\ref{expa}) which do not have any sense for $n=1$ (note $i\neq j$ in the sum, otherwise it is a single copy operator). The contribution of these operators overcomes the $(1-n)^{-1}$ because the number of terms in the sum is proportional to $n(n-1)$. The leading contribution will be dominated by products of two correlators, one in each of two different copies. The two-point function is proportional to $L^{-2 \Delta}$.  It is then simple to see that the general form of the leading long-distance expansion is of the form
\be
I(A,B)\sim C(A,B)\, L^{-4\, \Delta}\,,      
\ee
with $\Delta$ the lowest dimension of the theory. The coefficient $C(A,B)$ scales as the sizes $R_A^{2\Delta}R_B^{2 \Delta}$, and is a sum of products of coefficients for each of the regions independently. For a scalar field is just a product $C(A)C(B)$. The exact coefficient is known for spheres.

It is clear that the expansion continues with more negative powers of $L$ where the exponents are sums of the scaling dimensions $\Delta_i$ of the theory, and we must consider also the descendant fields (derivatives), which have dimensions $\Delta_i+1,\Delta_i+2,\cdots$. This means that from the complete knowledge of the mutual information we should in principle be able to reconstruct the full spectrum of scaling dimensions by looking at the different powers that appear in the long-distance expansion. This goes a good way into showing that mutual information uniquely determines the theory.  

This is also the place to remember that mutual information is careful enough to take into account all correlations between the two regions, and then be able to distinguish details. For example, some models are sometimes declared to be dual to each other without much care for details, as is the case of a free massless scalar and a Maxwell field in $d=3$. The duality is $F^{\mu\nu}=\varepsilon^{\mu\nu\sigma} \partial_\sigma \phi$. This duality then only shows that the Maxwell field and the derivatives of the scalar are the same theories with the same algebras and the same entropies. But the full scalar contains, in addition to the derivatives, the operator $\phi(x)$. It has a bigger algebra with larger mutual information. According to the above, the mutual information for the scalar decays as $\langle \phi(0)\phi(L)\rangle^2\sim L^{-2}$ while the one of the  Maxwell field decays much faster $\langle F_{\mu\nu}(0)F_{\mu\nu}(L)\rangle^2\sim L^{-6}$.

\begin{figure}[t]
\begin{center}  
\includegraphics[width=0.6\textwidth]{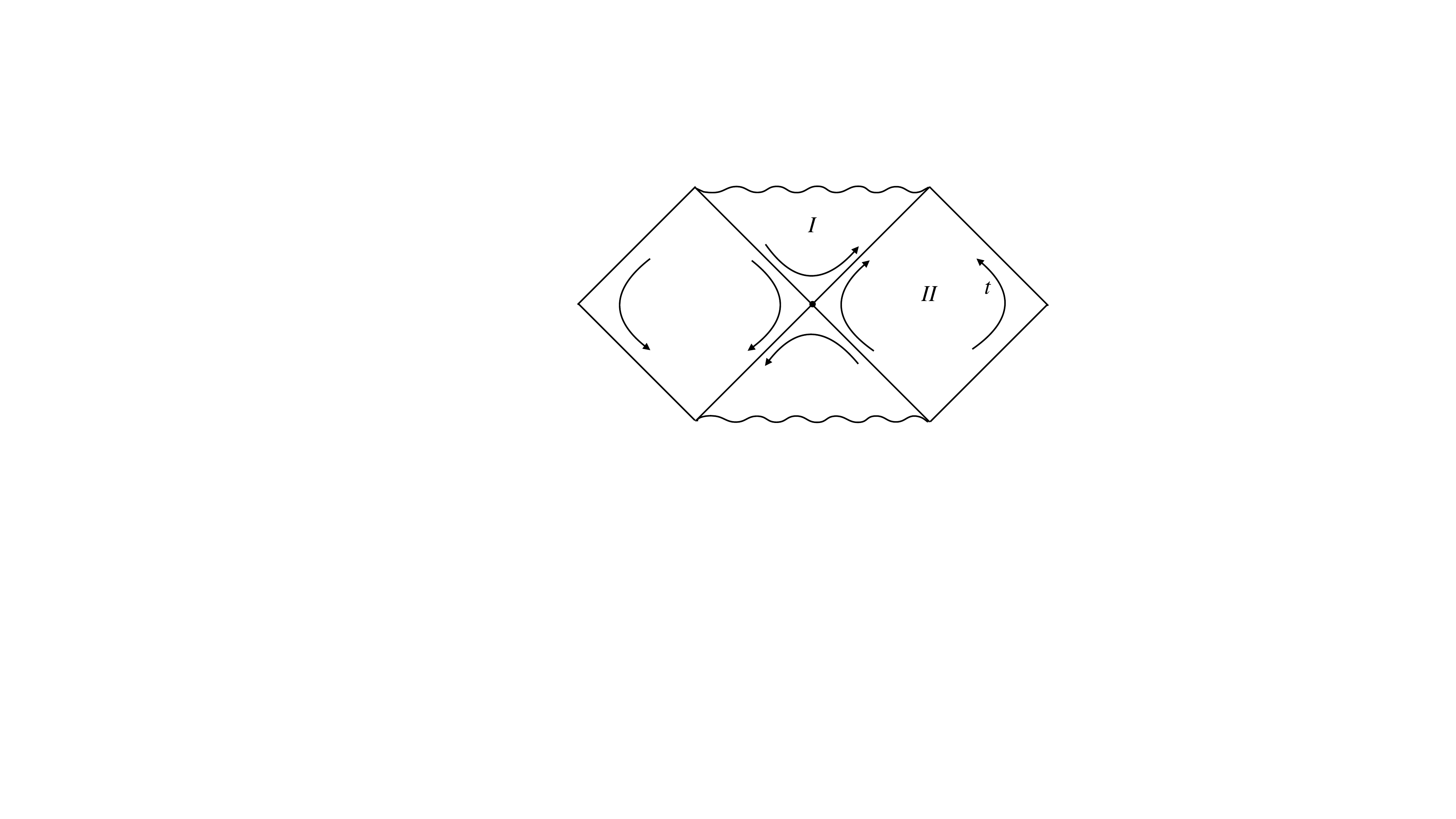}
\captionsetup{width=0.9\textwidth}
\caption{Eternal black hole space-time}
\label{bhh}
\end{center}  
\end{figure}
\subsection{A curved space-time excursion} 
\label{curved}

Figure \ref{bhh} shows the causal diagram corresponding to an eternal black hole. The arrows illustrate the flow of the killing symmetry that, for observers far away from the horizon,  looks like ordinary time translation symmetry. We can consider a QFT on this background and a state that is invariant under this symmetry. Consider the modular flow corresponding to this state and the region $II$. Quite near the horizon, at the boundary of this region, we expect the state and the flow to look like the ones for Rindler space. For a small region near the horizon, we should see not much difference with respect to flat space.  This should be so if the state looks like vacuum in small regions, which in this context is some form of the equivalence principle. Now we can expand the killing time $t$ near the horizon and compare this parameter with the boost parameter $\nu$. This is pure geometry. It turns out this relation is
\be
\nu= \kappa\, t\,,   \label{pop}
\ee       
where $\kappa$ is called the surface gravity.  
For example, for a Schwarzschild black hole
\be
ds^2=-\left(1-\frac{r_s}{r}\right) dt^2 + \frac{1}{\left(1-\frac{r_s}{r}\right)} dr^2 +r^2 d\Omega^2\,,
\ee
with $r_s=2 M G$ the Schwarzschild radius. Near $r=r_s$, making the coordinate  transformation $r=r_s+ z^2/(4 r_s)$, we get the metric transversal to the horizon
\be
ds^2= -z^2 \frac{dt^2}{4 r_s^2}+dz^2\,. 
\ee
Comparing this to the Minkowski metric in Rindler coordinates $ds^2= - d\nu^2 (x^1)^2 + (dx^1)^2$, with $\nu$ a boost parameter, we get $\kappa=(4 M G)^{-1}$. 

Equation (\ref{pop}) means the Killing symmetry that approaches the Hamiltonian $H$ at infinity approaches $\frac{\kappa}{2\pi} \hat{K}$ at the horizon, with $\hat{K}$ the modular Hamiltonian for Rindler wedge (which is $2\pi$ the boost operator).  
Since we expect  
 the usual modular Rindler flow corresponding to inverse temperature $1$ with respect to $\hat{K}$ near the horizon, we should have an ordinary  thermal state at  temperature 
\be
T= \frac{\kappa}{2\pi}\,  \label{dfgh}
\ee
for asymptotic observers. Therefore, it does not seem to be possible to put the black hole at thermal equilibrium unless the temperature is Hawking temperature (\ref{dfgh}), or, otherwise, something very violent happens at the horizon.

\begin{figure}[t]
\begin{center}  
\includegraphics[width=0.6\textwidth]{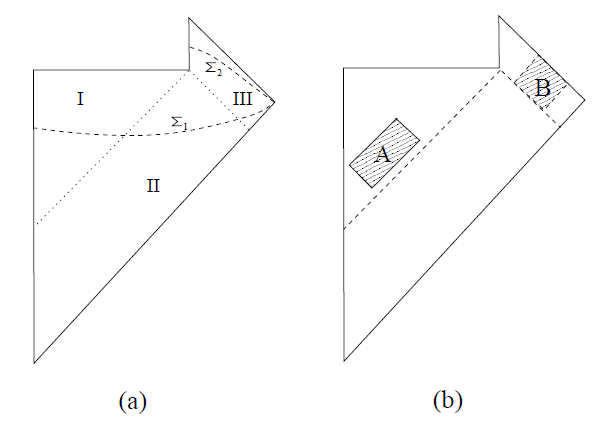}
\captionsetup{width=0.9\textwidth}
\caption{Space-time of an evaporating black hole.}
\label{bibo}
\end{center}  
\end{figure}

A black hole of a certain geometry is then a thermal object, and if not compensated by incoming radiation, it will evaporate. Figure \ref{bibo} shows a space-time where a black hole finally evaporates completely. In the semi-classical picture of QFT on a curved space one then expects a black hole generated by some pure matter to emit thermally, and in principle, the evolution will then not be unitary. This is called the black hole information paradox. Thermal radiation in the asymptotic region III is possible because of entanglement with the inside of the black hole region I.

 We do not say there is a paradox when we heat a rock at $T=0$ with a laser and it subsequently emits thermal radiation. In that case, we say we do not measure well enough to detect deviations from the thermal state such as to allow us to check the final radiation state is pure. But the black hole evaporation problem is a bit different from this rock, at least from the semi-classical perspective. Figure \ref{bibo} (right panel) shows two regions $A$ and $B$, one inside the horizon and one in the region of asymptotic observers. The mutual information between the two always increases with size, and then, it is not possible that once some mutual information is collected by $B$ in the form of Hawking radiation, this mutual information will go down to zero as required by a pure state in $B$ at the end of the evaporation process. This shows in a sharp way that no information can be retrieved semi-classically by asymptotic observers, no matter the interactions the QFT may have, and taking into account the full fine-grained information of the QFT.   

Then the evolution between the initial states at past infinity to the Cauchy surface $\Sigma_2$ in the figure is non-unitary. There is a unitary evolution, however, the one that goes to Cauchy surface $\Sigma_1$ and tracks this type of surface including the interior of the black hole as they move to the future. There are beautiful new results for holographic models that suggest that something like this later is actually happening in quantum gravity. In quantum gravity, we cannot play the game of taking algebras for regions of the space where the gravity lives. Gauge invariant operators have to be attached to something rigid, like the infinity, to be defined. So the algebra of the interior of the black hole has no gauge-invariant meaning and we cannot compute the mutual information between $A$ and $B$ in the figure. The asymptotically defined algebras are in principle well defined in quantum gravity, and represent the operator content accessible to asymptotic observers. Then, the trick that restores unitarity for asymptotic observers in the BH evaporation is that the space-time semiclassical picture of these algebras does look like taking the Cauchy surface that goes inside the horizon for late enough times, rather than $\Sigma_2$. The surprise here is that the semi-classical picture was not that wrong after all, and semiclassical unitarity of evolution does represent full quantum gravity unitarity.      

\subsection{Exercises}

\begin{itemize}

\item[1.-] (Luders's theorem) Consider the Hermitian operator $A=\sum_i a_i\, P^A_i$, where $a_i$ are the eigenvalues and $P^A_i$ the projections onto the space of eigenvectors for each eigenvalue. A measurement of $A$ transforms an arbitrary  state density matrix $\rho$ into an ensemble for the different eigenvectors given by the density matrix $\rho_A=\sum_i P^A_i \rho P_i^A$. Show that a measurement of another observable $B$ cannot distinguish between $\rho$ and $\rho_A$ for {\sl any} $\rho$ if and only if $[A,B]=0$.  

\item[2.-] a) Consider a real scalar field $\phi$ of dimension $\Delta$, such that $\langle \phi(0)\phi(x)\rangle=|x|^{-2\Delta}$. Define $\phi_\alpha=\int d^d x\, \alpha(x) \phi(x)$. Show that $\phi_\alpha|0\rangle$ has divergent norm if the smearing function $\alpha(x)$ has support in a piece of a $(d-2)$-dimensional spatial plane. The same is true for a support in a piece of a null plane, considering the unitarity bound $\Delta> (d-2)/2$. However, for a free scalar, the operator $\partial_+\phi$, where the derivative is in the null direction of the null plane, can produce localized operators by smearing on the null plane.  

\noindent b) Consider a smeared operator $\phi_\alpha=\int d^dx\, \alpha(x) \phi(x)$ formed out of a free scalar field $\phi(x)$.  $\phi_\alpha$ generates an Abelian algebra and the vacuum is Gaussian in this algebra. This can be represented as the algebra of functions on the real variable $\phi_\alpha$. How do you write the probability distribution of the Gaussian state on this variable $\phi_\alpha$? Show that the entropy for this classical continuous variable is badly defined: consider what happens to the entropy changing variables from $\phi_\alpha$ to a function of it. Show that the mutual information with other commuting smeared mode $\phi_\beta=\int d^dx\, \beta(x)\phi(x)$ is given by     
\begin{equation}
I=\frac{1}{2}\log \langle \phi_\alpha^2 \rangle +\frac{1}{2}\log \langle \phi_\beta^2 \rangle-\frac{1}{2}\textrm{tr}\log \left(\begin{array}{cc} \langle \phi_\alpha^2 \rangle& \langle \phi_\alpha \phi_\beta\rangle\\ \langle \phi_\alpha \phi_\beta\rangle &  \langle \phi_\beta^2 \rangle\end{array}\right)\,,
\end{equation}
and does not have the problems of the entropy. It can be used as a lower bound to the mutual information of any algebras where these modes are included. Show that for large distances between the support of the modes it has the right decay expected for the mutual information of distant regions in the free scalar. Show that if $\langle \phi_\alpha^2 \rangle$ diverges leaving $\langle \phi_\alpha \phi_\beta \rangle$ bounded the mutual information vanishes. Taking into account the item a) what does this suggest about the limit of mutual information when one of the regions collapses to a lower dimensional object?.  

\item[3.-] The heat kernel for the Laplacian in some Euclidean manifold is defined as $K(t)= \textrm{tr}\, e^{t \partial^2}$. For a free scalar field  we have the path integral partition function $\log Z=-(1/2)\textrm{tr} \log (-\partial^2)$. This can be written $\log Z= (1/2)\int_{\epsilon^{2}}^\infty \frac{dt}{t} \, K(t)$ where $\epsilon$ is a short distance cutoff. Why? Use the eigenvalues $\lambda_l$ and degeneracies $d_l$ of $-\partial^2$ on the spheres $S^2$ ($\lambda_l=l(l+1)$, $d_l=2 l+1$) and $S^4$ ($\lambda_l=l(l+3)$, $d_l=(1/6)(l+1)(l+2)(2 l+3)$) to write the heat kernel as an infinite sum. Expand this sum for small $t$ as an integral plus corrections given by the Euler MacLaurin formula. Use the result for computing the coefficient of the logarithmic term in the entropy of intervals in $d=2$ (answer $1/3$) and spheres in $d=4$ (answer $-1/90$). Beware: for $d\ge 3$ the massless free scalar can be conformally transformed to curved spacetimes provided we couple it with the curvature with some specific coefficient (this is called ``improving''). The equation of motion in curved space is 
\be
\left(-\partial^2 +\frac{(d/2-1)}{2(d-1)} R\right)\phi=0\,, 
\ee
with $R$ the scalar curvature of the space. This changes the operator to which we have to compute the heat kernel for $d=4$. The curvature scalar of the $S^4$ unit sphere is $R=12$.

\end{itemize}

\subsection{Notes and references}
Entanglement entropy in QFT was first considered by Sorkin and collaborators in \cite{Sorkin:2014kta,Bombelli:1986rw} in the search of a quantum origin to black hole entropy. This idea remained quite unnoticed until rediscovered by Srednicki in \cite{Srednicki:1993im}. 
There are many papers where entropy formulas for Gaussian states are obtained with different levels of generality. See for example  \cite{peschel2009reduced} for formulas and references. 
The free field lattice calculations are reviewed in \cite{Casini:2009sr}. The understanding of the structure of EE was made explicit in  \cite{Liu:2012eea,Grover:2011fa}. The replica method applied to EE was introduced by Callan and Wilczek in \cite{Callan:1994py}. The Renyi twist operators were introduced by Calabrese and Cardy in \cite{Calabrese:2004eu}. The paper of Bisognano and Whichmann 
 is \cite{Bisognano:1975ih}. This was published about the same time as Unruh's paper  \cite{unruh1976notes}, though the connection was made evident much later.  
 \cite{Hislop:1981uh} contains its generalization to spheres by Hislop and Longo, in a mathematically oriented context. A field theory proof for CFT's is in \cite{Casini:2011kv}, and this paper also contains its generalization to other curved spacetimes by conformal transformations. The computation of the universal terms for spheres in section \ref{tow} are from \cite{Casini:2011kv}. The case of $d=2$ \cite{Holzhey:1994we}, and $d=4$ \cite{Solodukhin:2008dh}, were previously known. In the paper by Solodukhin \cite{Solodukhin:2008dh} the expression of the logarithmic term for a generally smooth surface in $d=4$ was computed in terms of anomaly coefficients and intrinsic and extrinsic curvatures of the surface. The long-distance expansion was developed by Cardy in \cite{Cardy.esferaslejanas}. See also \cite{Agon:2015ftl}. The relation between Hawking radiation, Rindler modular flow and CPT theorem can be seen in its Lorentzian version in \cite{sewell1982quantum}. The equivalent argument in Euclidean space is older. The argument about information loss using the mutual information monotonicity in section \ref{curved} is from \cite{Casini:2007dk}.        The same idea was later rediscovered (expressed in terms of strong subadditivity instead of monotonicity of mutual information) in \cite{Mathur:2009hf} and \cite{Almheiri:2012rt}.

\newpage
 
\section{Irreversibility theorems}

\subsection{Changes with scale: the renormalization group}
One could have expected that the physics of a QFT at very different scales could be selected in a completely independent manner. However, this is not the case. There is an asymmetry between the UV and the IR, and this is the subject of the irreversibility theorems. They allow us to grasp some aspects of this asymmetry. 

\begin{figure}[t]
\begin{center}  
\includegraphics[width=0.3\textwidth]{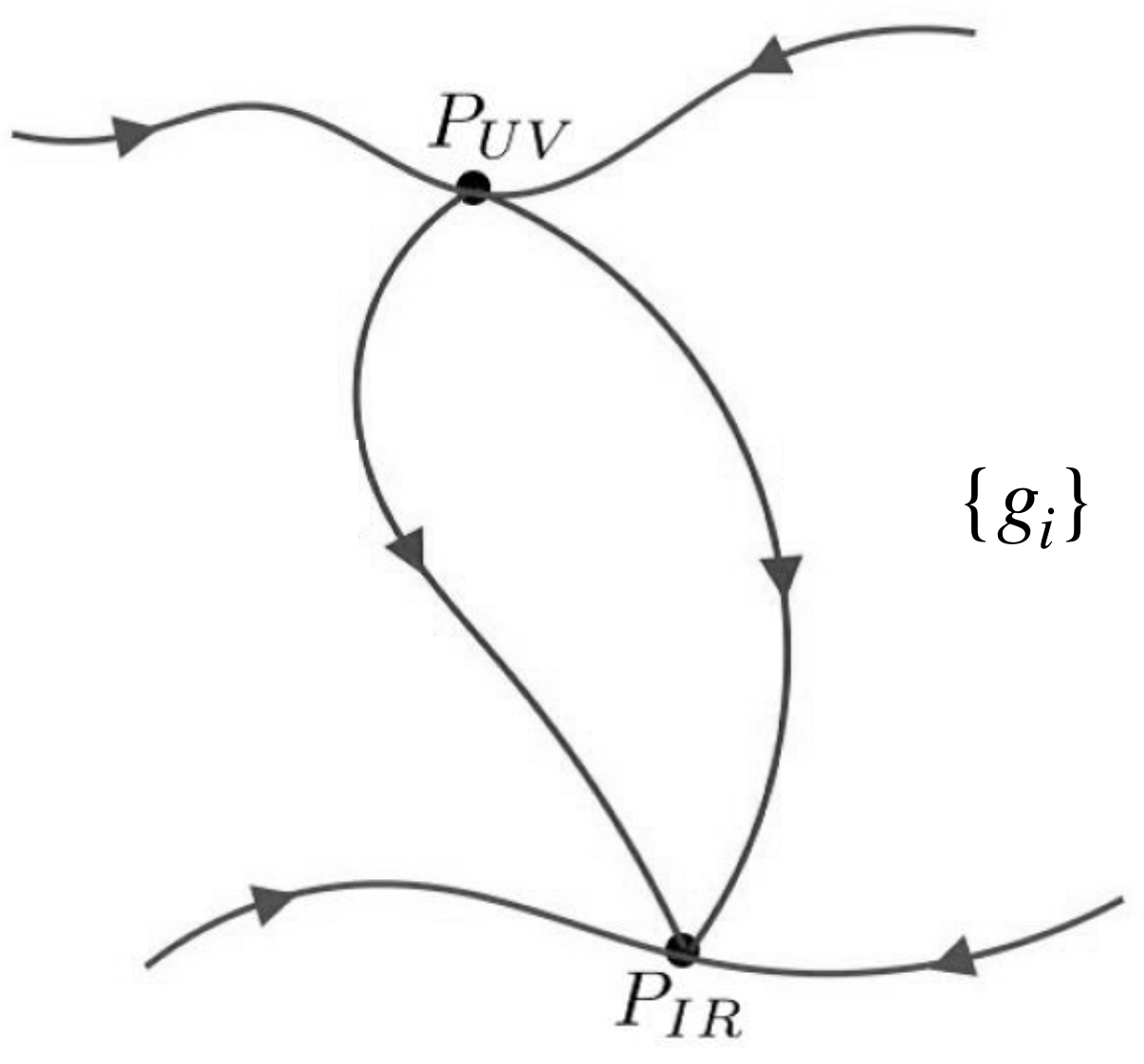}
\captionsetup{width=0.9\textwidth}
\caption{The fixed points $P_{UV}$ and $P_{IR}$ in the space of couplings $\{g_i\}$. The arrows indicate the direction of the RG and start at a UV fix point by a relevant deformation and end at an IR fix point by an irrelevant deformation.}
\label{rg}
\end{center}  
\end{figure}

In the perturbative approach, the change of physics with scale is called the renormalization group. It describes how theories change with some scale $\mu$ by describing the flow of the different coupling constants $g(\mu)$ determining the effective Lagrangian at scale $\mu$. See figure \ref{rg}. This flow in the space of couplings gives a flow in the space of theories. 
 This approach is tied to the perturbative description. The coupling constant is not a well-defined concept non perturbatively. 

Then, to check some form of irreversibility one could use other forms of testing the theory at different scales. The simplest one is to use correlators. The irreversibility theorems are general statements valid for all theories, and then it is not surprising that the first such theorem by Zamolodchikov uses stress tensor correlators.  This is an operator that can be defined for any theory (or at least for most of them). Some auxiliary field, a dilaton, is used in Komargodski-Schwimmer proof of the a-theorem in $d=4$.

Clearly, another quantity that is universally defined is the EE. It also has an air of knowing much about irreversibility because of its role in the second law. So we will be trying to understand irreversibility from EE.
  
One starts with a theory at the UV which is invariant under changes of scale and ends at the IR at another one. It is generally believed that scale invariance is generally enhanced in QFT to conformal symmetry. We assume then that the UV and IR fix points are CFT's. There is an asymmetry in the game in that the flow starts at the UV by turning on a relevant operator $\phi_\Delta$ (adding a term proportional to the integral of this operator to the action) with conformal dimension $\Delta< d$ and ends at the IR with an irrelevant perturbation, with $\Delta>d$. 
We will be searching for a quantity that is defined for CFT's and changes with a specific sign from the UV to the IR. It is important that this quantity has to be universal, in the sense that is regularization invariant, or an object of the continuum theory itself. As it characterizes a CFT it has to be dimensionless.   

The other point is how to describe different scales. For the entropy this is somewhat simple, we take scaled regions. The first idea that comes to mind is to take mutual information for scaled regions. This goes from some dimensionless number at the UV to another one at the IR, and this number can only depend on the CFT because the conformal geometry of the two regions has not changed under scaling. However, in scaling {\sl two} regions it is not possible to use monotonicity, and mutual information is in general not monotonic under scaling. For a single region, we can have an inclusion under scaling, but the entropy is not well defined (nor monotonous under inclusion, though the divergent area term renders it trivially monotonic and useless at the same time). For two states in the same region, the relative entropy is monotonous undertaking scaled (included) regions. However, the theory selects only one preferred state, there is only one vacuum. The theory is preventing us from getting a too easy result, no doubt from good reasons of its own. In particular, these simple and failed ideas did not use Lorentz symmetry.     

\subsection{The vacuum a Markovian state?}
\begin{figure}[t]  
\begin{subfigure}
\centering
\hspace{1.3cm}\includegraphics[width=0.4\textwidth]{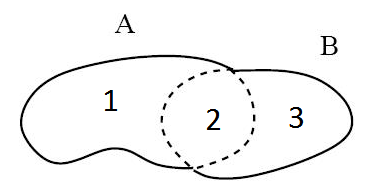}
\end{subfigure}
\hspace{1cm}
\begin{subfigure}
\centering
\includegraphics[width=0.4\textwidth]{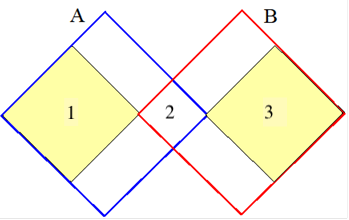}
\end{subfigure}
\captionsetup{width=0.9\textwidth}
\caption{Strong subadditivity of two intersecting regions. Left: spatial picture, right: space-time picture.}
\label{ssa}
\end{figure}

We start with the inequality we are going to use to test irreversibility. This is the strong sub-additive inequality (SSA) of the entropy. It is the deepest inequality the entropy posses, and it is an aspect of the monotonicity of relative entropy. For algebras attached to regions it acquires a geometric form (see figure \ref{ssa}),
\be\label{markov0}
S_A+S_B-S_{A\cap B}-S_{A\cup B}\ge 0\,. 
\ee

Now, for our purposes, we need that this inequality, at least when applied to some specific geometrical configuration, to saturate for a CFT. This is so because for a CFT, the dimensionless quantities to which the irreversibility would apply, whatever they would be, must not change under scaling. Then our inequality, which would drive the change of the ``RG charges'' for different sizes, should be saturated for the configuration of interest and a CFT, where no RG flow is taking place. This will guide us much in the following.

 Then we need to find some regions $A,B$ such that for a CFT 
\be\label{markov}
S_A+S_B-S_{A\cap B}-S_{A\cup B}= 0\,. 
\ee
When a state satisfies this equation it is called a Markovian state between $1$, $2$, and $3$ in figure \ref{ssa}. As this is a saturation of an inequality it must be that the state is rather special. The name comes from the Markov process in classical probability. There we have three parts  $1,2,3$ of a system whose possibilities are described by some set of indices $i,j,k$ for each of the parts. The state is given by $p(ijk)$, determining the joint probability of getting $ijk$ for the full system. The state (or the process) is Markovian if the conditional probability $p(i|jk)$ of getting $i$ given that we have $jk$, is equal to the conditional probability $p(i|j)$ of getting $i$ given that we have $j$. In equations, this gives 
\be
p(i|jk)=\frac{p(ijk)}{p(jk)}=p(i|j)=\frac{p(ij)}{p(j)}\,.   \label{esq} 
\ee
In words, a state is Markovian when nothing new about $1$ can be known knowing $23$ that was not known knowing only $2$. The relation between $1$ and $3$ is determined by the relation between $1$ and $2$, and the relation between $2$ and $3$. Markovian probabilities are used to describe stochastic processes where the evolution is local in time, here the three times $1,2,3$. Equation (\ref{esq}) gives
\be
p(ijk)=\frac{p(ij)p(jk)}{p(j)}\,\iff S(12)+S(23)=S(123)+S(2)\,.
\ee

Then, we see the Markov equation for the entropy is giving us restrictions on the possible form of the correlations between different times, which are to be intermediated by the intermediate times. For our purposes, this relation is now in space rather than time, as in figure \ref{ssa}.

At the quantum level, a saturation of strong subadditivity also constraints the structure of the state $123$ to be rather special. The general result is that the space ${\cal H}_2$ could be written in the form of a sum over factors  
\be
 {\cal H}_2=\bigoplus_k  {\cal H}_{2 L}^k \otimes{\cal H}_{2 R}^k\,,
\ee
such that
\be
\rho_{123}= \bigoplus_k p_k\, \rho_{1, 2 L}^k\otimes \rho_{2 R, 3}^k\,,
\ee
where $p_k$ are probabilities.

This structure has an important consequence. If we restrict attention to the system $(13)$, tracing over $(2)$ we get
 \be
\rho_{13}= \bigoplus_k p_k\, \rho_{1}^k\otimes \rho_{3}^k\,.
\ee
This is precisely the general form of unentangled states between $(1)$ and $(3)$. Then a condition to have saturation of SSA is the absence of quantum entanglement between these two regions. 

\subsection{Quantum entanglement and the Reeh-Schlieder theorem.}

The Reeh-Schlieder theorem is an old theorem in QFT that says something a bit startling. It says that given any region $W$ containing some space-time volume, however small, acting with operators in $W$ on the vacuum we can get a dense set of vectors in the Hilbert space. Even far away particles can be approximated by acting with operators in a small region $W$. Technically, it is said that the algebra on $W$ acts cyclically on the vacuum: it generates a dense set of states out of the vacuum. Reeh-Schlieder property is a bit counter-intuitive, but the same happens for a simple finite bipartite system where the two-factor spaces are of the same dimension and the density matrix has no zeros. Then, it is a property that requires entanglement, but it can be a small amount of entanglement.   

\begin{figure}[t]  
\begin{subfigure}
\centering
\hspace{1.3cm}\includegraphics[width=0.32\textwidth]{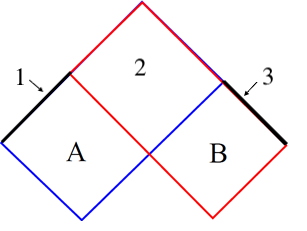}
\end{subfigure}
\hspace{2cm}
\begin{subfigure}
\centering
\includegraphics[width=0.32\textwidth]{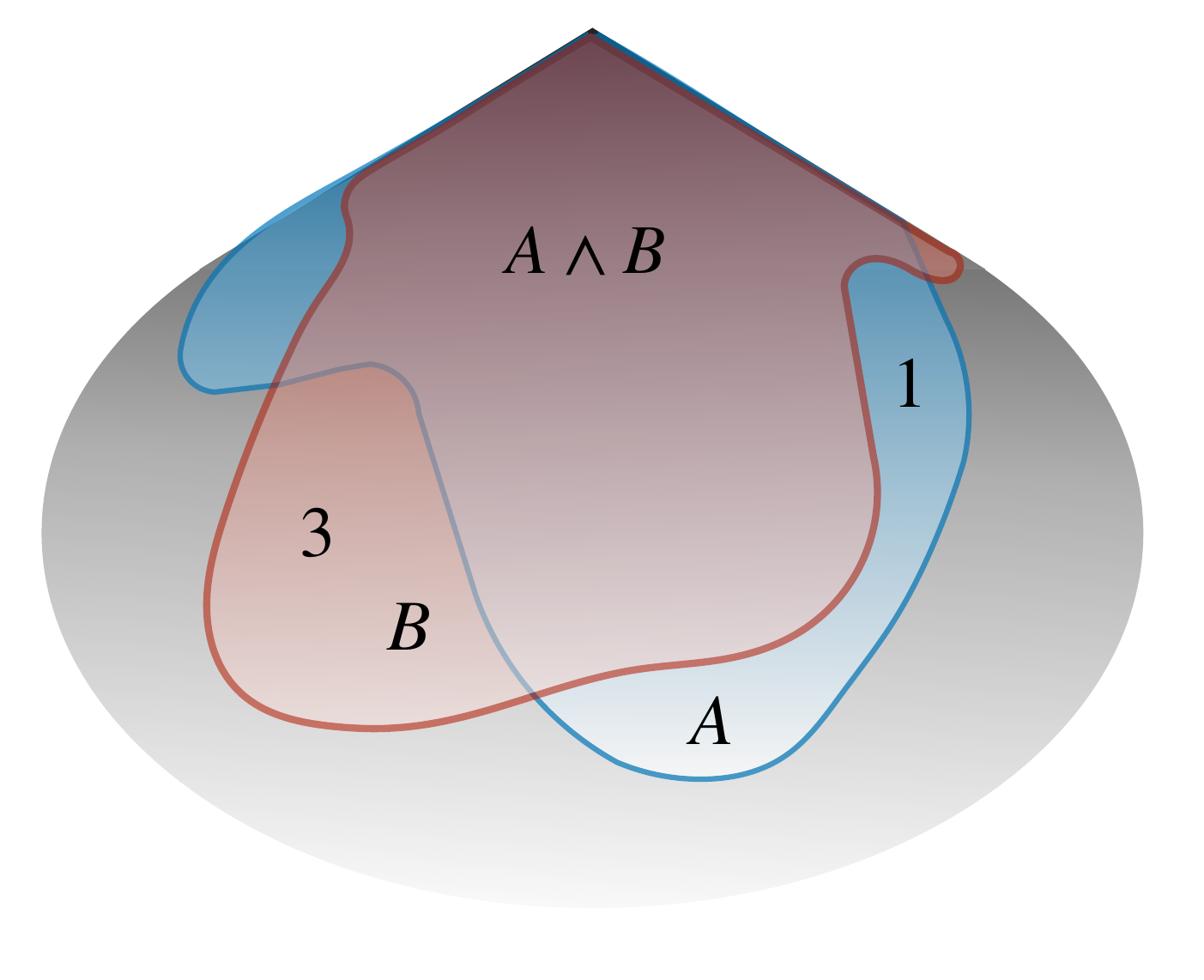}
\end{subfigure}
\captionsetup{width=0.9\textwidth}
\caption{Left: regions $1$ and $3$ are null and do not have space-time volume. Right: two regions with boundary in a null cone. }
\label{cono1}
\end{figure}

In QFT, the reason for this property is the analyticity of correlation functions, which in turn is a consequence of the positivity of energy and Lorentz invariance. Suppose that acting with the field operators in $W$ on the vacuum we do not obtain a dense set of vectors. Then there should be some vector $|\psi\rangle$ orthogonal to the generated set
\be 
\langle \psi|\phi(x_1)\,\cdots \phi(x_n) |0\rangle=0\,,\hspace{.6cm} x_1\,\cdots\,x_n\in W\,. 
 \ee
As these correlation functions are analytic and vanish on $W$, it turns out that they have to vanish for every $x_1\,\cdots\,x_n$ now not restricted to $W$. But the operators on the full space acting on the vacuum do generate a dense set in the Hilbert space. This in fact defines what is the full Hilbert space in analogy with Fock space for free fields. Therefore the only possibility is $|\psi\rangle=0$.

There is another interesting consequence of this theorem. The vacuum is not only cyclic for the operators in $W$ but also separating. This means that one cannot annihilate the vacuum with any non-trivial operator in $W$. To annihilate the vacuum one needs non-local operators, that cannot belong to the algebra of any bounded region, such as an annihilation operator for a free field, or the Hamiltonian or a charge generator. This property can be deduced from the other: it is not difficult to realize that if the vacuum is cyclic for an algebra it will be separating for the commutant of that algebra.  
 
Returning to the original purpose of looking at regions with no quantum entanglement in QFT, we see that the Reeh-Schlieder theorem basically forbids it. If we have regions $(1)$ and $(3)$, acting with operators in $(1)$ on the vacuum we could produce a state resembling an entangled pair of qubits in $(13)$. We could produce it for example using some projector in $(1)$, and obtain the entangled pair with generally very small, but non zero probability. Therefore the vacuum state could not be unentangled between $1$ and $3$, because local operations in $1$ or $3$ cannot create entanglement, though they can destroy it. It is possible to show rigorously that there is always non zero distillable entanglement for any two regions with non zero space-time volume in QFT.

\subsection{Markovianity for regions on the null plane}

The same Reeh-Schlieder theorem that forbids saturation of SSA for general regions gives us a hint on a special situation where this can be avoided. To apply the Reeh-Schlieder theorem we need regions with space-time volume because a unique analytic extension of an analytic function from its values in a neighbourhood is important in the argument. This is prevented if regions $1$ and $3$ do not have any space-time volume.

For causal regions, the only possibility is the regions $1$ and $3$ are null. This can be achieved if $A$ and $B$ have their boundaries in the same null surface (see figure \ref{cono1}). In most cases, however, this will not be enough to make SSA vanish. For example, this is the case of an ordinary QFT (not conformal) for an arrangement like the one of figure \ref{cono1}, right panel, where the regions have a boundary on a null cone. 

As a side remark, we recall that algebras on a null surface as the corresponding ones to regions $1$ and $3$ in this figure are non-trivial only if the theory is UV free (for $d>2$). For other cases these algebras are trivial, but that does not prevent the SSA to make sense. The SSA inequality in fact does not involve any region with zero volume, even if the regions $1$ and $3$ are null and have zero volume. 

  \begin{figure}[t]
\begin{center}  
\includegraphics[width=0.4\textwidth]{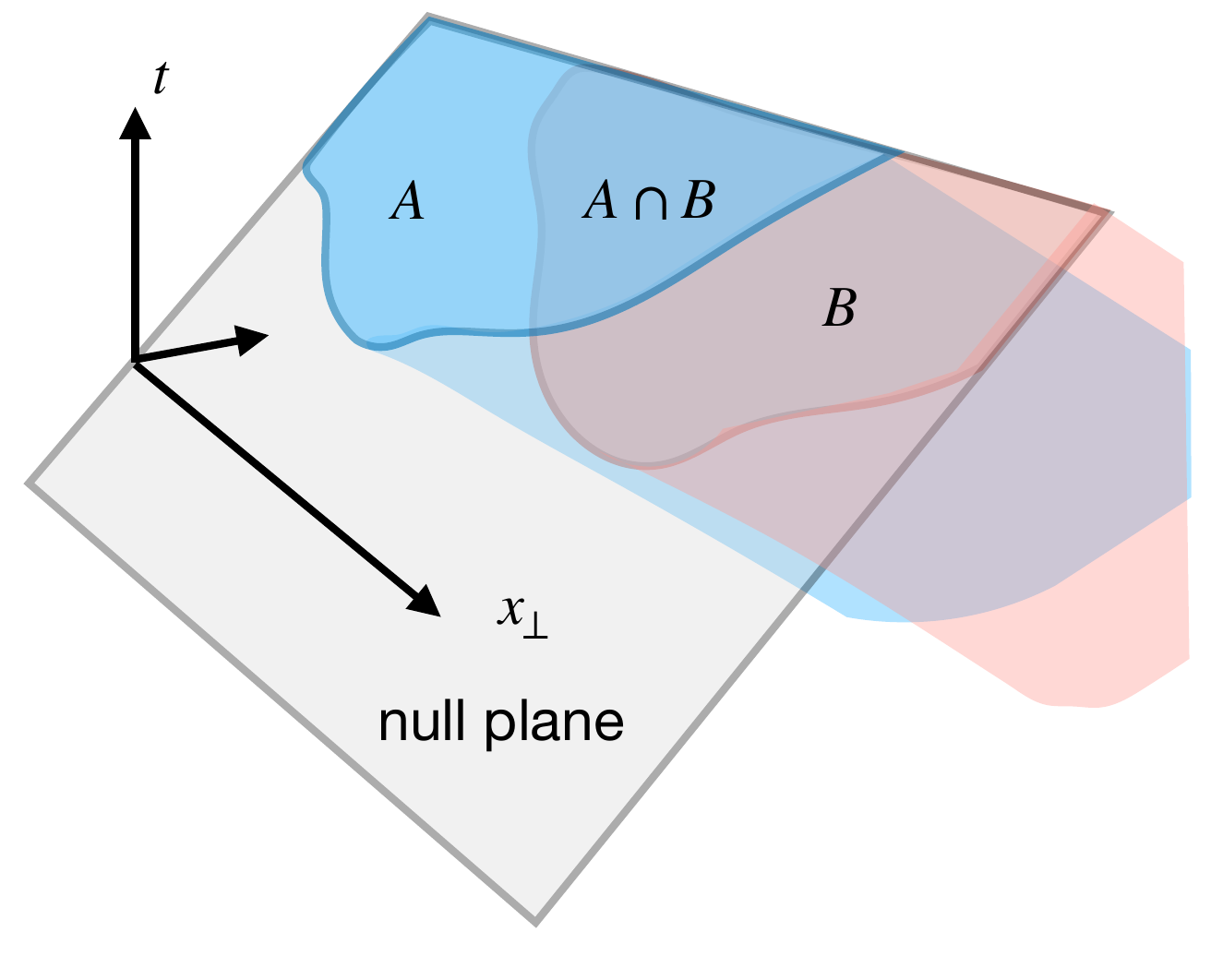}
\captionsetup{width=0.9\textwidth}
\caption{Regions with boundary in a null plane. }
\label{planee}
\end{center}  
\end{figure}

\begin{figure}[t]
\begin{center}  
\includegraphics[width=0.45\textwidth]{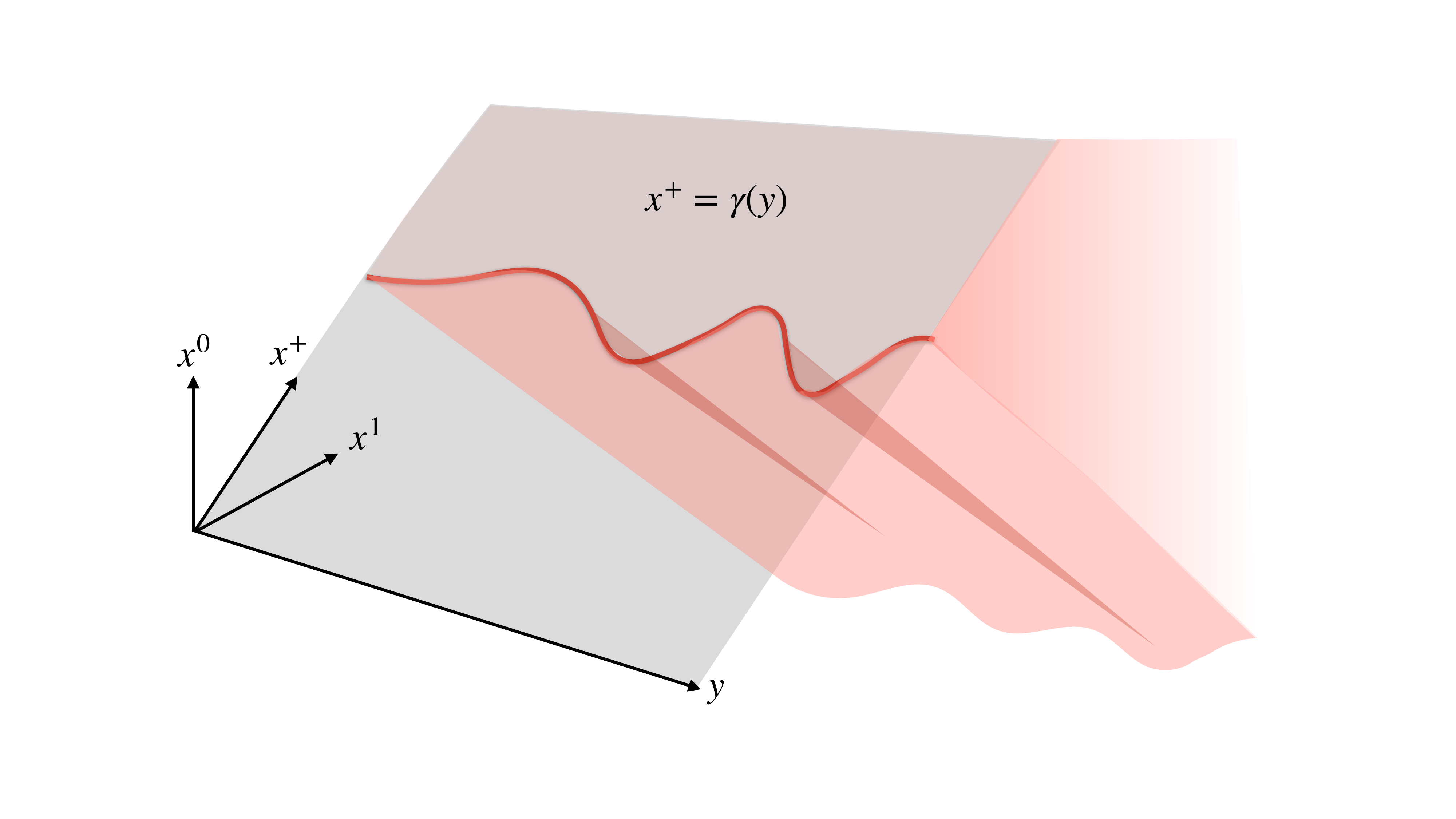}
\captionsetup{width=0.9\textwidth}
\caption{Space-time region with boundary $x^+=\gamma(y)$ (red curve) on the null plane  $x^-=0$.
}
\label{plapla}
\end{center}  
\end{figure}

Surprisingly, there is a special case where SSA vanish for all QFT. This is when the regions have the boundary in the same null plane, as in figure \ref{planee}.   
 The reason is very simple and geometrical. Let us describe a causal region with boundary on the null plane by writing the boundary with the function $x^+=\gamma(y)$, where $x^+=x^0+x^1$ and $y$ represents the vector of spatial coordinates parallel to the null plane, $y=(x^2,\cdots,x^{d-2})$. See figure \ref{plapla}. For a cut-off theory, the entropy of this region is a function of $\gamma$. If we choose a Lorentz invariant cut-off (what can be done for example taking limits of the mutual information and subtracting divergences), we can make a boost keeping the null plane invariant and we will get the same entropy. The boost just rescales $x^+\rightarrow \lambda\, x^+$, with $\lambda>0$. We have
 \be
 S(\lambda \,\gamma)=S(\gamma)\,.
\ee 
Taking the limit $\lambda\rightarrow 0$ the region approaches the Rindler wedge. As we can do this for all regions $\gamma$ it means the entropies for all these regions are the same with a Lorentz invariant cutoff. Another way to say this is that on the null plane there are no Lorentz invariant geometric features on which our entropies could depend. 

This gives us trivially Markovianity (\ref{markov}) for any arrangement as the one on figure \ref{planee}. Once we have this relation we can lift the Lorentz invariant cutoff and use any other cutoff and the Markov property will still be obeyed, because divergent terms are always local and extensive, and therefore they are cancelled in the SSA inequality.\footnote{One should take configurations with smooth intersection and union to avoid singular terms from corners. On the null plane the possible new terms do not affect the Markovianity.}  

Notice this result is very general and geometric. It does not depend on any peculiar property of the entropy but on the Lorentz symmetry. It will also hold for Renyi entropies, or more generally, free energies of insertions of $d-2$ dimensional surface operators. It will also hold in curved space-times with a bifurcate Killing horizon, such as eternal black holes, or de Sitter space, provided the state is invariant under the Killing symmetry, and we take our regions with boundary on the Killing horizon.

\subsection{General form of the EE for a CFT on the null cone}
\begin{figure}[t]
\begin{center}
\includegraphics[width=0.45\textwidth]{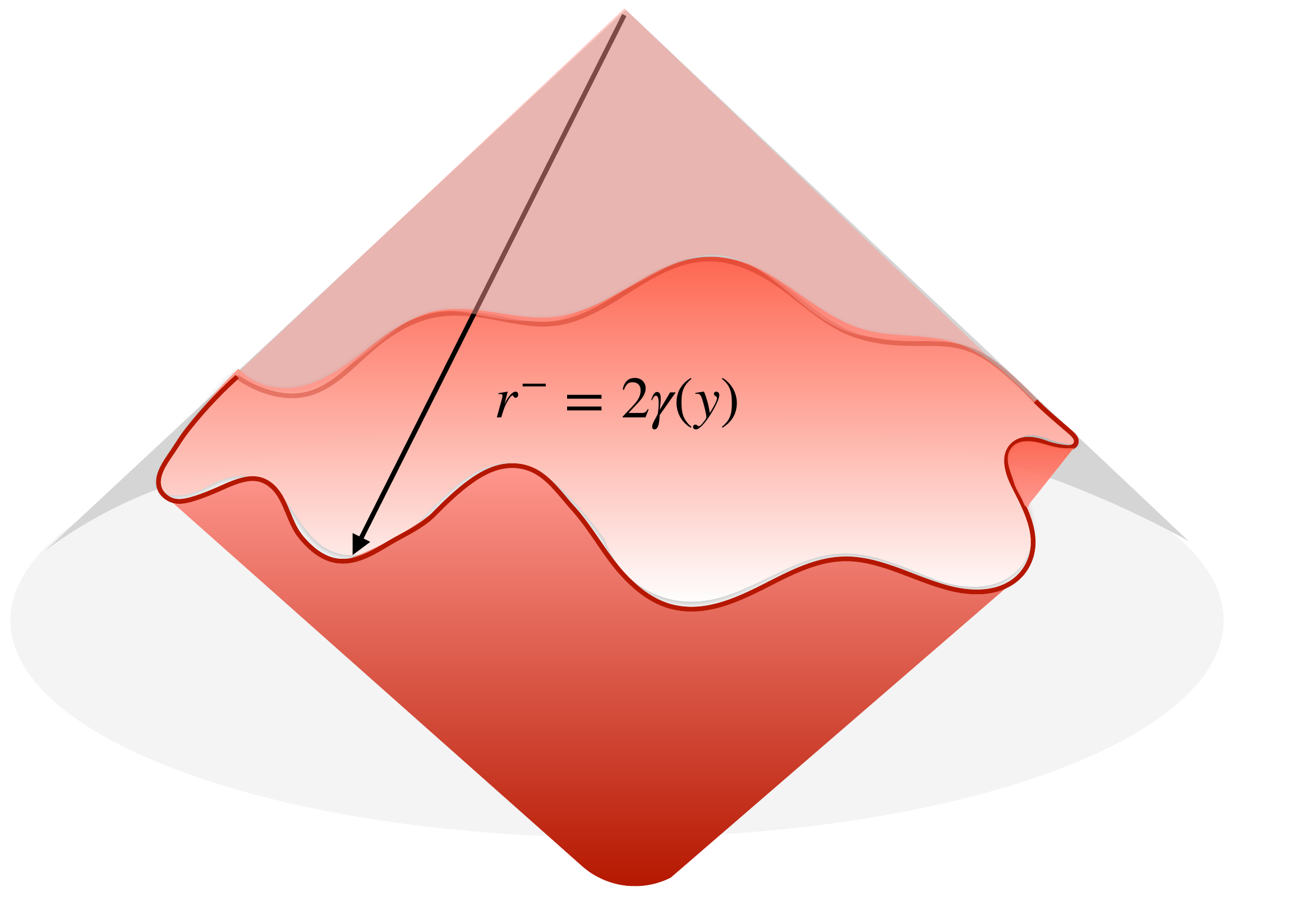} 
\caption{A region with boundary on the light-cone. This setup applies for CFTs.}
\label{rmenos}
\end{center}
\end{figure}

The previous result on the null plane applies to all QFT. For a CFT a null plane can be mapped into a null cone. We can then conformally map this result from regions with boundaries on the null plane to regions with boundaries on the null cone. These are described by a function $r^-=2 \,\gamma(y)$ where $r^-=r-x^0$, see figure \ref{rmenos}. The coordinates $y$ can be taken as angular variables. 
 And this is exactly the setup we were looking for: regions where SSA saturates for a conformal theory but does not saturate away from the fixed point. 

What does the Markov property tell us about the EE of a CFT for regions with boundaries on the null cone? It is not difficult to convince oneself that the equation (\ref{markov}) tell us the EE is a local functional of $\gamma(y)$. In other words, $S(\gamma)$ has to depend on $\gamma$ additively in the angular variables. In this way, the SSA vanishes null ray per null ray, since for each null ray labelled by $y$  the maximum between $\gamma_A(y)$ and $\gamma_B(y)$ will appear in $\gamma_{A\cup B}(y)$, and the minimum will appear in $\gamma_{A\cap B}(y)$. Of course, this locality of the entropy always holds for the divergent terms, but the special feature here is that possible finite terms must be also local in the angular variables.

Then one can think of the EE on the null cone as an ``action'' depending on the dimensionless field $\gamma(y)/\epsilon$ on the $d-2$ dimensional unit sphere. This action has to be local, and further, one can take it to be invariant under Lorentz transformations, which maps the null cone in itself. If $\hat{g}_{ab}$ is the induced Minkowski metric on the surface $\gamma$, and $g_{ab}$ the one of the unit sphere, we have
\be
\hat{g}_{ab}=\gamma^2(y)\, g_{ab}\,.
\ee
The physical metric on the surface $\hat{g}_{ab}$ is invariant under Lorentz transformations. Lorentz transformations induce a multiplicative change on the curve $\gamma\rightarrow e^{A(y)} \gamma$. Therefore, Lorentz transformations act like a conformal map of the unit sphere into itself. 
Then we have to write an action for $\gamma(y)$ on the unit sphere that is local and conformally invariant.

The solution to this problem was studied previously in relation to dilaton fields. Here $\log(\gamma)$ is a dilaton field, in the sense that it transforms additively under conformal transformation since $\gamma$ transforms multiplicatively. We will not describe the details on the derivation of the possible actions since in fact, we will not need this result to prove the irreversibility theorems. We just quote the general form of the EE for $d=3$ and $d=4$. 

For $d=3$ we have 
\be
S(\gamma)=\int d\theta\,\left (c_1\, \frac{\gamma(\theta)}{\epsilon} -\frac{F}{2\pi} \right)\,.\label{tresd}
\ee
The first term is of course the area term, while the second one is the constant term (or $F$-term). We see the constant term that we computed for the sphere is the same for any curve in the light cone. 

For $d=4$ we have
\be
S(\gamma) = \int d^2 \Omega \left \lbrace  c_2\, \frac{\gamma^2}{\epsilon^2}- \frac{A}{\pi} \,  \left(\log \frac{\gamma}{\epsilon} + \frac{1}{2}\left(\frac{\nabla_\Omega \gamma}{\gamma} \right)^2 \right) \right \rbrace\,.\label{cuatrod}
\ee
We recognize the logarithmic term proportional to the $A$ anomaly coefficient. This term comes together with another one depending on the derivatives of the function $\gamma$. This is because it is the only form of preserving Lorentz invariance and locality, and having a logarithmic term at the same time. The whole term proportional to $A$, interpreted as an action, is called the Wess-Zumino action for the dilaton (for a theory in $d=2$). The universal part of the entropy depends only on the coefficient $A$, as happens for spheres.   

\subsection{SSA on the null cone}
Now we continue with the idea of testing the RG using SSA. We have discovered that SSA vanishes for a CFT on the null cone, and we have to see precisely what can be obtained from it when the theory is not a CFT. 

\begin{figure}[t]
\begin{center}
\includegraphics[width=0.37\textwidth]{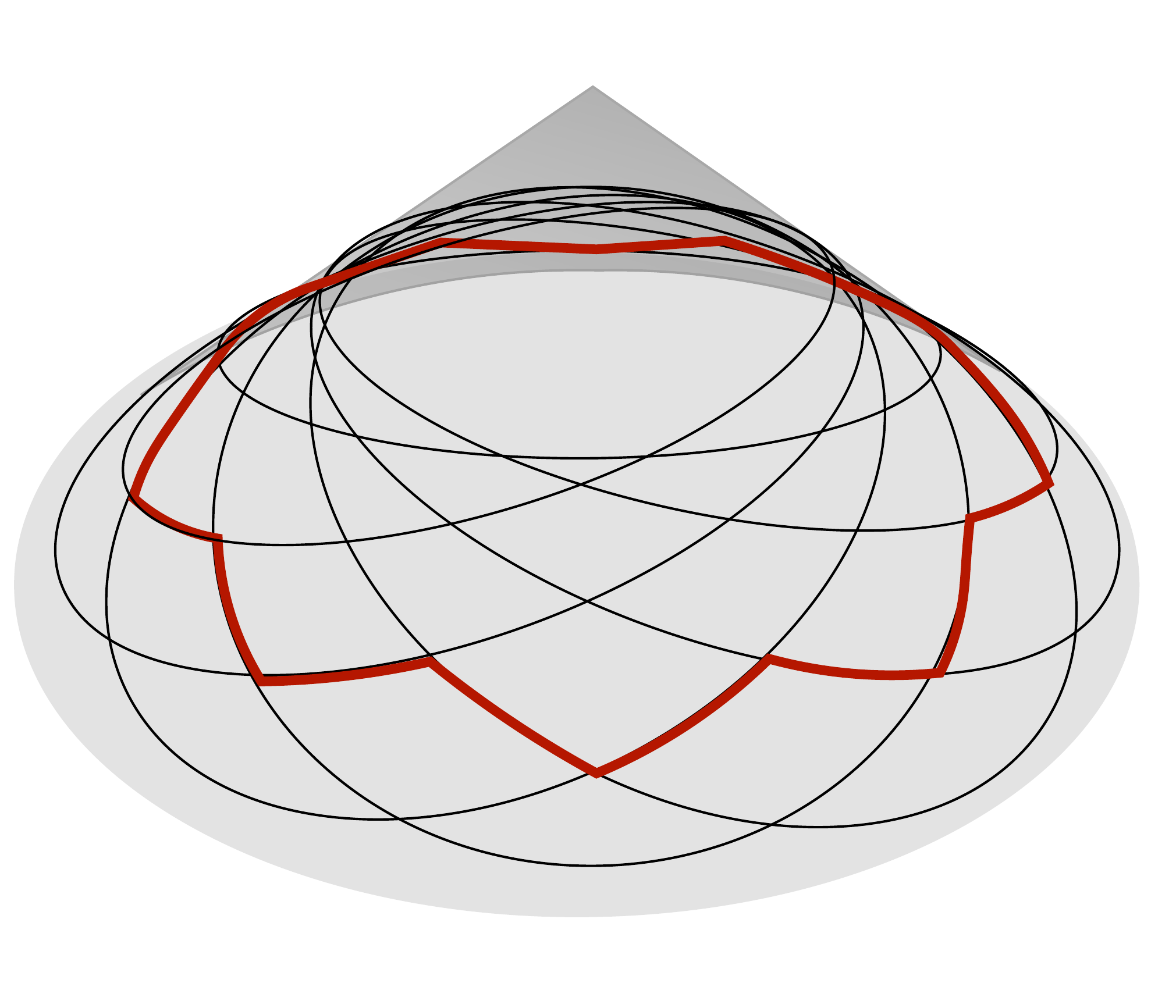} 
\caption{Boosted circles lying on the null cone in $d=3$. The vertical axis of the cone gives the time direction. The red curve is a ``wiggly sphere''. }
\label{boosted}
\end{center}
\end{figure}

We start with a boosted sphere of radius $\sqrt{r R}$ lying on the null cone between the time slices at time $|t|=r$ and $|t|=R>r$. We then take a large number $N$ of rotated copies $A_i$ of this sphere, as equally distributed on the unit sphere of directions as possible. See figure \ref{boosted}. The details of this distribution on the unit sphere of directions turn out to be irrelevant as far as a uniform distribution is approached for large $N$. We are using a large number of rotated spheres because we want that applying SSA many times we end up with regions that approach to spheres too. Repeatedly applying SSA to $N$ regions $A_1,\cdots ,A_N$ the following inequality can be derived
\be
\sum S(A_i)\ge S(\cup_i A_i)+ S(\cup_{i\neq j} A_i\cap A_j)+S(\cup_{i\neq j\neq k} A_i\cap A_j\cap A_k)+\cdots +S(\cap_i A_i)\,.
\ee
The $N$ regions on the right-hand side are ordered by inclusion and look like ``wiggly spheres'' as shown in the figure \ref{boosted}. The larger $N$ the wiggles along the null cone get smaller.

To write the inequality now is pure geometry. We have to count the density of wiggly spheres as a function of their radius. This is readily done and we get, dividing the inequality by $N$, 
\begin{equation}
S(\sqrt{r R})\ge  \int_r^R dl\ \beta(l) \tilde{S}(l)\,. \label{41}
\end{equation}
In this expression $ \tilde{S}(l)$ are the entropies of wiggly spheres. The wiggly spheres have an approximate radius $l\in (r,R)$, and lie around the surface of equal time $|t|=l$; the deviations from the perfect sphere of radius $l$ at $|t|=l$ form the wiggles, that lie on the null cone, and have a typical width $\sim l/N^{1/(d-2)}$ that tends to zero for large $N$.  
$\beta(l)$ is the density of wiggly spheres as the number of boosted spheres $N\rightarrow \infty$, divided by $N$.\footnote{Strictly speaking the integral in (\ref{41}) is a sum over $N$ wiggly sphere entropies divided by $N$. The notation with an integral and a density of wiggly spheres is a convenience here, that will make sense for later expressions when we take the limit $N\rightarrow \infty$, and more information about the entropies of the wiggly spheres is introduced.} It is given by    
\begin{equation}
\beta(l)=\frac{\text{Vol}(S_{d-3})}{\text{Vol}(S_{d-2})}\,  \frac{2^{d-3} (r R)^{\frac{d-2}{2}} \left( (l-r)(R-l) \right)^{\frac{d-4}{2}}    }{ l^{d-2}  (R-r)^{d-3}}\,,
\end{equation}
 and has unit integral, 
\be
\int_r^R dl\, \beta(l)=1\,.\label{normali}
\ee

\subsection{Converting wiggly spheres into spheres}

The inequality (\ref{41}) is much like what we wanted. But still, it relates spheres to wiggly spheres. The size of the wiggles goes to zero with large $N$ but still, we have to investigate if that produces non-trivial contributions. If we assume we can replace wiggly spheres with spheres on the right-hand side (in $d=2$ this is immediate because there are no wiggles) we get an inequality for spheres. Upon taking the limit $R\rightarrow r$, and expanding for small $R-r$, we get
\be
r\, S''(r) -(d-3) S'(r)\le 0\,. \label{teraaa}
\ee
To have a check, we can replace this for a CFT and we know that due to Markov property the left-hand side should be identically zero. For $d=2$ we have $S=c/3 \log (r/\epsilon)$ for intervals, and for $d=3$ we have $S= c_1 R/\epsilon-F$ for circles. Upon replacing in (\ref{teraaa}) we get the expected equation. But for $d=4$ we have 
\be
S(r)=c_2 \frac{r^2}{\epsilon^2}-4 A \log(R/\epsilon)\,.
\ee
 This gives
\be
r\, S''(r) - S'(r)=\frac{8A}{r}\ge 0 \,.
\ee
Then, it does not vanish, nor has the correct sign. Hence, there is a contribution from the wiggles that do not disappear taking $N\rightarrow \infty$. 

The issue here is that there is a non-trivial contribution to the wiggly sphere entropy from the term with derivatives in (\ref{cuatrod}) that comes together with the logarithmic term. This invalidates the replacement of wiggly spheres by spheres. We will now see that taking this difference into account correctly restores the Markov equation.

With $l=\sqrt{x^2+y^2+z^2}$, and $\theta$ the usual polar angle, the equation for the boosted sphere of radius $\sqrt{r R}$ is 
\be
|t|=l=\frac{2 r R}{r+R-(R-r)\cos(\theta)}\,.
\ee
We have 
\be
\frac{1}{2}\frac{(\nabla_\Omega \gamma)^2}{\gamma^2}= \frac{1}{2}\left(\frac{1}{l}\partial_\theta l\right)^2=\frac{(R-l)(l-r)}{2 r R}\,. 
\ee
We get a constant integrand in the term with derivatives in (\ref{cuatrod}) (except for higher-order terms in $1/N$) on the surface of the wiggly sphere of approximate radius $l$. 
Taking into account this term, the Markov equation for the finite terms
\be
\log(\sqrt{rR})=\int_r^R dl\, \beta(l)\,\left(\log(l)+\frac{(R-l)(l-r)}{2 r R}\right)\,,
\ee
is now satisfied, once we replace $\beta=\frac{r R}{l^2 (R-r)}$ corresponding to $d=4$.

Therefore, a finite term coming from the wiggles obstructs replacing the wiggly spheres by spheres directly for $d\ge 4$.

To get rid of the wiggles, 
the idea is to take advantage of the Markov property of a CFT and subtract from the inequality for the entropies $S$ the equation corresponding to the entropies $S_0$ of the UV CFT. This can be done at no cost since the SSA of $S_0$ vanishes exactly. We have shown that, in addition, the divergent terms coming from massive deformations at the UV are also Markovian and cancel in the SSA inequality; we can subtract them as well, without spoiling the inequality.     
Then, in any dimensions, we safely replace 
\be
S(l)\rightarrow \Delta S(l)=S(l)-S_0(l)-\textrm{massive divergent terms}\,,
\ee
 in (\ref{41}). Now the finite terms of the wiggles coming from the UV fixed point disappear in the subtraction, and we are free to replace subtracted wiggly spheres by subtracted spheres, taking the limit $N\rightarrow \infty$, and getting the inequality
\begin{equation}
\Delta S(\sqrt{r R}) \ge   \int_r^R dl\ \beta(l) \Delta S(l)\leftrightarrow r\, \Delta S''(r) -(d-3) \Delta S'(r)\le 0\,.  \label{111}
\end{equation}
 
We still have to check that there are no finite terms induced by a mass parameter that gives a contribution for the wiggles that survive in the limit of small wiggles. In fact, the difference in the EE from a wiggly and non-wiggly sphere is controlled by the UV. These terms should be proportional to some mass scale set by the coupling constant $g$ of the theory deformation at the UV, compensated by positive powers of the distance scale set by the size of the wiggles. In consequence, they do not contribute to the large $N$ limit. In more detail, the contributions of the wiggles can produce a local term, in which case they should be of the same form as the ones encountered for CFTs but where a power of the cutoff has been replaced by one of a mass parameter. In any case, a local term is always Markovian and disappear from the inequality. In the case that the term induced by the massive deformation of the UV is non-local, the change from the wiggly sphere to the sphere is suppressed by powers of the wiggly size and does not contribute in the limit.   

Note that for $d=3$ the formula (\ref{tresd}) gives no contribution for the wiggles, and we can safely replace wiggly circles with circles without subtracting the CFT entropies. But this is not the case in higher dimensions.

\subsection{Irreversibility theorems}

We then have (\ref{111}) for spheres in any dimension, where the UV CFT entropy along with other possible divergent contributions have been subtracted.
Writing the entropy as a function of the area $a=c \,r^{d-2}$ rather than the radius $r$, we get the compact expression
\be
\Delta S''(a)\le 0\,, \label{sisi}
\ee 
valid in any dimension. Thus the constraint for $\Delta S(a)$ is that it must be concave as a function of the area.

 With our definition of $\Delta S$, that has the entropy with the UV CFT terms and other possible divergent terms subtracted, in the UV limit of small $r$ all local geometric terms vanish, and we get the leading ``nonlocal'' term 
\be
\Delta S_{UV}(r) \sim  c_0 \,g^2 r^{2(d-\Delta)}+\ldots =c_0\, g^2 a^{\frac{2(d-\Delta)}{d-2}}+\ldots\,,\label{oyo1}
\ee
where the ellipsis are higher powers in $r$. In the IR fixed point all contributions (except the universal term) are local (proportional to integral of curvatures on the surface) and we have
\bea 
\Delta S_{IR}(r)&=&\Delta \mu_{d-2}\,r^{d-2}+\Delta \mu_{d-4}\, r^{d-4}+\ldots  + \left\lbrace \begin{array}{l} (-)^{\frac{d-2}{2}} 4\,\Delta A\, \log(m r)\,\, d \,\, \textrm{even}\\ (-)^{\frac{d-1}{2}} \Delta F \hspace{1.9cm}d\,\,\textrm{odd }  \end{array}\right.\,\nonumber\\
\label{even1}
&=&\Delta \mu_{d-2}\, a +\Delta \mu_{d-4}\, a^{\frac{d-4}{d-2}}+\ldots  
  + \left\lbrace \begin{array}{l} \frac{(-)^{\frac{d-2}{2}} 4}{(d-2)}\Delta A\, \log(m^{d-2}a)\,\, d \,\, \textrm{even}\\ (-)^{\frac{d-1}{2}} \Delta F \hspace{1.9cm}d\,\,\textrm{odd }  \end{array}\right.\,,
\eea
with $m$ a characteristic energy scale of the RG flow.    
The coefficients $\Delta \mu_{d-k}$ have dimension $d-k$ and have the interpretation of a finite renormalization of the coefficient of the term $r^{d-k}$ in the entropy of spheres,  between the UV and IR fixed points. The last term gives the change in the part of the EE which only depends on the fixpoints: $\Delta A=A_{IR}-A_{UV}$, with $A$ the Euler trace anomaly coefficient for even dimensions, and $\Delta F=F_{IR}-F_{UV}$, with $F$ the constant term of the free energy of a $d$-dimensional Euclidean sphere, for odd $d$. 

Concavity, Eq.~(\ref{sisi}), implies two relations between the short and long-distance expansions for $\Delta S(a)$: 1) The slope of the $\Delta S(a)$ curve is bigger at the UV than at the IR; 2) Given that $\Delta S(0)=0$, the height at the origin of the tangent line at the IR has to be positive. 

\bigskip 

The first requirement, comparing (\ref{oyo1}) and (\ref{even1}), and provided $\Delta < (d+2)/2$, gives place to the ``area theorem" in any dimensions, that is, the decrease along the RG of the coefficient of the area term,\footnote{If 
$\Delta > (d+2)/2$ the area term at the $UV$ can be considered infinite because the slope of (\ref{oyo1}) diverges as $r\rightarrow 0$.}  
\be
\Delta \mu_{d-2}\le 0\,.\label{dfghj}
\ee
 The meaning of this inequality is that massive modes do not contribute to entanglement across the boundary for large regions as they would do if they would have remained massless. 
In $d=2$ the area coefficient is dimensionless ($a \sim \log(r)$)  and the area theorem coincides with the $c$-theorem, $c_{UV}\ge c_{IR}$. 

\bigskip

The second requirement gives for $d=3$ the $F$-theorem,
\be
\Delta F\le 0\,,
\ee
and for $d=4$ the $a$-theorem,
\be
\Delta A\le 0\,.\label{mia}
\ee
For higher dimensions $d>4$ it gives
\be
\Delta \mu_{d-4}\ge 0\,.\label{tuya}
\ee 
The inequality does not constraint the sign of the subleading terms, in particular the change of $A$ or $F$, for $d>4$.  

In addition to these constraints that come from the comparison of the UV and IR expansions, we have to check (\ref{sisi}) at the UV and infrared expansions themselves. At the IR we get again (\ref{mia}) and (\ref{tuya}) for $d\ge 4$. For $d=3$ we get information on the sign of the first subleading correction to the constant
\be
\Delta S^{d=3}_{IR}= \Delta \mu_1 r -\Delta F - \frac{k}{r^{\alpha}}+\ldots\,,
\ee
where the last term is purely infrared in origin and $\alpha$ is related to the leading irrelevant dimension of the operator driving the theory to the IR. We get $k>0$ from (\ref{sisi}).  
 At the UV we get that the sign of the coefficient $c_0$ in (\ref{oyo1}) is the same as the one of $\Delta-(d+2)/2$. 

Notice that while the inequality (\ref{sisi}) saturates at the UV, it does not saturate at the IR for $d\ge 4$. The SSA inequality always saturates at the IR for regions smooth enough (with IR size curvatures) but this does not allow us to derive (\ref{sisi}) precisely because we are not allowed to convert wiggly spheres into spheres for these large wiggles.  

\bigskip

For $d=2,3$ we can spare of subtracting the UV CFT entropy. 
Let us write again the inequalities for the entropy of an interval in $d=2$, or a circle in $d=3$,
\bea
d=2:\hspace{.5cm}r S''(r)+S'(r)\le 0 \rightarrow c(r)=r S'(r)\,, \,\,c'(r)\le 0\,,\\
d=3:\hspace{.5cm}S''(r)\le 0 \rightarrow F(r)=r S'(r)-S(r)\,,\,\, F'(r)\le 0\,.
\eea
These monotonic functions $c(r)$ and $F(r)$ interpolate between $c_{UV}$ and $c_{IR}$ in $d=2$ and $F_{UV}$ and $F_{IR}$ in $d=3$. No such interpolating function depending only on the entropy at scale $r$ exist for $d=4$.

\subsection{SSA plus Lorentz symmetry imply divergent entropy}

From these results, we can see the divergence of the entropy in QFT in a different light, as implied by Lorentz symmetry and the SSA inequality. 
 Let us consider $d=2$. The entropy of an interval as a function of the ``area'' $a=\log(r)$ is concave $S''(a)\le 0$. 
 Suppose we assume the entropy is finite. $S(a)$ would then be a finite positive concave function of the line $a\in (-\infty,\infty)$.  
 If $S'(a_0)$ were positive at some $a_0$, going for smaller $a$ there would be a point in which $S$ would turn negative, because the slope of the curve always increases towards smaller $a$. If  $S'(a_0)\le 0$ the same reasoning shows $S$ would become negative for large enough $a$. Therefore the only possibility is the trivial $S(a)$ constant. Similar reasoning using SSA and Lorentz invariance shows that the only finite entropy possibility in $d=2$ for multi-interval regions is that $S$ is proportional to the number of intervals, and independent of any other geometrical feature. This gives zero mutual information. The conclusion is that finite positive entropy, Lorentz invariance, and SSA lead to the trivial result of a theory without correlations. The same purely geometric ideas and conclusions apply to any dimension.

\subsection{Modular Hamiltonians on the null plane and the null cone}
The strong subadditive equation, rather than inequality, holds for regions in the null plane. As we have seen this is a very geometrical fact and extends to almost anything that depends on regions with boundaries in the plane, including the Renyi entropies. However, the Markov equation for the entropy is especially powerful, because it is the saturation of an inequality. It implies a particular form of the density matrices. Interestingly, it implies the same Markovian  equation for the modular Hamiltonians
\be
S_A+S_B-S_{A\cap B}-S_{A\cup B}=0 \,\leftrightarrow K_A+K_B-K_{A\cap B}-K_{A\cup B}=0\,.
\ee
This is now an equation for operators rather than numbers. 

The content of this equation is unfolded if we reason as we did for the entropy. If $\gamma_A(y)$ and $\gamma_B(y)$ are deformed around two separated points $y_1,y_2$ the equation tells the cancellation takes place at $y_1$ and $y_2$ independently of the rest of the curve. As a consequence. the cancellation has a natural explanation in the locality of $K_A$ as a function of $\gamma_A(y)$, that is, $K_A$ is an integral over $y$ of a quantity that depends on the null rays alone. We can think heuristically that the density matrix is a tensor product over the different null rays such that the modular Hamiltonians are additive. This idea agrees with the extensivity of the Renyi entropies too.   

To get the Modular Hamiltonian we then just have to calibrate the contribution of a single null ray with the result for Rindler space. Remember that Rindler modular Hamiltonian is a conserved charge, and then is of the form 
\be
2 \pi \int_\Sigma d\sigma\,j_\mu\, \eta^\mu \,,
\ee
 where $j_\mu$ is the Noether current of the boost symmetry, and $\eta$ is the unit normal vector to the arbitrary surface $\Sigma$ forming a Cauchy surface for the wedge. Taking this surface to the null plane we get
\be
K_{\textrm{Rindler}}=2\pi\, \int d^{d-2}y\, \int_0^\infty dx^+\,x^+\, T_{++}(x^+,y)\,.
\ee
 We then get for an arbitrary curve  
\be
K_\gamma=2\pi \int d^{d-2} y\, \int_{\gamma(y)}^\infty dx^+\, (x^+-\gamma(y))\, T_{++}(x^+, y)\,. \label{eq:DeltaH2}
\ee
 $y$ denote the transverse coordinates $(x^2, \ldots, x^{d-1})$, and $x^+= \gamma(y)$ parametrizes the boundary of the region on the null plane.  

This expression is again local and depends exclusively on $T_{\mu\nu}$. However, there is an important difference with Rindler's case. This modular Hamiltonian is not a conserved charge coming from a Noether current. We cannot write it as a local expression on any other surface except the null one. Written on other surfaces it will be non-local. In a like manner, the modular flow acts locally as Rindler's on each null ray but is non-local on the bulk of the region. 

For a CFT the modular Hamiltonian can be transformed to the cone obtaining  
\be
K_\gamma=2 \pi \int d\Omega\, \int_0^{\gamma(\Omega)} dr \,r^{d-1}\,\frac{ \gamma(\Omega)-r}{\gamma(\Omega)}\, T_{+ + }\,,
\ee
with the surface described by $x^0=r=\gamma(\Omega)$ and where the $+$ components correspond to the null coordinate $r^+=(r+x^0)/2$.

\subsection{Interpretations of irreversibility}
There are other, independent, proofs of the irreversibility theorems, based on more conventional quantities,  in dimensions $d=2$ ($c$ theorem) and $d=4$ ($A$ theorem), but the only one available up to present for dimension $d=3$ is entropic. This may not be so surprising, given the very non-local nature of $F$ which is not measured by an anomaly. 

It is important to step back for a moment and try to understand the results conceptually with words rather than equations. A heuristic interpretation of irreversibility is especially expected in the present case where we have a unification of the theorems in different dimensions in terms of the entropy and QI properties. However, in this important point, the focus on the entropy has not yet given a final word. 
Sometimes the irreversibility theorems are worded in terms of a loss of degrees of freedom with scale. This is not very correct, we always have infinitely many degrees of freedom, and there is no change in the RG charges with scale for a CFT, for example. A better description is a loss on the number of field degrees of freedom. But this is a notion that has sense only for free fields, and further, the central charges measure very differently free fields with a different spin, giving a very non-classical notion of the weight of each field.   

\begin{figure}[t]
\begin{center}
\includegraphics[width=0.35\textwidth]{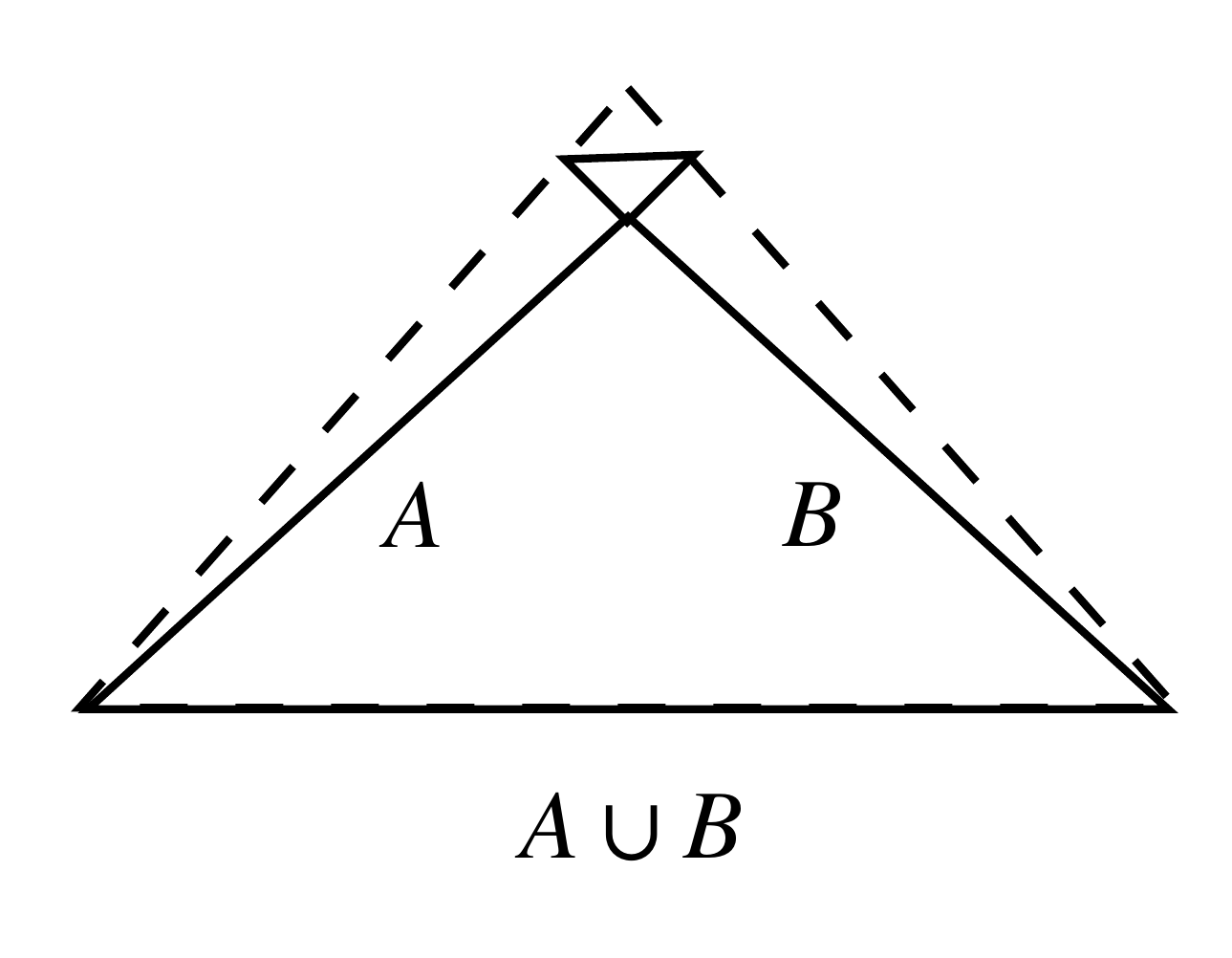} 
\caption{Strong subadditivity of boosted intervals in $d=2$.}
\label{triangle}
\end{center}
\end{figure}

A heuristic interpretation of the entropic irreversibility theorems goes along the following lines. Minkowski space violates the triangle inequality for distances very badly, and we can draw very small regions $A$, $B$, with small $A\cap B$, such that $A\cup B$ is very large, as in figure \ref{triangle}. The question is whether we can reconstruct a state in a large region $A\cup B$ from the knowledge of the reduced states in small ones $A$, $B$, $A\cap B$.
Notice that the knowledge of the state in the big region clearly gives the one in the smaller ones, the question is the opposite.  
 For the conformal case this reconstruction is indeed quite direct due to the Markov property. We have from $K_A+K_B-K_{A\cap B}-K_{A\cup B}=0$, where $K_X=-\log \rho_X$,
\be
\rho_{A\cup B}=e^{\log \rho_A+\log \rho_B -\log \rho_{A\cap B}}\,.
\ee 
This simple reconstruction is prevented when the state is non-Markovian, and we are not in a CFT, leading to the idea that the reconstructibility from different scales is what is lost in the RG flow. However, it would be desirable to have a sharper idea of reconstructibility and how much it fails. For example, the analyticity of correlators tells that the complete knowledge of the correlation functions in any small region should in principle give the unique correlators to all scales. This too strong and operationally non-practicable notion of reconstruction is evidently not what is required.

On the other hand, the area theorem, which includes the c-theorem in $d=2$, does have a more clear QI interpretation. It can be shown that this theorem follows just from the monotonicity of relative entropy between the vacuum of the theory and the one of its UV CFT fixpoint when compared in the null Cauchy surfaces of diamonds. The relative entropy monotonicity tells about increasing distinguishability and this is measured by the change of the area term (or the central charge in $d=2$). The same idea also explains the g-theorem about defects on CFT's.       

To have such a heuristic interpretation made more concrete would be important for different reasons. It could help in generalizing the theorems to other situations like non-relativistic systems, or higher dimensions. 
 It is clear that the SSA analysis gives the same result in any dimensions, but the power of the inequality is not enough to reach the last term of the expansion of the entropy of spheres (which depend on the dimensionless RG charges). Inequalities involving more regions, and consequently giving place to more derivatives of the entropy, would be needed to reach the RG charges for $d\ge 5$. But these are not known. Holographic irreversibility theorems hold for any dimensions but it is not clear whether this is the result of the peculiar nature of these theories. There is also the fact that we are much less certain of the possible QFT zoo in higher dimensions. Some people argue all QFT should be free for $d\ge 7$.

\subsection{Unitarity bounds from mutual information}
This is an application of the Markov property on the null cone for CFT's. We will see it implies unitarity bounds.  

Let us take a null cone and two regions $A_1, A_2$, with boundary in this cone. We also take a spatially separated region $B$. We compute
\bea
 && \hspace{-.7cm}I(A_1,B)+I(A_2,B)-I(A_1\cap A_2,B)-I(A_1\cup A_2,B)=(S(A_1)+S(A_2)-S(A_1\cap A_2)-S(A_1\cup A_2))\nonumber\\
 && - (S(A_1\cup B)+S(A_2\cup B)-S((A_1\cap A_2)\cup B)-S((A_1\cup A_2)\cup B)\,. 
\eea
The first bracket on the right-hand side vanishes because of the Markov property for a CFT. The second is an SSA combination of entropies, with a minus sign. Therefore we get the superadditivity of mutual information in this case
\be
I(A_1,B)+I(A_2,B)-I(A_1\cap A_2,B)-I(A_1\cup A_2,B)\le 0\,.
\ee

\begin{figure}[t]
\begin{center}
\includegraphics[width=0.55\textwidth]{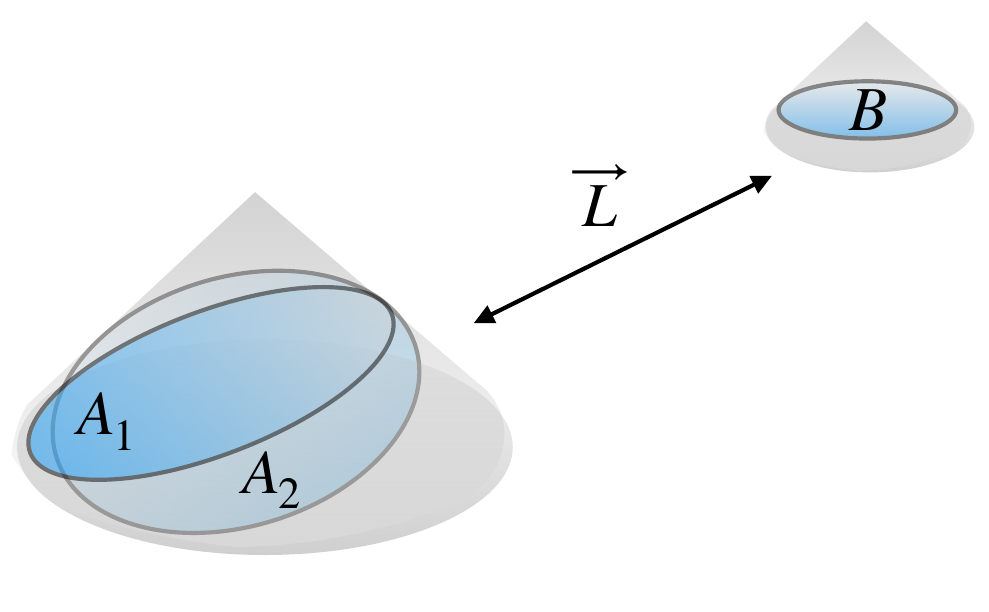} 
\caption{Two far away regions with boundary on null cones.}
\label{unitarity}
\end{center}
\end{figure}

To get more tangible information about this inequality we evaluate it for two distant spheres $A$ and $B$, where the separation $L$ is much larger than the radius. When the mutual information is dominated by the interchange of a scalar field at large distances (see section \ref{long}), the mutual information between two spheres has an expansion
\be
I(A,B)\sim c^2\,\frac{R_A^{2\, \Delta} R_B^{2\, \Delta}}{L^{4 \,\Delta}}\,. \label{hjhj}
\ee

We can now play the same game as in figure \ref{boosted}, considering multiple rotated boosted spheres for the first region, and using strong super additivity as we did with the proof of the irreversibility theorems using strong subadditivity. The result is again the same inequality (\ref{teraaa}) involving derivatives with respect to the radius of the first sphere but with opposite sign,
\be
R_A\, \frac{d^2 I(A,B)}{dR_A^2} -(d-3) \frac{d I(A,B)}{dR_A}\ge 0\,.
\ee
Using (\ref{hjhj}) we get
\be
\Delta \ge \frac{d-2}{2}\,.\label{ub}
\ee
This is precisely the unitarity bound for the scaling dimension of scalar fields. Note this is supposed to apply to the smallest scaling dimension of the theory that dominates long-distance mutual information, and therefore the bound applies to all other scalar fields as well. The unitarity bound in the usual field theory derivation comes simply from the positivity of the Hilbert space scalar product. We have, writing $\phi_\alpha=\int d^d x\, \alpha(x)\, \phi(x)$ for a smeared operator,
\be 
\langle 0|\phi_\alpha^\dagger | \phi_\alpha |0\rangle= \int d^d x\, d^d y\, \alpha^*(x)\,\alpha(y) \langle 0|\phi(x)\phi(y)|\rangle \ge 0\,.
\ee
Inserting the adequate correlator, the analysis reveals that this holds for any smooth $\alpha$ if and only if (\ref{ub}) holds.

Independently of the interpretation in terms of fields, this shows on general grounds that the mutual information of a CFT cannot decay at a slower rate than $L^{-2(d-2)}$. Since this is valid for spheres, by monotonicity, it is valid for any shapes.

\subsection{Notes and references}
The structure of quantum Markovian states was found in \cite{hayden2004structure}. For a nice account of the Reeh-Slieder theorem see Witten's review \cite{Witten:2018lha}. A proof that there is always distillable entanglement for regions with non zero volume is by Verch and Werner \cite{verch2005distillability}. Markovianity for regions on the null plane (and CFT's on the cone) was shown in \cite{Casini:2017roe}. This paper also gives the structure of the modular Hamiltonians. Previous analogous results for free fields by Wall are in \cite{Wall:2011hj}. The presentation here follows the line of \cite{Casini:2018kzx}. This last paper also shows the general structure of EE on the null cone for a CFT. 

The irreversibility theorems have a long history. The theorem in $d=2$ was proved by Zamolodchikov in \cite{Zamolodchikov:1986gt} using stress tensor correlators. For even dimensions it was then conjectured by Cardy the $A$ anomaly should be the monotonic quantity \cite{Cardy:1988cwa}. For odd dimensions, the conjecture that the constant term in the entropy was the correct quantity to look at was given by Myers and Sinha in \cite{Myers:2010xs}. This paper also gave a holographic proof in any dimensions. An apparently different conjecture for irreversibility about the constant term in the free energy of a sphere was given by Jafferis, Klebanov, Pufu and Safdi in \cite{Jafferis:2011zi}. That these two quantities coincide was shown in \cite{Casini:2011kv}. The interpolating $F(r)$ function was conjectured by Liu and Mezei in \cite{Liu:2012eea}. The $A$ theorem in $d=4$ was proved by Komargodski and Schwimmer using dilaton effective actions in \cite{Komargodski:2011vj}.    

Entropic irreversibility theorem in $d=2$ was proved in \cite{Casini:2004bw}. The generalization to $d=3$ is in \cite{Casini:2012ei}. The case of $d=4$ is in the paper \cite{Casini:2017vbe}, where a unification for the consequences of SSA in all dimensions was given. The presentation here follows \cite{Casini:2018kzx}.   

Strong subadditivity plus finite positive entropy and Lorentz symmetry give a trivial entropy function proportional to the area for all polyhedra \cite{casini2004geometric}. This implies zero mutual information for any region and no correlations. 

A nice wording on the interpretation of the entropic irreversibility theorems is in \cite{Lashkari:2017rcl}. A proof of the area theorem in any dimensions, and the c-theorem in $d=2$, based on monotonicity of relative entropy is in \cite{Casini:2016udt}. This follows ideas on the entropic proof of the g-theorem  \cite{Casini:2016fgb}. 

Unitarity bounds from mutual information superadditivity were found in a  work with G. Torroba and E. Teste \cite{Casini:2021mutual}. 

\newpage

\section{Energy-entropy  bounds}
 In relativistic quantum field theory, the energy has to be positive for stability reasons. Any negative energy could be boosted to an arbitrarily large negative energy, and there would not be a lower bound on energies. As a consequence, the final products of scattering would have an infinite volume of phase space available and the theory would easily not make sense. We want our quantum theory to have a fundamental state. However, energy density can and must be negative for some states. For example, the classical energy density of a scalar field $T_{00}=\frac{1}{2}(\dot{\phi}^2+(\vec{\nabla} \phi)^2+m^2\phi^2)$ is manifestly positive, but in the quantization the subtraction of zero-point energy renders the energy density operator indefinite. 
  The question is how much negative there can be, or how much can it be concentrated or spread in space and time. The role of the energy bounds is to inform us about this, and as it happens, the quantity bounding the energy is often an entropy, these can also be seen as entropy bounds. The bounds can be alternatively be seen as lower bounds for the energy or as upper bounds for some entropic quantity. 
   
Though we are going to focus mainly on QFT aspects, this subject is very much entangled with gravity. We will see there are thought experiments involving black holes behind our QFT results. Some form of positivity of energy density is used in many theorems about classical gravity: singularity theorems, the horizon area increase for black holes, the non-traversability of wormholes, the boundary causality of AdS, etc. 
 QFT necessarily violates (some) of these energy conditions, and allows for example, for the black holes to evaporate. Energy-entropy bounds are seen in this light as a necessity for quantum corrected semi-classical version of the gravity theorems.

\subsection{Bekenstein bound. I}

Bekenstein proposed the following experiment with black holes. Take a small probe of matter outside of the black hole, quite near the horizon of a large black hole, and let it fall and be swallowed by it. If the second law is valid in presence of black holes (generalized second law), the final entropy should be greater than the initial one. This entropy has to count both the entropy $S$ of the object outside the black hole and the one of the black hole itself, given by $A/(4G)$, where $A$ is the horizon area, and $G$ Newton's constant. We then should have for the changes of these terms
\be
\frac{\Delta A}{4G}\ge \Delta S\,.\label{ta}
\ee 
Computing $\Delta A$ is computing a change in the geometry when the object crosses the horizon. This is governed by Einstein equations. It turns out that for small changes like the one of this experiment Einstein equations just reproduces what is expected from thermodynamics: the change of the black hole entropy is given by the change of energy over the temperature,
\be
\Delta S_{BH}=\frac{\Delta A}{4G}=\frac{\Delta E}{T}\,.\label{te}
\ee
This is called the first law of black hole thermodynamics, which is geometrized by Einstein equations. 
Here $T$ is the black hole temperature $\frac{\kappa}{2\pi}$ (see section \ref{curved}), where $\kappa$ is the surface gravity of the black hole, and $E$ is the ordinary energy as measured by asymptotic observers. This is the Killing energy corresponding to the time like symmetry of the black hole space-time. As we recalled in section \ref{curved}, very near the horizon, the Killing symmetry generator approaches $\frac{\kappa}{2\pi} K$, with $K$ the modular Hamiltonian of the Rindler wedge. We get
from (\ref{ta}) and (\ref{te})
\be
\frac{\Delta A}{4G}=\Delta K= 2\pi \int_{x^1> 0} x^1\, \Delta T_{00}(x) \ge \Delta S\,. \label{BB0}
\ee
This is written in near horizon coordinates, with $x^1=0$ the horizon position. Notice that $G$ has disappeared from the inequality between the two last terms in this expression because the $G^{-1}$ in the entropy formula has cancelled with the $G$ in the change of geometry given a source. Therefore we can safely take the limit $G\rightarrow 0$, and think this is now a problem in Rindler space.      

Bekenstein's original formulation was classical. He thought of a compact object of energy $E$ at a distance $R$ from the horizon (with $R$ greater than the object size), and let the object has some entropy $S$. After it is swallowed by the black hole no entropy should remain outside, and he obtained
\be
S\le 2 \pi\, R \, E\,. \label{BB}
\ee
As it stands this formula suggests that one could not make the entropy increase too much for a fixed energy. It is then some kind of universal bound on the number of degrees of freedom, implied by black hole physics. 

However, the bound (\ref{BB}) is ill-defined in the quantum case. The entropy, if localized, should include vacuum entanglement, which makes it divergent; the energy, if not global, could be negative (while the entropy is always positive), the size of the object is ill-defined, etc. These problems were historically preventing any better interpretation of the bound. To understand what the bound says in QFT it is better to return to the version (\ref{BB0}) that is really what follows from Bekenstein thought experiment. Then, we should replace $\Delta T_{00}$ by $\Delta \langle T_{00}\rangle $ and interpret the change $\Delta S$ as the change of the entropy in half-space, between the state with matter, to the vacuum state. Both states, the initial and final one, have large amounts of entanglement due to vacuum fluctuations, or near horizon correlations. However, the difference between entropies is now finite. So (\ref{BB0}) makes perfect sense in QFT, and now reads
\be
\Delta \langle K_R \rangle= 2\pi \int_{x^1> 0}d^{d-1}x\,  x^1\,  \langle T_{00}(x)\rangle \ge \Delta S_R\,,           \label{alo}
\ee
where the subscript $R$ means quantities referred to the Rindler wedge. We have used $\langle 0|T_{00}(x)|0\rangle=0$. 

This inequality is precisely what is implied by the positivity of relative entropy between the state $\rho$ with matter and the vacuum $\rho_0$ when both are reduced to half-space. We have
\be
S(\rho|\rho_0)=\textrm{tr} \rho (\log \rho -\log \rho_0)= \textrm{tr} \left(\rho \log \rho -\rho_0\log \rho_0 + \rho_0\log \rho_0-\rho \log \rho_0\right)=\Delta \langle K_R \rangle - \Delta S_R\,,
\ee
where we have used $K_R=-\log \rho_0$. Rindler's modular Hamiltonian $K_R$ has its usual expression proportional to the boost operator, giving (\ref{alo}) from positivity of relative entropy.

This proves Bekenstein bound for any state in any theory. In conclusion, Bekenstein bound does not imply new physics coming from black holes, though it is remarkable that the consistency with gravity hinges on the pre-established relationship between the vacuum state on half-space with energy density.      

Under this light, the interpretation of the bound is very different from what it was originally thought. There is no bound on degrees of freedom, but the bound is rather related to the idea of distinguishability: when restricted to a region, fluctuations can be as large as to make it hard to distinguish the vacuum from another state if the energy (times distance to the boundary) of this other state is not big enough. The relative entropy measures this distinguishability in a precise operational way and is always positive.     

Now there is no problem with the energy being negative in half-space, but this must be accompanied by a decrease of entanglement for the vacuum state.  

Relative entropy between two states $\omega_1$, $\omega_0$, in any region $V$, can always be written 
\be
S_V(\omega_1|\omega_0)=\Delta \langle K_V \rangle - \Delta S_V\,,
\ee
where $K_V$ is the modular Hamiltonian of $\omega_0$ in $V$. One can produce other Bekenstein type bounds involving energy and entropy whenever $K_V$ is given in terms of the stress tensor, as in the case of spheres or, more generally, regions with boundary in the null cone, in CFT's.

\subsection{The first law of EE}

If we perturb a density matrix $\rho \rightarrow \rho+\delta \rho$, with $\textrm{tr} \delta \rho=0$ to keep normalization, we get to first order
\be
S(\rho+\delta\rho)= - \textrm{tr}  (\rho+\delta\rho) \log(\rho+\delta\rho)
=S(\rho)- \textrm{tr}  \delta\rho \log(\rho)+{\cal O}((\delta \rho)^2)\,.
\ee
To first order in the deviation, we can then write, using the language of the previous section, 
\be
\Delta S=\Delta \langle K\rangle\,.\label{lalad}
\ee
This gives us the change in the entropy from the knowledge of the change of the expectation value of the modular Hamiltonian. Notice that this implies the relative entropy $S(\rho+\delta \rho |\rho)=\Delta \langle K \rangle-\Delta S$ is of second order in the perturbation. This must be so because it is always positive and is zero for $\delta \rho=0$.

This has been called the first law of entanglement entropy because it is a generalization of 
\be
dS=\beta dE
\ee
in thermodynamics. Indeed, this last equation is (\ref{lalad}) for the special case of small changes with respect to a thermal state.    

The importance of (\ref{lalad}) is that it allows evaluating changes in the entropy by the knowledge of the expectation value of an operator. As in thermodynamics, it opens the way for the entropy to be measured or computed from the knowledge of these expectation values. Of course, the formula is useful provided the modular Hamiltonian is known (as in Rindler for example). Formulas for expansions to higher order in terms of the modular Hamiltonian are known. For higher orders, the possible non-commutativity between $\rho$ and $\delta \rho$ should be taken into account.  

The first law has an interesting interpretation in holographic theories. It relates the variation of entropy, which is given by a variation of a minimal area in the bulk and depends on the variations of the metric, with a variation of energy. To satisfy this equation the metric of the holographic theory has to satisfy Einstein equations, which are then encoded in the boundary QFT in the distribution of entanglement for different regions.   

\subsection{Area term in EE and the Newton constant}

As an application of the first law of entanglement entropy, we compute the general form of the massive corrections to the area term in EE. That is, we want to compute how the area term is dressed as we go from small regions to large regions. The area theorem says it must be a negative correction, here we will evaluate it in terms of correlation functions of the stress tensor.

In order to address this  consider a sphere of large radius $R\rightarrow \infty$, approaching  half-space, so the modular Hamiltonian is given by
\be\label{eq:rindlerH}
 K =2\pi\,\int_{x^1 \ge 0}\, d^{d-1}\vec x\,x^1\,T_{00}(x)\,,
\ee
for any QFT (not necessarily at a fixed point). $S(\rho_R)$ is dominated by the IR fixed point, and hence
\be
\Delta S = S(\rho_R)-S(\rho^{UV}_R) \approx ( \mu_{d-2}^{IR}- \mu_{d-2}^{UV})\,R^{d-2}\,;
\ee
all the other terms are subleading. The change $\Delta  \mu_{d-2}$ can be obtained by performing a small variation of $R$,
\be\label{eq:change1}
R\,\frac{d \Delta S}{dR}=(d-2)\,\Delta S\,.
\ee

Under a small change of state $\delta \rho$, the first law allows to relate the variation in the entropy to the change in the modular Hamiltonian, $\delta S =  {\rm tr}(\delta \rho\, K )$. 
 A variation of the size of the region can be obtained by changing $g_{\mu\nu}\rightarrow \lambda^2 g_{\mu\nu}$, with $\lambda=(1+\delta \lambda)$,  in the Euclidean path
integral representation of the density matrix, and keeping all mass parameters and the coordinate size of the region fixed. Using $\delta \log Z/\delta g_{\mu\nu}= \langle T_{\mu\nu}\rangle/2$, this step is equivalent to an insertion of $\Theta(x)=T_{\mu}^\mu(x)$ in every point on the $d$-dimensional path integral.
 By applying the first law to the variation of entropy in each of the states,   we then get
\be\label{eq:DeltaS1}
\Delta S = -\frac{1}{d-2}\,\int d^dx\,\langle \Theta(x) K \rangle+\frac{1}{d-2}\,\int d^dx\,\langle \Theta_{UV}(x) K_{UV} \rangle\,,
\ee
where we still have formula (\ref{eq:rindlerH}) for $K$ but there is an additional  minus sign from the rotation to imaginary time of the two indices of $T_{00}$. 
Using that the trace of the stress-tensor vanishes in the UV CFT, and (\ref{eq:change1}), 
\be\label{eq:DeltaS3}
\Delta S=-\frac{2\pi}{d-2}\,\int d^dy\,\,\int_{x^1\ge 0} d^{d-1}x\,\,x^1\,\langle \Theta(y) T_{00}(x) \rangle\,.
\ee

Translation invariance along the boundary of the Rindler space implies that the integrand is independent of the spatial coordinates $(x^2-y^2,\,\ldots,\,x^{d-1}-y^{d-1})$ parallel to the Rindler edge. These integrals give simply a factor of the boundary area $R^{d-2}$, and we arrive to
\be\label{eq:DeltaS4}
\Delta  \mu_{d-2}=-\frac{2\pi}{d-2}\,\int d^{d}y\,\int_{x^1>0} dx^1\,x^1\,\langle \Theta(y) T_{00}(x) \rangle\,.
\ee
Because of conservation, translation and rotation invariance  $\langle \Theta(y) T_{\alpha \beta}(x) \rangle$ must have the following Lorentz-covariant structure:
\be
\langle \Theta(y) T_{\alpha \beta}(x) \rangle= (g_{\alpha \beta} \nabla^2-\partial_\alpha \partial_\beta) F(s)\,,
\ee
with $s=y-x$. In particular, taking the trace 
\be
\langle \Theta(y) \Theta(x) \rangle = (d-1) \nabla^2 F(s)\,.
\ee

Let us now perform the integral,
\be
\int d^{d}y\,\int_{x^1>0}\,dx^1\,x^1\, \langle \Theta(y) T_{00}(x) \rangle =\int d^{d}y\,\int_{x^1>0}\,dx^1\,x^1\, \vec \nabla^2 F(s) 
= \int d^{d}x\,F(x)\,,
\ee
where we integrated by parts twice. On the other hand,
\be
\int d^{d}x\,x^2\,\langle \Theta(x) \Theta(0)\rangle = (d-1)\int d^{d}x\,x^2\, \nabla^2 F(x)
=2\,d\,(d-1) \int d^{d-1}x\,F(x)\,,
\ee
after integration by parts.

Comparing both expressions, we derive the sum rule
\be
\Delta \mu_{d-2}=-\frac{\pi}{d (d-1)(d-2)}\,\int d^{d}x\,x^2\,\langle \Theta(x) \Theta(0)\rangle\,. \label{popop}
\ee
This sum rule agrees with the sign $\Delta  \mu_{d-2} \le 0$ deduced from the irreversibility theorems, since the correlator $\langle \Theta(x) \Theta(0)\rangle >0$ by reflection positivity. Notice however that the change can be divergent depending on the detail on the approach to the UV fixpoint. If $\phi_\Delta$ is the field producing a relevant perturbation in the action at the UV, we have $\Theta\sim \phi_\Delta$ for small scales and (\ref{popop}) diverges when $\Delta \ge (d+2)/2 $. This is in accordance with the estimations at section \ref{ge}.

The case $d=2$ can be obtained by taking into account that in this case the behaviour is logarithmic instead of a power law, or alternatively from the equation above by adimensionalizing the integral using a mass scale $m$
\be
\Delta S=-(m R)^{d-2}\,\frac{\pi}{d (d-1)(d-2)}\,\int d^{d}x\,m^{-(d-2)}x^2\,\langle \Theta(x) \Theta(0)\rangle\,,
\ee
and expanding for  $d\to 2$. We have
\be
\Delta S=-\frac{\pi}{2} \log(m R) \,\int d^2 x\,x^2\,\langle \Theta(x) \Theta(0)\rangle\,.
\ee
This gives the change in the logarithmic term of the entropy, coming from one boundary in Rindler space. We have two boundaries for an interval, and the change in the central charge can be obtained using that for an interval in a CFT  $S\sim c/3 \log (R)+ \textrm{cons}$. We get  
\be
c_{UV}-c_{IR}=3 \pi \,\int d^2 x\,x^2\,\langle \Theta(x) \Theta(0)\rangle\,.\label{ccc}
\ee
This coincides with the known sum rule from Zamolodchikov c-theorem.

\bigskip

Interestingly, this same change in the area term should be reflected in the entropy of a black hole as we change the size of the black hole from a small to a large radius. The area term has its origin in short distance entanglement across the horizon and should not be changed by large scale details of the geometry. From the Bekenstein-Hawking entropy formula, the only way to achieve this matching is that Newton's constant itself renormalizes as $\Delta (4G)^{-1}=\Delta \mu_{d-2}$, in the way predicted by (\ref{popop}). This gives
\be
\Delta \left(\frac{1}{G}\right)= -\frac{4\pi}{d (d-1)(d-2)}\,\int d^{d}x\,x^2\,\langle \Theta(x) \Theta(0)\rangle\,.
\ee
This formula is called the Adler-Zee formula and was obtained long ago by computing corrections in the gravity effective action when integrating out quantum fields. It tells that the contribution of the dressing of massive quantum fields makes gravity stronger at the IR, probably reflecting the fact that any matter gravitates attractively.     

\subsection{Consequences for the irreversibility theorems}
Eq. (\ref{popop}) shows that what drives the total change of the area term under the RG is the operator $\Theta=T_\mu^\mu$. This is zero precisely for CFT's where no RG running takes place. By reflection positivity $\langle \Theta(x) \Theta(0)\rangle>0$, and in fact the correlator cannot be zero unless the operator $\Theta\equiv 0$ (this follows from Kallen-Lehmann representation of two-point functions). Since $\Theta\equiv 0$ is equivalent to say the theory is a CFT and never moves away from the fixpoint, any non-trivial RG running gives a non zero negative change on the area term. 

For the case $d=2$ this is a well-known consequence of Zamolodchikov's theorem. Eq. (\ref{ccc}) prevents any RG to connect two theories with equal central charge.     

For $d=3$ we can obtain the same conclusion. The area term is going to change for any non-trivial RG flow. In terms of the entropies, the change in the area term is
\be
\Delta \mu_{1}= \int_0^\infty dr\, S''(r)<0\,.
\ee
This means $S''(r)$ is strictly negative for at least some $r$ and is always non-positive by the $F$ theorem. The change in $F$ is
\be
 F_{UV}-F_{IR}= -\int_0^\infty dr\, r\, S''(r)>0\,,
\ee
which has then to be strictly positive. Notice that because of the suppression by the power of $r$, $\Delta \mu_1$ can be divergent (if the relevant deformation at the UV has $\Delta> 5/2$) while $\Delta F$ can still be finite. 
\subsection{Monotonicity of modular Hamiltonians}

Now we are going to use relative entropy again to prove an inequality for modular Hamiltonians. The setup is completely general and does not restrict to QFT.

Let us consider a bipartite system $G=AA'$, where $A'$ is the complement of $A$. Let us take $A\subseteq B\subseteq G$, where $G$ is the global system. We have another bipartition $G=BB'$ with $B'\subseteq A'$.  Then, let us take a pure reference state $\rho_0=|\psi\rangle\langle \psi|$ in the global system $G$ and another state $\rho^1$.  
From the monotonicity of relative entropy we have
\bea
S_A(\rho^1|\rho^0) &=& \Delta \langle K_A\rangle -\Delta S(A)\le  S_B(\rho^1|\rho^0)= \Delta \langle K_B\rangle -\Delta S(B)\,,\\
S_{B'}(\rho^1|\rho^0) &=& \Delta \langle K_{B'}\rangle -\Delta S(B')\le  S_{A'}(\rho^1|\rho^0)= \Delta \langle K_{A'}\rangle -\Delta S(A')\,.
\eea
All the differences are between the quantities for the state $\rho^1$ minus the ones for $\rho^0$. All the modular Hamiltonians are determined by $\rho^0$. 

The entropies for complementary regions in a pure state are equal, $S_0(B)- S_0(B')=S_0(A)- S_0(A')=0$. We will use the notation $\hat{K}_X=K_X-K_{X'}$ for the full modular Hamiltonian of $X$. This is the one implementing the modular flow in the full space and is a well-defined operator in the continuum limit. He also have $\hat{K}_X |\psi\rangle=0$ for any $X$. See section \ref{density}. 
 
Adding the two inequalities and rearranging terms we get
\be
\langle \hat{K}_B-\hat{K}_A\rangle_1 \ge  S_1(B)- S_1(B')- S_1(A)+ S_1(A')\,.\label{pos1}
\ee
Note the only place where the state $\rho^0$ enters in this inequality is in the choice of the modular Hamiltonian. If the state $\rho^1$ is pure the right-hand side vanishes. If it is not pure, we can add a new separated Hilbert space on a system $Z$ to purify it. Then, we can write 
\be
\langle \hat{K}_B-\hat{K}_A\rangle_1 \ge  I_1(B,Z)-I_1(A,Z)\ge 0\,.\label{pos}
\ee
Both mutual information on the right-hand side is for the state $\rho^1$. Monotonicity of mutual information shows the right-hand side is positive. The quantity on the right-hand side can be thought a form of entropy that is in the complement of $A$ in $B$ but is not produced by entanglement with the rest of the system $G$ but with some other universe $Z$.  

As (\ref{pos}) is positive for any state $\rho^1$ we conclude the difference between the full modular Hamiltonians of included regions is a positive operator
\be
A\subseteq B\rightarrow \hat{K}_B-\hat{K}_A\ge 0\,.\label{mono}
\ee
\subsection{Bekenstein bound II }

A different form of Bekenstein bound follows from this inequality for modular Hamiltonians. Take two parallel Rindler wedges, $B$ is the region $x^1>0$, and $A\subset B$ is the region $x^1>L$, and the region $B-A$ is a strip of width $L$. The difference between translated Rindler modular Hamiltonians (which are proportional to boost generators) is proportional to the standard Hamiltonian. Eq. (\ref{pos1}) gives in this case
\be
2 \pi \,L \,E\ge S(x^1>0)-S(x^1>L)+S(x^1<L)-S(x^1 <0)\,.    
\ee

All the quantities involved are for an arbitrary state in the QFT. This has a more striking resemblance with the original version of the Bekenstein bound (\ref{BB}) than the one we have previously derived, though it is not what follows from Bekenstein thought experiment, and has a less local flavour. Here the energy is the global energy, which is positive, and the entropy combination on the right-hand side vanishes for pure global states, making the inequality trivial in that case. The inequality shows that $2\pi L E $ bounds the failure of saturation of monotonicity of relative entropy. As we have seen, the combination of entropies on the right-hand side is a positive quantity conveying a meaning of entropy in the strip which cannot be associated with entanglement with other regions of the space. 

Suppose we live in a meta-theory and can manipulate degrees of freedom of the QFT and another system $Z$ of our laboratory. We could make a unitary transformation of the vacuum involving fields in the strip and some external operators, entangling $Z$ and the QFT. The result of this operation could make the entropy $S(Z)$ increase without limit without changing the energy if the QFT has enough degrees of freedom, say many copies of a given field. However, subtracting the trivial inequality for the vacuum, and considering that unitary transformations in an algebra do not change the entropy, we get 
\be
2\pi\,L\,  E\ge \Delta S_{x^1>0}+\Delta S_{x^1<L}\,,    
\ee
which has the form of the previous Bekenstein bound, and we know will not be violated thanks to vacuum entanglement.  

\subsection{The ANEC}
Now we show another application of the monotonicity of modular Hamiltonians. Let us recall the expression of the (full) modular Hamiltonians for regions with boundaries in a null plane in vacuum. For any QFT we have
\be
\hat{K}_\gamma=2\pi \, \int d^{d-2}y\, \int dx^+\, (x^+-\gamma(y)) \, T_{++}(x^+,y)\,. 
\ee

Now consider another curve $\gamma'$ which is changed only around a single light ray of coordinate $y=0$; $\gamma'(y)=\gamma(y)+\delta\gamma(y)$, with $\delta \gamma>0$ positive and vanishing outside an interval $y\in (-\epsilon,\epsilon)$. Since the region $\gamma'\subseteq \gamma$, from monotonicity we have
\be
\hat{K}_\gamma-\hat{K}_{\gamma'}=2 \pi\int d^{d-2}y\, \delta \gamma(y) \,\int dx^+\,  T_{++}(x^+,y)\ge 0\,.
\ee   

This has to hold for any positive smearing $\delta\gamma$, leading to the averaged null energy condition (ANEC)
\be
\int dx^+\,  \langle T_{++}(x^+)\rangle\ge 0\,,
\ee
where the state producing the expectation value is arbitrary. 

The ANEC in flat space implies Hoffmann-Maldacena inequalities between trace anomaly coefficients in a CFT. For QFT in curved space-time, the same inequality holds for space-times with a bifurcate Killing symmetry, where the integration is over affine parameter along a light ray of the Killing horizon. Due to the Gao-Wald theorem, this ensures boundary causality around the AdS vacuum in holographic theories.  

\subsection{Generalized second law}
The generalized second law is the statement that the area of a black hole horizon over $4G$ plus the entropy of the matter outside of it never decreases in time. In classical gravity  Hawking's area theorem shows the area of the horizon never decreases provided matter satisfy the null energy condition. When quantum fields are coupled to gravity this energy condition can fail and the black hole is allowed to evaporate. The question is what is ensuring the generalized second law in the semi-classical regime where the horizon is nearly stationary and the quantum fields produce small back-reaction.      
To have control over the problem we think in the semi-classical limit in which gravity is weak, curvature small, and the black hole radius is very large. This is the limit taken in the Bekenstein bound described above, in which the horizon approaches the null plane. To express the second law we need to define what is time. This can be taken as any foliation of space-time by spatial surfaces ordered in time, and their intersection with the horizon provides us with the area of the horizon as a function of time.  
As we have seen there is a natural quantity expressing the idea of what is the entropy outside the black hole, and this is the EE. This is again a function of the cut of the horizon because the choice of Cauchy slice outside of the black hole does not change the entropy.      

\begin{figure}[t]
\begin{center}
\includegraphics[width=0.4\textwidth]{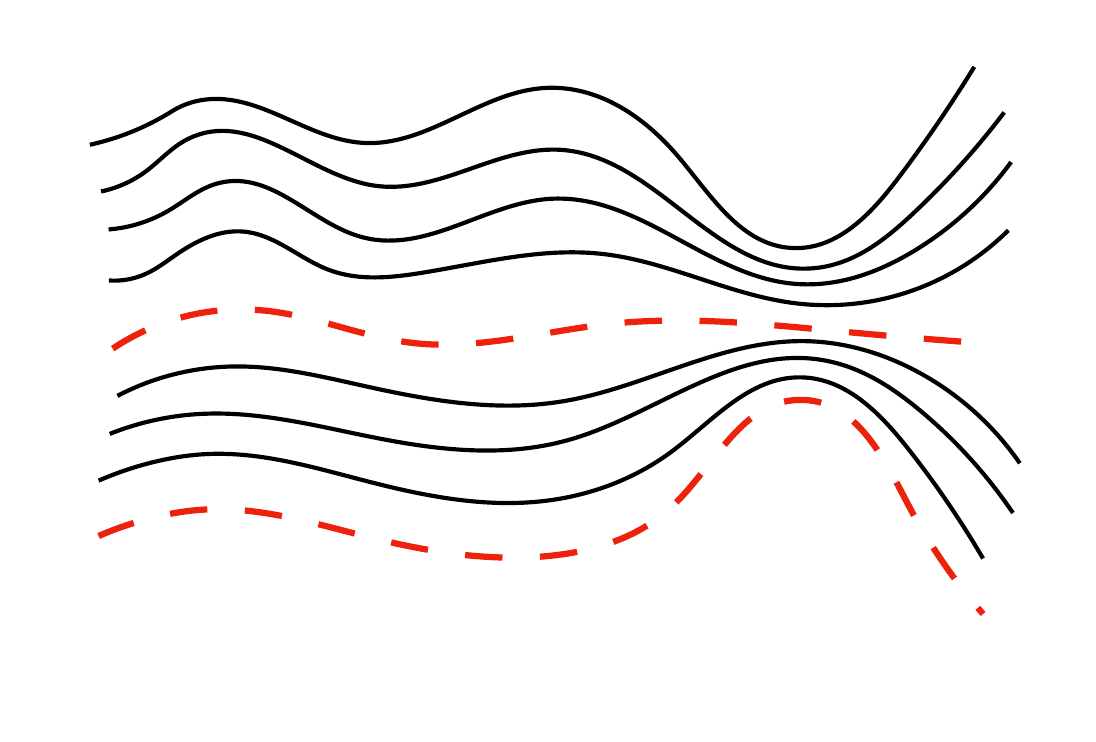} 
\caption{Foliation of the horizon; in red, two different times.}
\label{curvas}
\end{center}
\end{figure}

Then we are in a situation like figure \ref{curvas}, with a foliation of the horizon that in the limit $G\rightarrow 0$ is a null plane in Minkowski space. We need to show that the generalized entropy
\be
S_{\textrm{gen}}=\frac{A}{4G}+ S_{\textrm{out}}\label{tata}
\ee 
always increases in the direction of the null rays towards the future. 

First we have to evaluate the change in the area. Raychaudhuri equation gives 
\be
\nabla_\lambda \theta=-(d-2)^{-1} \, \theta^2-\sigma_{ab}\sigma^{ab}-R_{\lambda\lambda}\,.\label{churi}
\ee
Here $\theta=d \Delta A/(\Delta A d\lambda)$ is the expansion, the rate of change of the area $\Delta A$ in a small interval $\Delta y$ around a single light ray of transversal coordinate $y$ with the affine parameter $\lambda$. $\sigma_{ab}$ is called the shear, measuring deformations in the shape of the light-front. In the limit we are interested here,  the classical unperturbed horizon is stationary, and changes are induced by quantum matter. The first two terms in the Raychaudhuri equation are second-order and can be neglected. The last term is $R_{\lambda\lambda}=8 \pi G T_{\lambda\lambda}$, through Einstein equations. It follows that, to the lowest order,
\be
\frac{\partial_\lambda^2 \Delta A}{\Delta A}=-8 \pi G \,   T_{\lambda\lambda}\,.
\ee
After integrating this equation gives for the area of a slice $\gamma$
\be
A(\gamma)= -8 \pi G \,  \int d^{d-2}y \int_{\gamma}^\infty d\lambda\, (\lambda-\gamma(y))\,    \langle T_{\lambda\lambda}\rangle +\textrm{const}= -4 G \langle K_\gamma\rangle  +\textrm{const}  \,.
\ee
Notice this is proportional to the expectation value of the modular Hamiltonian $K_\gamma$ for the region determined by $\gamma$. 

Eq. (\ref{tata}) gives
\be
S_{\textrm{gen}}=   -\langle K_\gamma\rangle + (S_{\textrm{out}}-S^0_{\textrm{out}}) + \textrm{const}= \textrm{const}-S_\gamma(\rho_{\textrm{out}}|\rho^0_{\textrm{out}})\,.
\ee
We have added and subtracted the vacuum entropy $S^0_{\textrm{out}}$ which is constant for slices of the Rindler space as argued when we discussed the Markov property. It is clear that the increase of the generalized entropy will then be a consequence of the monotonicity of relative entropy as we decrease the size of the region moving $\gamma$ to the future.

\subsection{Local temperatures}

All results of this chapter depend on the fact that the modular Hamiltonian, for certain regions and states enjoying particular time-like symmetries, turn out to be local on the energy density operator. Naturally one may wonder if there is a way to generalize this result for other regions and states. For these other cases, the modular Hamiltonian is non-local, so the generalization has to go by a different route. 

\begin{figure}[t]  
\begin{subfigure}
\centering
\hspace{1.3cm}\includegraphics[width=0.26\textwidth]{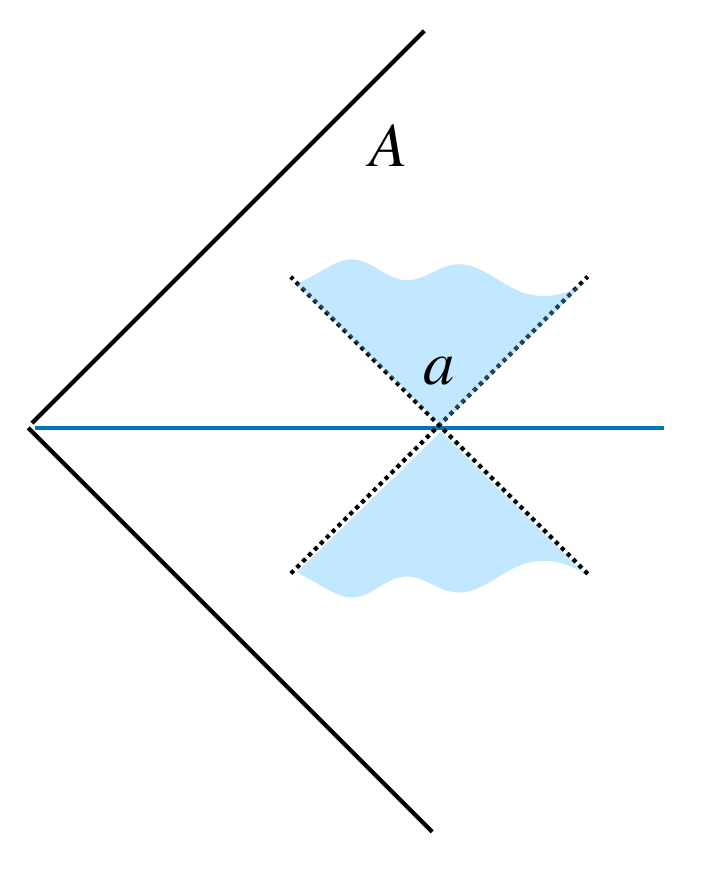}
\end{subfigure}
\hspace{2cm}
\begin{subfigure}
\centering
\includegraphics[width=0.3\textwidth]{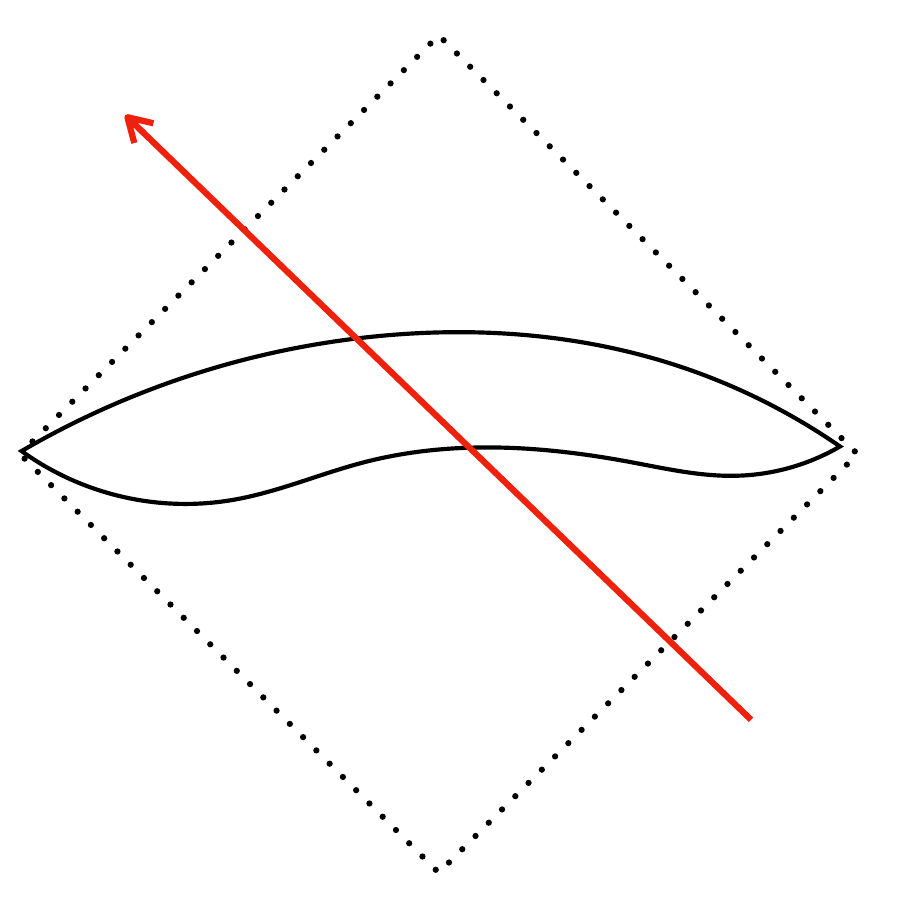}
\end{subfigure}
\captionsetup{width=0.9\textwidth}
\caption{Left: an excitation that is spatially localized at $t=0$ in the Rindler wedge. Right: A light like laser excitation (red) crossing two Cauchy surfaces.}
\label{rayo}
\end{figure}  

Let us consider again the case of a Rindler wedge. We try to find a definition of Unruh temperature that does not involve an observer with a particular trajectory given by the modular flow because these trajectories and observers are not going to be available in the general case.  The idea is to take a unitary operator $U$ well-localized around a point $x^1=l, x^0=0$ inside the wedge $W$ (figure \ref{rayo}) and compare the vacuum $\rho^0$ with a new state $\rho^1$ formed by acting with $U$ on the vacuum. The entropy in the wedge does not change and the relative entropy is
\be
S_W(\rho^1|\rho^0)=\Delta \langle K\rangle\sim 2 \pi \, l\, E\,. 
\ee
The relative entropy grows proportional to the energy. This result is independent of the ``chemical composition'' of the excitation, and the coefficient $(2\pi l)$  defines an inverse local temperature at the given point in the wedge $W$. This of course coincides with Unruh temperature for acceleration $a=l^{-1}$ as corresponds to boost orbits passing through the point.  One could consider localizing the excitation around other points $x$ inside the wedge, not lying at $t=0$, and the result is again proportional to the momentum $P^\mu$ of the excitation
\be
 S_W(\rho^1|\rho^0)\sim  \eta_\mu(x) \, P^\mu\,,\label{fif}
\ee
where $\eta_\mu(x)=2\pi x^\alpha (v_\alpha u_\mu- u_\alpha v_\mu)$ is proportional to the boost Killing vector of the wedge, where $v,u$ are two orthogonal unit vectors normal to the edge of the wedge.  

For a CFT one can produce the same type of energy increasing relative entropy for spherical regions. For a region $A$ that is not a sphere nor a wedge, and a point $x\in A$, we can place $A$ inside a wedge $W$,  and take a sphere $S$ inside $A$, including the point $x$, to give purely geometrical upper and lower bounds from monotonicity
\be
   \beta_S(x)\,  E \le S_A(\rho^1|\rho^0)\le \beta_W(x) \, E\,.    
\ee
This suggests a degree of universality that would hold for any region. In particular, one could ask whether in the limit of large energies we could have $S_A(\rho^1|\rho^0)\sim \beta_A(x) E$ for some geometrically determined $\beta_A(x)$, independently of composition. Formula (\ref{fif}) already shows this is not possible, because in general, the coefficient $\beta$ will depend on the direction of the excitation too. In figure \ref{rayo} right panel it is shown a light-like high energy excitation with almost null momentum. This is localized along different points on the causal set $A$ and we must have the same $\beta$ for all these points. In particular, for regions $\gamma$ with boundary on the light cone in a CFT, we can read off $\beta$ from the local form of the modular Hamiltonian on the null surface. It is easily seen that we can change $\beta$ for the same point $x$ and different directions at will by changing the form of the curve $\gamma$. Then, the conjecture would rather be that for the limit of high energy, almost null momentum, we could have
\be
S_A(\rho^1|\rho^0)\sim \beta_A(x,\Omega)\, E\,,
\ee
with the direction-dependent local inverse temperature $\beta_A(x,\Omega)$ determined geometrically, and independent of excitation composition. 

This in fact turns out to be correct for free massive scalar or fermion fields  ($\beta$ is independent of the mass, and the same for bosons and fermions) and one can argue the same local temperatures occur for superrenormalizable theories, since this is a large energy limit. $\beta$ is determined by an eikonal-type equation with Rindler like boundary conditions on $\partial A$. However, this equation is not easy to solve. In $d=2$ we have two null directions, and the explicit solution for $n$ intervals given by the null coordinates of their end points $(a_1^\pm,b_1^\pm),\cdots, (a_n^\pm,b_n^\pm)$ is
\be
\beta^\pm(x)=\frac{2\pi}{\sum_i \left(\frac{1}{x^\pm-a_i^\pm}+\frac{1}{b_i^\pm-x^\pm}\right)}\,.  
\ee
This can be thought of as produced by a local term in the modular Hamiltonian of $n$ intervals of the form $K^\pm_{\textrm{loc}}= \int dx\, \beta^\pm(x)\, T_{\pm\pm}(x)$. In more dimensions, however, the local temperatures cannot be exclusively produced by local terms proportional to the stress tensor. These terms do not have enough parameters to encode a $\beta(x,\Omega)$ with arbitrary dependence on the angular variables $\Omega$. 

Local temperatures display a universal connection between entanglement, energy and geometry in QFT. The interpretation of this connection is rather subtle though. The temperature measures the statistics of entanglement on high energy modes around a point. However, this entanglement has a long-distance origin, being sustained with the complement of $A$.     

\subsection{The QNEC}
The quantum null energy condition (QNEC), as the Bekenstein bound, was discovered from a gravity conjecture, restricted to some flat space limit. The conjecture, in this case, is called the quantum focusing conjecture. 

\begin{figure}[t]
\begin{center}
\includegraphics[width=0.65\textwidth]{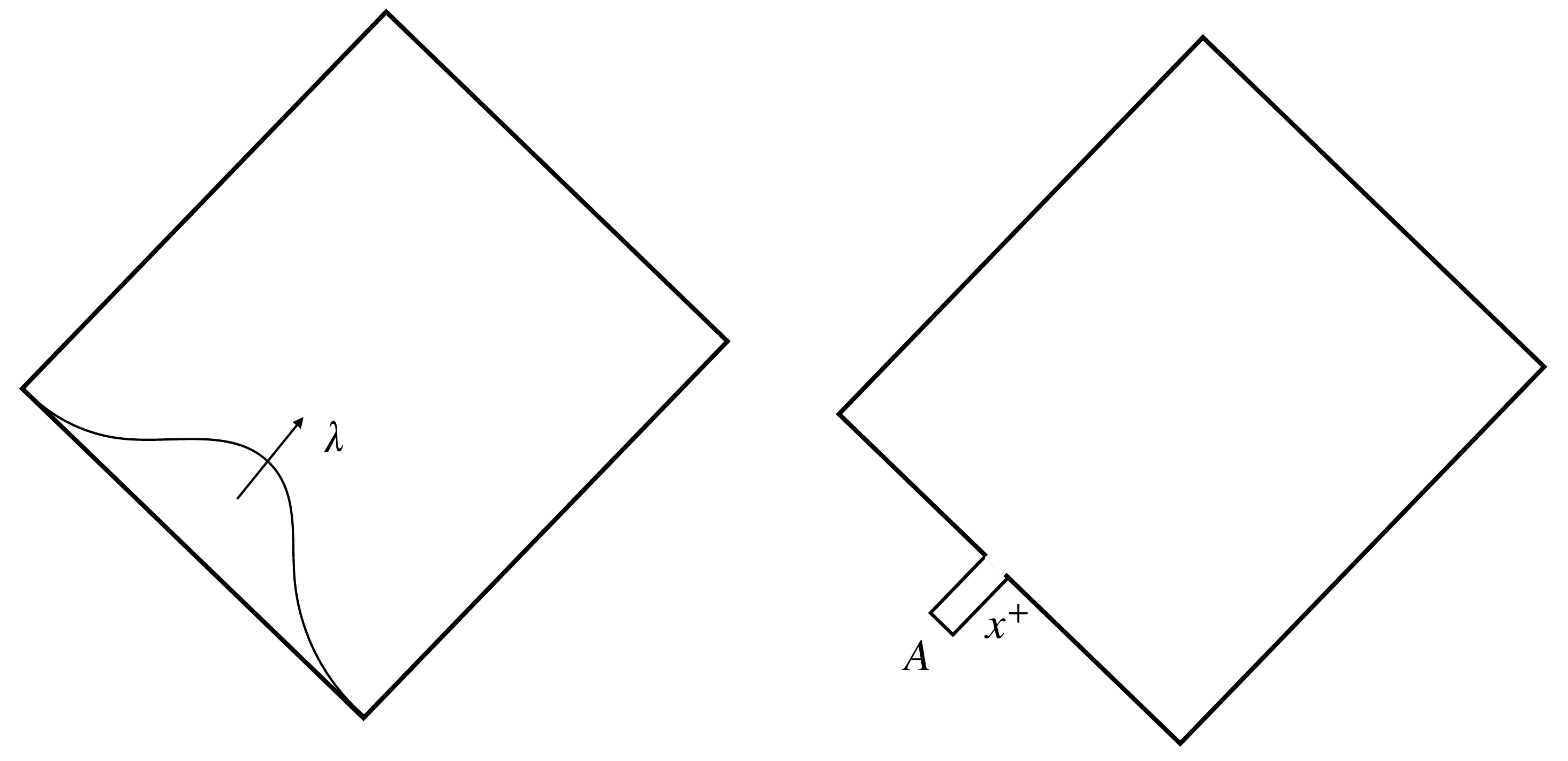} 
\caption{Null surfaces with boundary deformed. On the right a thin deformation.}
\label{qnec}
\end{center}
\end{figure}

The idea starts with Raychaudhuri equation (\ref{churi}). In classical gravity, using Einstein equations, and assuming the null energy condition (NEC), $T_{\mu\nu}k^\mu k^\nu\ge 0$ for all null vectors $k^\mu$, the equation (\ref{churi}) implies that the expansion decreases
\be
\frac{d\theta}{d\lambda}<0\,,
\ee
with $\lambda$ parametrizing the null direction. 
 This is the focusing property of gravity, which accelerates null rays to each other. Quantum fields can violate the NEC, making the expansion positive, and allowing, for example, black holes to evaporate. Though the classical focusing does not hold anymore, the idea is that a quantum corrected expansion $\Theta$ should still be focusing. In the generalized second law, the area that increases in classical Hawking theorem is replaced by the generalized entropy, $S_{\textrm{gen}}=A/(4G)+S_{\textrm{out}}$, containing also the quantum entropy of fields for a Cauchy slice ending at the null surface. The quantum focusing conjecture is then
 \be
 \frac{d\Theta}{d\lambda}<0\,,\hspace{.7cm} \Theta= \frac{dS_{\textrm{gen}}}{d\lambda}\,,   
\ee
where $\lambda$ is a parameter describing the advance of the cut on the null surface (figure \ref{qnec}). 

In the limit $G\rightarrow 0$ this conjecture gives a statement about QFT in Minkowski space. This follows again from writing the derivative of the area in terms of the stress tensor using Raychudhuri equation. The QNEC is the resulting inequality
\be 
\int d^{d-2}y\, \langle T_{ab}(x)\rangle\, \frac{dx^a}{d\lambda}\frac{d x^b}{d \lambda}\ge \frac{1}{2\pi}\frac{d^2 S_{\textrm{out}}}{d\lambda^2}\,.\label{qqne}
\ee
In this expression, $\lambda$ is a parameter of null deformations $x^{a}(y,\lambda)$ of the boundary of the null surface (see figure \ref{qnec}), and $y$ represents the transversal coordinates of the null surface.   

Contrary to the Bekenstein bound or the ANEC, this inequality remarkably can give a lower bound to the stress tensor at a point. This is achieved by taking a localized variation of the surface. On the other hand, the lower bound is in terms of a second derivative of a global entropy of regions with boundaries passing over the point. One technical point is that the null surface has to have zero expansion at the point where the variation is made. 

More remarkable still is that the QNEC is in fact true in QFT. The known proofs are however not so simple. You can see the references below. We limit ourselves here to give an argument based on the replica trick.  

The two sides of the inequality (\ref{qqne}) look quite different in nature. The left-hand side is additive in the $y$ direction over the deformation, while the right-hand side is not. Deformations on different points will contribute to the second derivative of the entropy. However, these second derivatives at different points are in fact a particular case of SSA and are negative. Therefore, the first observation is that (\ref{qqne}) can be reduced to the case of deformations in an infinitesimal interval in the coordinate $y$, plus SSA. Hence we focus on the so-called local QNEC, the situation of figure \ref{qnec}, right panel. Calling $x_0^+$ to the size of the spike deformation, we should have
\be
\lim_{x^+_0 \rightarrow 0}\lim_{ A\rightarrow 0}\frac{1}{A}  \frac{d^2 S_{\textrm{out}}}{ dx_0^{+\,2}}\le  2 \pi \, \langle T_{++}(x)\rangle\, \,,\label{equality}
\ee
where $ A$ is the small transversal area. In this limit, we take $A$ to zero first than the size of the spike $x_0^+$.    

Now we want to compute the change of entropy and for that use the replica trick. The deformation of the twist operator $\tau_n$ along the spike in figure \ref{qnec} is very near annihilating itself and leave a twist operator $\tau_n^0$ without spike. Since this limit is taken keeping the state fixed, an operator product expansion of this twist operator should live us with the insertion of local operators of the replicated theory along the line of the spike. The expansion should have the form 
\be
\tau_n=(1+(1-n)\, O_\Delta \, A^{\frac{\Delta-h-h_0}{d-2}}\, (x^+)^h\,(x_0^+)^{h_0}+\cdots ) \,\tau_n^0\,.
\ee
The operators can be chosen to sit at $x^+=0$, and
 we expect positive powers of $A$ and positive integer powers of $x^+$ and $x_0^+$ in the expansion. The form of the term in the expansion is only dictated by dimensional analysis and Lorentz invariance. The powers of $x^+,x^+_0$, have to be compensated by the spin $s=h+h_0$ of the field $O_\Delta$ by boosting invariance. We have also placed a $(1-n)$ factor for the single copy operators. 
 
 Let us analyse first the case of the single copy operators. They will contribute to the entropy as 
\be 
 \langle O_\Delta\rangle \,A^{\frac{\Delta-h-h_0}{d-2}} \,(x^+)^h(x_0^+)^{h_0}\,.
\ee 
 Only terms with $h_0=2$, corresponding to spin $s=2+h$ for the operator, contribute to the second derivative with respect to $x_0^+$ in (\ref {equality}) after the limit $x^+_0\rightarrow 0$ is taken. Only $(\Delta-h-h_0)=(\Delta-s)\le d-2$ contribute to (\ref{equality}) after the limit $A\rightarrow 0$ is taken.  For operators with spin $s\ge 2$ there is a minimum value of $\Delta-s=d-2$ (usually called the twist of the operator) and this is the case of the stress tensor. This implies we have to look at the term with $h=0$, $h=2$, that is linear with $A$. Therefore, the only contribution to the left hand side of (\ref{equality}) coming from single copy operators has to be proportional to  $\langle T_{++}(x^+=0)\rangle$.  
 
 For multiple copy operators we need to produce $(x_0^+)^2$ still, but this time two fields of spin $1$ in different copies can do the work. Their minimal dimension is the one corresponding to the derivative $\partial_+ \phi$ for a free scalar field $\phi$, which is $\Delta=d/2$. These two fields will again imply a minimal linear power for $A$ for dimensional reasons. So they can contribute, but any other operator or products of operators in more than two copies could not contribute. If the theory does not contain an exactly free scalar field there is no contribution at all from the multiple copies. Let us restrict ourselves to this simpler case.     

Single copy operators are the ones that contribute to the entropy linearly in the state, and therefore we are in the limit of the first law of entanglement entropy, with $\Delta S=\Delta K$, as far as (\ref{equality}) is concerned. Therefore to find out the coefficient of the entropy term in (\ref{equality}) we simply calibrate with the modular Hamiltonian of a deformed null plane. We have
\be
\partial_{x_0^+}^2 \,2 \,\pi\,\int_{x_0^+}^\infty dx^+\, (x^+-x_0^+)\,T_{++}(\lambda)=2 \pi \, T_{++}(x_0^+)\,.
\ee
This shows the local QNEC (\ref{equality}) actually saturates (this is not so for the free scalar). 
This is confirmed by holographic calculations in holographic theories. The saturation of the local QNEC plus SSA gives a proof of the QNEC. 

Notice that if we were not interested in second derivatives with respect to $x^+_0$ for $x_0^+\rightarrow 0$ a whole series expansion in $A \,(x^+)^{n+2} (\partial_+)^n T_{++}(0)$ would be allowed, and we could actually produce contributions to the single copy operators non-local in the light ray. This is indeed what actually happens, the spike deforms the modular Hamiltonian non locally along the null ray. This is naturally expected. The OPE should live a local contribution of operators, but in this null setup, all the null line is local with respect to the deformation. There could not be operators for $x^+< x^+_0$ though, because they do not commute with the complement of the region.        

\subsection{Exercises}

\begin{itemize}

\item[1.-] Using the first law of entanglement compute the correction to the vacuum entropy of a ball of radius $R$ in a CFT due to a small temperature $R T\ll 1$. Compare with the result for a thermal state in an interval in $d=2$: $S(L)=\frac{c}{3} \log\left(\frac{\beta}{\pi\epsilon}\,\sinh\left(\frac{\pi L}{\beta}\right)\right)$.

\item[2.-] Compute the change of the area term due to a mass for a free fermion field in $d=3$ using the formula in terms of correlators of the stress tensor traces. You can also compute the entropy directly using the heat kernel, see \cite{Hertzberg:2010uv}. 

\end{itemize}

\subsection{Notes and references}
A simple construction of states with negative energy density for any theory can be found in \cite{Witten:2018lha}.  The paper by Bekenstein is \cite{bekenstein1981universal}. A nice historical account on Bekenstein bound and its ramifications is the paper by Bousso \cite{Bousso:2018bli}.
 The proof of the bound is from \cite{Casini:2008cr}. Marolf, Minic and Ross introduced the idea that the change in entanglement entropy was fundamental to understand the bound, especially to deal with the so-called species problem \cite{Marolf:2003sq}. 
 
The first law of EE was introduced in \cite{Blanco:2013joa,Wong:2013gua}. Its application to derive Einstein equations in holographic theories was developed in \cite{Lashkari:2013koa,Faulkner:2013ica}.   

The derivation of the area terms in EE is from \cite{Casini:2014yca}. Analogous calculations were done by Rosenhaus and Smolkin  \cite{Rosenhaus:2014ula}. For an holographic proof see \cite{Casini:2015ffa}. Adler and Zee papers on the renormalization of Newton's constant are \cite{Adler:1982ri,Zee:1980sj}. The relation between QFT entanglement and black hole entropy has been much debated. See for example \cite{Cooperman:2013iqr} and references therein. 

The presentation of monotonicity of modular Hamiltonians from the monotonicity of relative entropy as well as the second version of Bekeinstein bound comes from \cite{Blanco:2013lea}. Monotonicity of modular Hamiltonians was a result known by mathematicians about modular theory.     

The proof of the ANEC is from \cite{Faulkner:2016mzt}. An argument not involving modular Hamiltonians has been given \cite{Hartman:2016lgu}.  
 
The proof of the semi-classical GSL is from Wall's paper \cite{Wall:2011hj}. The generalized second law as well as the Bekenstein bound have a long history. The physical ideas of the proofs in both instances have been previously advanced with different degrees of precision. See for example \cite{Marolf:2003wu,Marolf:2003sq,sorkin1986toward,Sorkin:1997ja,Frolov:1993fy,zurek1985statistical}.

The local temperatures were introduced and studied in \cite{Arias:2016nip,Arias:2017dda}. 

The quantum focusing conjecture and the QNEC were proposed in \cite{Bousso:2015mna}. The proof for free fields is in \cite{Bousso:2015wca}, for holographic models  \cite{Koeller:2015qmn,Leichenauer:2018obf}, and for interacting theories \cite{Balakrishnan:2017bjg,Balakrishnan:2019gxl}. 
 
\newpage

\section{Entanglement and symmetries}
In this chapter, we enter more into the details on how algebras are assigned to regions and their mutual relationship. It will turn out that some features in these relations encode symmetries of the theory.  

There are models in which the possible assignations of algebras to regions is non-unique. These break two basic properties of the net of algebras: duality and additivity. The models showing these features contain some operators which, though can be assigned to a region, are non locally generated by field operators in the same region.  The restitution of duality or additivity cannot be done simultaneously by changing algebra-region assignations within the model. These non locally generated operators are called intertwiners and twists for the case of global symmetries and Wilson or 't Hooft loops for generalized global symmetries of the type that appear in gauge theories.

It is clear from what we have learnt that the UV ambiguities in the EE can be cured by considering relative entropy quantities such as mutual information. But this is not the whole story. For the QFT where the assignation of local algebras to regions is not unique, the different choices are physical possibilities of the continuum QFT and change the universal pieces of the entropy or the mutual information. These issues are behind topological terms in the entropy and the mismatch of the logarithmic coefficient in EE for the Maxwell field with respect to the anomaly coefficient. 

This same non-uniqueness allows us to define a relative entropy measuring the statistics of non-local operators. This gives a frame to describe the physics of generalized symmetries and phases in QFT using relative entropy order parameters.

\subsection{Harmony of algebras and regions. }
In the algebraic approach, a QFT is described by a net of von Neumann algebras. This is an assignation of an operator algebra to causal regions of space-time. The particular QFT model is determined by how the algebras in the net relate to each other and the vacuum state.

In this chapter, in contrast to the preceding ones, we will be less interested in the geometry of Minkowski space, and restrict attention to consider multiple causal regions based on the same Cauchy surface ${\cal C}$, that can be taken to be the surface $t=0$. These causal regions will have in general non-trivial topologies, whose topological properties are the same as the ones of subregions of ${\cal C}$  in which they are based. Hence, we will often make no distinction between a $d-1$ dimensional subset of ${\cal C}$ and its causal $d$-dimensional completion.

The algebras ${\cal A}(R)$ attached to regions $R$ satisfy the basic relations of {\sl isotony} 
\be
{\cal A}(R_1)\subseteq  {\cal A}(R_2)\,,\hspace{1cm}R_1\subseteq R_2\,, \label{isotonia}
\ee
and {\sl causality},
\be
{\cal A}(R) \subseteq  ({\cal A}(R'))'\,,\label{causality}
\ee
where $R'$ is the causal complement of $R$, i.e. the space-time set of points spatially separated from $R$, and  ${\cal A}'$ is the algebra of all operators that commute with those of ${\cal A}$. We always have von Neumann's relation ${\cal A}''={\cal A}$. Causality (\ref{causality}) is another way to say that observables based at spatially separated regions must commute. 

Some extensions of these relations are expected to hold for sufficiently complete models but are not granted on general grounds. This is exactly what this is all about! For example, (\ref{causality}) could be extended to the relation of {\sl duality} (also called Haag's duality)
\be
 {\cal A}(R)= ({\cal A}(R'))'\,,   \label{duality}
\ee
and we could also expect a form of {\sl additivity}
\be
{\cal A}(R_1 \vee R_2)={\cal A}(R_1 )\vee{\cal A}( R_2)\,, \label{additivity}
\ee 
where $R_1 \vee R_2=(R_1\cup R_2)''$,  ${\cal A}_1 \vee{\cal A}_2=({\cal A}_1 \cup {\cal A}_2)''$ are the smallest causal regions and von Neumann algebras containing $R_1, R_2$ and ${\cal A}_1, {\cal A}_2$ respectively. The heuristic meaning of duality is that local algebras contain a maximal set of operators compatible with causality. Additivity is the fact that the algebras of two regions generate the algebra of their union. It is naturally expected when we think the QFT algebras are generated by quantum fields.

When we have the most harmonious relation between algebras and regions and the net satisfies (\ref{duality}) and (\ref{additivity}) for all $R$ based on the same Cauchy surface we can call the net {\sl complete}. The main focus of this discussion concerns nets that are not complete in this sense and how this is related to generalized symmetries in the QFT.

 The de Morgan laws 
\bea
({\cal A}_1\vee {\cal A}_2)'={\cal A}_1'\cap {\cal A}_2'\,,\\
(R_1\vee R_2)'=R_1' \cap R_2' \,,
\eea
 are universally valid for regions and algebras. 
From these relations it follows that if we have unrestricted validity of duality (\ref{duality}) and additivity (\ref{additivity}), we have the {\sl intersection property}
\be
{\cal A}(R_1 \cap R_2)={\cal A}(R_1 )\cap {\cal A}( R_2)\,. \label{intersection} 
\ee
Conversely, additivity follows from unrestricted validity of duality and the intersection property. From this perspective, the intersection property is then another aspect of duality and additivity.

It is expected that algebras for topologically trivial regions $R$, such as a ball, satisfy duality, and that we have additivity for topologically trivial regions whose union is also topologically trivial. This last statement means the algebra of $R$ is generated by the algebras of any collection of balls (of any size) included in $R$ and whose union is all $R$. This accounts for the idea that the operator content of the theory is formed by local degrees of freedom. The additivity property then can be summarized in that any localized operator of the theory is locally generated (in a topologically trivial space).\footnote{These statements should apply for most theories though there are counterexamples. The orbifold of theories with spontaneously broken global symmetries gives place to examples where balls do not obey duality. This fact does not interfere with our description of global symmetries below. Some generalized free fields are not additive for topological trivial regions. However, this is a problem more related to the absence of a stress tensor (or the so-called time slice axiom) in these theories: In these theories, the algebras are not uniquely associated to causally complete regions (or equivalence classes of pieces of Cauchy surfaces with the same causal completion) but depend on the space-time geometry of the region beyond causality.}

However, when the region is topologically non-trivial,  whether any operator of a certain algebra  ${\cal A}(R)$ is locally generated inside $R$ itself becomes a more subtle question.  We will see in the next section some examples that will show the existence of non locally generated operators in such ${\cal A}(R)$ is not an uncommon phenomenon. 

Let us be more precise. Given a net, we can always construct an additive algebra for a region $R$ as
\be
{\cal A}_{\textrm{add}}(R)= \bigvee_{B \,\textrm{is a ball}\,, \,B\subseteq R} {\cal A}(B)\,. 
\ee
This gives us an algebra that is minimal, in the sense that it contains all operators which must form part of the algebra because they are locally formed in $R$. The assignation of ${\cal A}_{\textrm{add}}(R)$ to any $R$  gives the minimal possible net 
and if ${\cal A}_{\textrm{add}}(R)\subsetneq {\cal A}(R)$ it follows that we can have different nets with the same operator content of the full theory. 

On the other hand, in this freedom of choosing the operator content of different regions, the greatest possible algebra of operators that can be assigned to $R$ and still satisfy causality must correspond to a minimal one assigned to $R'$,
\be
{\cal A}_{\textrm{max}}(R)= ({\cal A}_{\textrm{add}}(R'))'\,.
\ee
Evidently if ${\cal A}_{\textrm{add}}(R)\subsetneq {\cal A}_{\textrm{max}}(R)$ it follows that the additive net does not satisfy duality. In this situation, one can enlarge the additive net by adding non locally generated operators, to generate a net satisfying duality~(\ref{duality}). In general, this may be done in multiple ways.  By construction, these nets satisfy duality
\be
{\cal A}(R)=({\cal A}(R'))'\,,
\ee
but in general, will not satisfy additivity. Therefore there is a tension between duality and additivity which cannot be resolved in these incomplete theories. 

All these algebraic features have consequences on the EE. If the net does not satisfy duality we cannot expect the equality of entropies for complementary regions $S(R)=S(R')$ to hold. This equality holds for pure states for complementary algebras, and as the algebras of $R$ and $R'$ are not complementary the entropies will be generally different. If we have two different possible algebras for the same region, there will be more than one possible entropy for the same region too. 

To be more concrete, let us call $a \in {\cal A}_{\textrm{max}}(R)$ to a  collection of non locally generated operators in $R$ such that 
\be
{\cal A}_{\textrm{max}}(R)=({\cal A}_{\textrm{add}}(R'))' = {\cal A}_{\textrm{add}}(R)\vee \{a\}\,.
\label{a}\ee
In the same way we have operators $b \in {\cal A}_{\textrm{max}}(R')$ non locally generated in $R'$ such that
\be
{\cal A}_{\textrm{max}}(R')=({\cal A}_{\textrm{add}}(R))' = {\cal A}_{\textrm{add}}(R')\vee \{b\}\,.
\label{b}\ee
Evidently, the {\sl dual} sets of operators $\{a\}$ and $\{b\}$ cannot commute to each other. 
Otherwise it would be  ${\cal A}_{\textrm{max}}(R) \subseteq ({\cal A}_{\textrm{max}}(R'))'= {\cal A}_{\textrm{add}}(R)$ and these operators would be locally generated. Given the existence of non locally generated operators $a$ in $R$,
the necessity of the existence of dual complementary sets of non locally generated operators $b$ in $R'$  is  due to the fact that for two different algebra choices ${\cal A}_{1,2}$ for $R$ there are two different  choices ${\cal A}_{1,2}'$ associated to $R'$. The latter cannot coincide because of von Neumann's relation ${\cal A}''={\cal A}$.    
 
In constructing nets ${\cal A}(R)$  we have to resign some operators from either  ${\cal A}_{\textrm{max}}(R)$ or ${\cal A}_{\textrm{max}}(R')$ since, because of causality, we cannot take all possible ones both for $R$ and $R'$.
In particular, a possible choice is  
${\cal A}_{\textrm{max}}(R)$ for $R$ and ${\cal A}_{\textrm{add}}(R')$ for $R'$ or vice-versa, and usually there are  some intermediate choices. If the topologies of $R$ and $R'$ are the same, both of these choices are not very natural and may break some spatial symmetries.\footnote{When referring to the topology of an infinite region, i.e. the complement of a bounded one, we will assume the space is compactified in a sphere.} 

An important remark is that even if some non locally generated operator is excluded from the algebra of $R$ it does not mean this operator does not exist in the theory. All non locally generated operators that could be assigned to $R$ are always formed locally in a ball containing $R$ and thus its existence cannot be avoided. They will always belong to the algebra of this ball. The sets $\{a\}$ and $\{b\}$ form complementary sets of observables based on complementary regions. This does not violate causality because to construct $\{a\}$ in a laboratory from microscopic operators we need to have access to a ball including $R$ which non trivially intersects $R'$.       
 
The multiplication of non local operators by additive ones cannot change the non local nature of the operator. Therefore non local operators can be grouped into equivalence classes $[a]$ under the action of the additive operators. Because the algebras are closed under multiplication the classes have to close a fusion algebra between themselves $[a]\, [a']=\sum_a'' n^{a''}_{a a'} \,[a'']$. The same will happen with the complementary operators $[b] \,[b']=\sum_{b''} \tilde{n}^{b''}_{b b'} \,[b'']$.   
These dual fusion rules are sometimes associated with groups representations or their conjugacy classes. In fact, this structure is directly linked to the {\sl symmetries} involved in the model. 

The particular topology of $R$ where the additivity or duality fail may restrict the type of symmetry involved. Global symmetries (in the orbifold theory --- see below) give algebra-region ``problems'' when some of the homotopy groups $\pi_0(R)$ or $\pi_{d-2}(R)$ are non-trivial. The case of ordinary gauge symmetries might give problems for regions with non-trivial $\pi_1(R)$ or $\pi_{d-3}(R)$. Higher homotopy groups correspond to the case of gauge symmetries for higher forms gauge fields.

\subsection{Example: a chiral free scalar.}
A simple example where these features appear is a free chiral scalar field in $d=2$. For this model, we can make an exact computation of the EE too. This model is the derivative of a free massless scalar $j(x^+)=\partial_+ \phi$ (the full massless scalar in $d=2$ is not a well defined QFT). The field only depends on the null coordinate $x^+$, which is an operator in a line. We will suppress the $+$ index in what follows. The model is conformal invariant too. It is Gaussian with Hamiltonian and  commutation relations  
\be
H=\frac{1}{2}\int dx\, j(x)^2\,,\hspace{.5cm} [j(x),j(y)]=i\delta'(x-y)\,.\label{conr}
\ee
This defines the theory completely. 

\begin{figure}[t]
\begin{center}  
\includegraphics[width=0.55\textwidth]{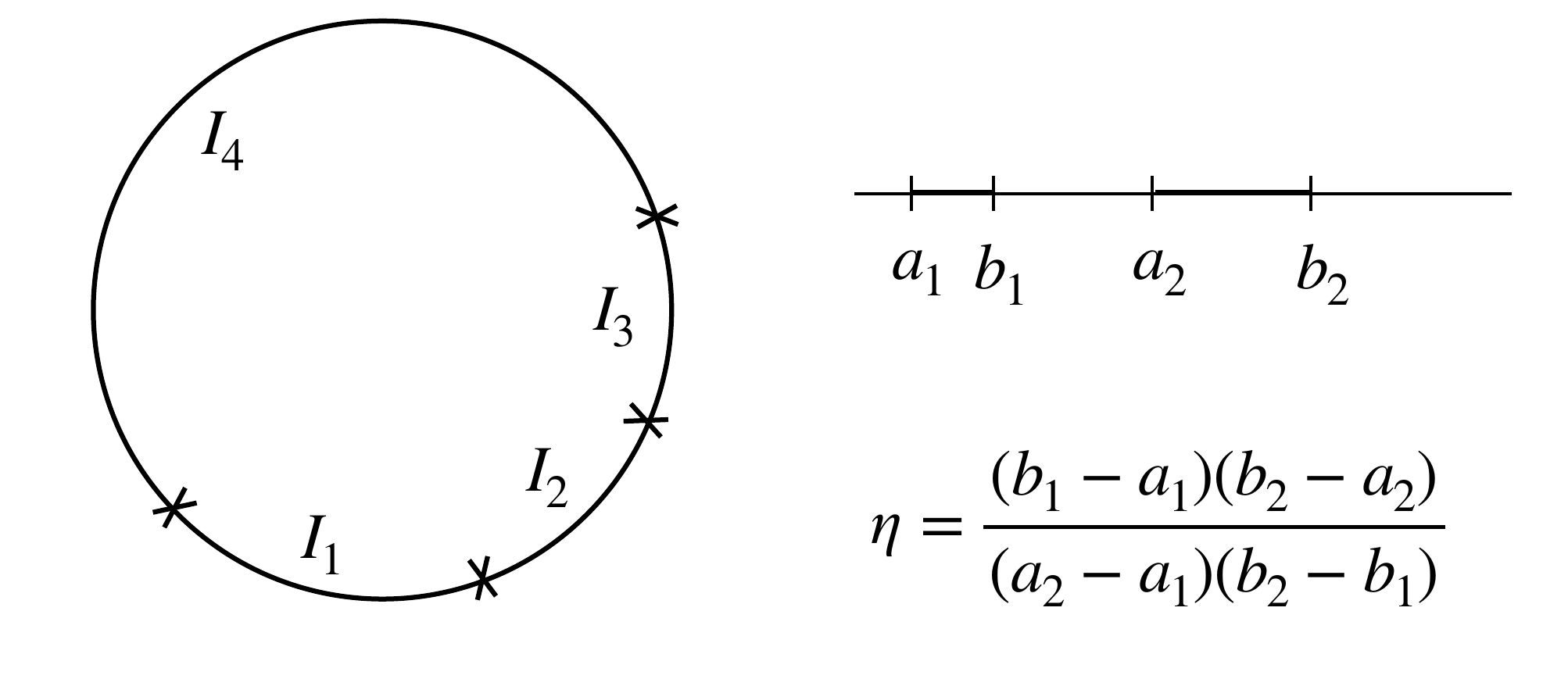}
\captionsetup{width=0.9\textwidth}
\caption{Left panel: Intervals in a circle. The region $R=I_1\cup I_3$ and its complement $R^{\prime}=I_2\cup I_4$. 
Right panel: Cross ratio $\eta$ for four ordered points $a_1,b_1,a_2,b_2$.}
\label{dualityexample}
\end{center}  
\end{figure}

Consider the case of a general CFT on a line, and the case of a region topologically non-trivial: the simplest case is two intervals. We can compactify the theory in a circle as shown in figure \ref{dualityexample}. This is equivalent to the line after a conformal transformation. In the circle we have four intervals $I_1,I_2,I_3,I_4$. Assume we have Haag duality for two intervals. The entropies will satisfy 
\be
S(I_1\cup I_3)=S(I_2\cup I_4)\,.\label{iden}
\ee
From that, adding single interval entropies we get
\be
 I(I_1,I_3)=I(I_2,I_4)+S(I_1)+S(I_3)-S(I_4)-S(I_2)\,.
 \label{1.13}
\ee
A conformal invariant quantity depending on four points $a_1,b_1,a_2,b_2$ (here the end points of the two intervals) on the line must be a function of a single number, the cross ratio, formed out of the coordinates of the four points as
\be
\eta=\frac{(b_1-a_1)(b_2-a_2)}{(a_2-a_1)(b_2-b_1)}\in(0,1)\,.
\ee
The mutual information is then a function of the cross-ratio and we write it for convenience as 
\be
I(\eta)=-\frac{c}{3} \log(1-\eta)+U(\eta)\,,\label{uu}
\ee
with $c$ the central charge. This explicitly extracts the logarithmic divergence in the case of two intervals touching each other. 

If the cross ratio of $I_1 I_3$ is $\eta$ the one of $I_2 I_4$ is $1-\eta$. Using this in  eq (\ref{1.13}), and the single interval entropies $S(r)=c/3 \log(r)$, we get
\be
I(\eta)=I(1-\eta)-\frac{c}{3} \log(\frac{1-\eta}{\eta})\leftrightarrow U(\eta)=U(1-\eta)\,.
\ee
This symmetry of $U(\eta)$ expresses Haag duality for two intervals in a regularization independent way. 

\begin{figure}[t]
\begin{center}  
\includegraphics[width=0.55\textwidth]{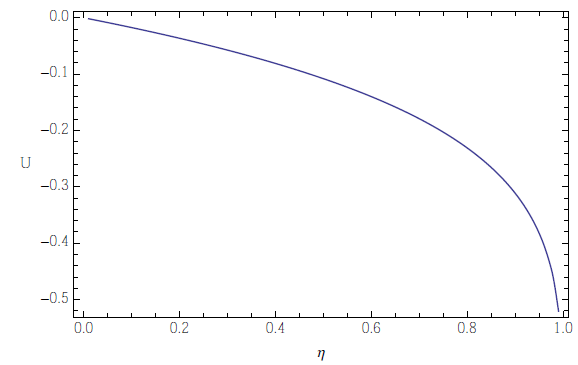}
\captionsetup{width=0.9\textwidth}
\caption{The function $U(\eta)$ for a chiral scalar.}
\label{curr}
\end{center}  
\end{figure}

For the chiral scalar we have to use $c=1/2$ (has $c=1$ for only one chirality). The function $U(\eta)$ can be computed exactly in this model. It has a complicated expression
\be
U(\eta)= -\frac{i\pi}{2}\int_0^\infty ds\frac{s}{\sinh^2(\pi s)}\, \log\left(\frac{ _2F_1(1+i s,-i s,1,\eta)}{  _2F_1(1-i s,is,1,\eta)}\right)\,,
\ee
which is shown in figure \ref{curr}. This is not symmetric. We then conclude for this model
\be
S(I_1\cup I_3)\neq S(I_2\cup I_4)\,.\label{neq}
\ee

Why is this happening? This shows there is a violation of Haag duality for two intervals. The reason is easy to figure out. In computing the mutual information and the entropies in (\ref{neq}) we have used the additive algebras, generated by the operators $\partial \phi(x)$ with $x$ in each of the two intervals. This must be so because in the mutual information the algebra of the two intervals must be the tensor product of the algebras of single intervals. However, for each two-interval, there is still another operator we can assign to it. If $x_1\in I_1$ and $x_3\in I_3$ we have the operator
\be
O_{13}=\phi(x_1)-\phi(x_3)=\int_{x_1}^{x_3}dx\, \partial_x \phi(x)\,,\label{gll}
\ee 
or a smeared version of it. It is clear that $O_{13}$ commutes with all operators additively generated in $I_2,I_4$. It also belongs to the full model because is a linear combination of $\partial_x\phi(x)$. However, it does not belong to the additive algebra of $I_1, I_3$ because to construct it we needed operators $\partial \phi(x)$ outside these two intervals (see (\ref{gll}) and notice a single $\phi(x)$ is not a well-defined operator). This is an example of non locally generated operator in the two intervals. There is an analogous non-local operator $O_{24}$ for the intervals $I_2,I_4$. These two operators do not commute. From (\ref{conr}) we get
\be
[O_{13},O_{24}]=i \,,
\ee
independently of the smeared function, we could have used to define the operators. The structure of the algebras is summarized by
\bea
({\cal A}_{\textrm{add}}(I_1I_3))'=({\cal A}(I_1)\vee {\cal A}(I_3))'={\cal A}(I_2)\vee {\cal A}(I_4)\vee O_{24}={\cal A}_{\textrm{add}}(I_2I_4)\vee O_{24} \,,\\
({\cal A}_{\textrm{add}}(I_2I_4))'=({\cal A}(I_2)\vee {\cal A}(I_4))'={\cal A}(I_1)\vee {\cal A}(I_3)\vee O_{13}={\cal A}_{\textrm{add}}(I_1I_3)\vee O_{13}\,.
\eea

There is a different way to look at this problem that is very enlightening. The chiral scalar is in fact a subalgebra of the chiral free massless fermion. The fermion theory is complete, it does not have any algebra-region problems. The mutual information for the fermion is
\be
I(\eta)=-\frac{c}{3} \log(1-\eta)\,,
\ee
with $U(\eta)=U(1-\eta)=0$, so there is no problem with Haag duality. 
The chiral scalar $\partial \phi(x)=j(x)$ is equivalent to the subalgebra generated by the electric current $j(x)=\psi^\dagger\psi$ of the fermion. These two operators have exactly the same $n$-point functions, a relation that is called bosonization. It is special of $d=2$ (notice the surprising fact that a bilinear of a fermion satisfies the bosonic Wick's theorem of a free scalar). Since the chiral scalar is a subalgebra of the fermion, its mutual information has to be smaller, and $U(\eta)\le 0$ in (\ref{uu}).
 The algebra of the current (or the chiral scalar) is exactly formed by the operators of the fermion algebra that are invariant under charge transformations $\psi(x)\rightarrow e^{i\alpha} \psi(x)$. So there is a $U(1)$ symmetry in the fermion such that the {\sl orbifold}, the part of the algebra invariant under the symmetry, is the scalar. This is one of the generic ways in which algebra-region problems might appear. If there is a theory with a global symmetry, there is another one with algebra region problems associated with it.

\subsection{Global symmetries: Regions with non trivial $\pi_0$ or $\pi_{d-2}$.}
\label{global}
The main features shown in the example above can be formalized in general for the case of global symmetries. 
We consider the subalgebra ${\cal O}$ of a theory ${\cal F}$, consisting of operators invariant under a global symmetry group $G$ acting on ${\cal F}$. The theory ${\cal O}={\cal F}/G$ is called an orbifold. The orbifold can always be constructed when there is a global symmetry and is not a merely academic construction. When there is global symmetry the algebra ${\cal O}$ is the one accessible to local observers: one cannot have in a laboratory an operator that changes, say, the baryon number.     

The regions $R$ with non trivial homotopy group $\pi_0(R)$, that is, disconnected regions, and their complements,  are the relevant ones here. The complement $R'$ will have non trivial $\pi_{d-2}(R')$.  The simplest example is two disjoint balls $R_1$, $R_2$, and its complement $S=(R_1\cup R_2)'$, which is topologically a ``shell'' with the topology of $S^{d-2}\times R$. 

\begin{figure}[t]
\begin{center}  
\includegraphics[width=0.45\textwidth]{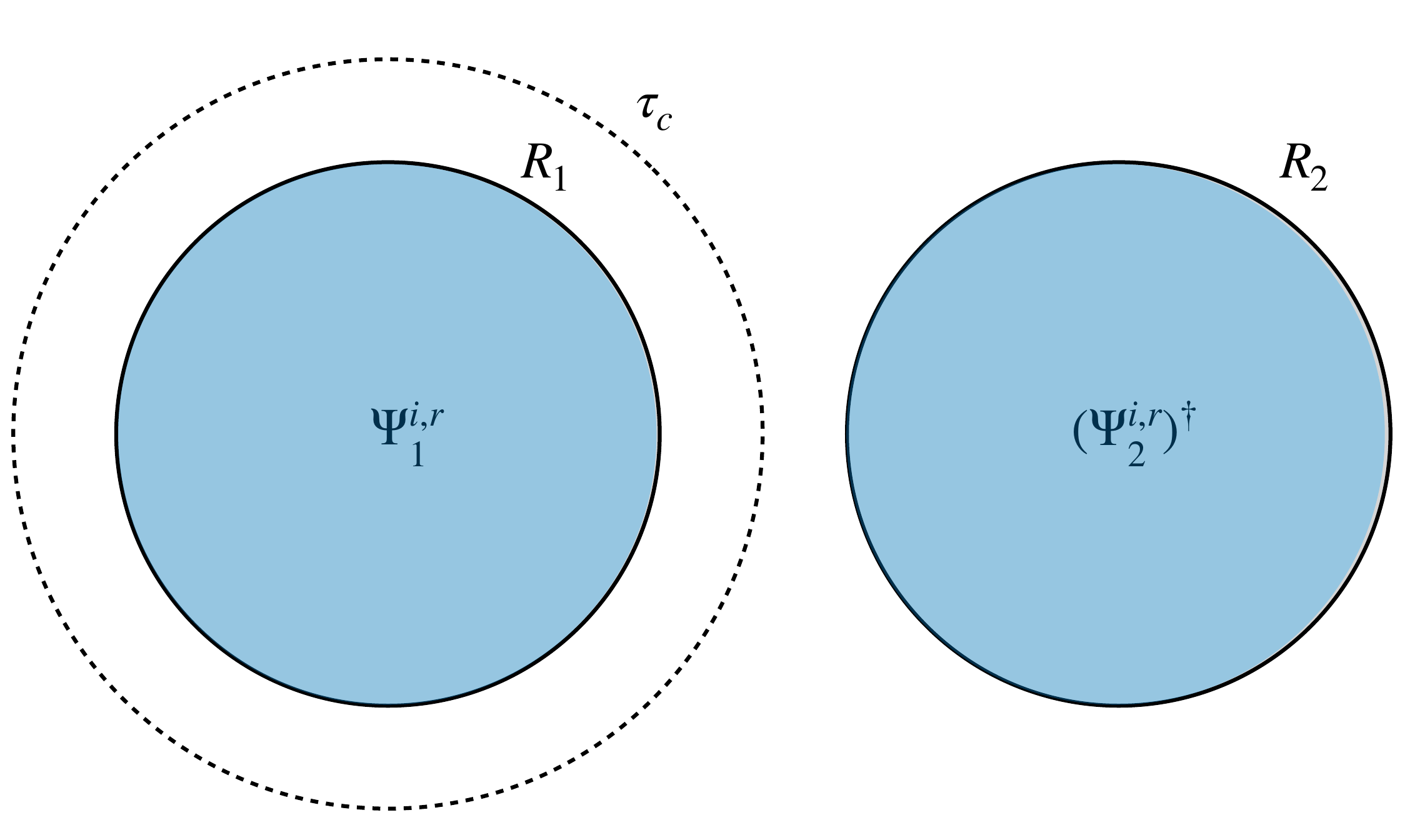}
\captionsetup{width=0.9\textwidth}
\caption{A region formed by two disjoint balls $R_1$ and $R_2$ (blue region) containing the intertwiner formed by a charge-anticharge operator. In the complement $S$ of the two balls, which has a non contractible $d-2$ dimensional surface, lives the twist operator.}
\label{int-twist}
\end{center}  
\end{figure}

Let $\psi^{i,r}_1$, $\psi^{i,r}_2$ be charge creating operators in $R_1,R_2$  in the theory ${\cal F}$, corresponding to the irreducible representation $r$, and where $i$ is an index of the representation.  The {\sl intertwiner} corresponding to this representation
\be\label{Ir}
{\cal I}_r= \sum_i \psi^{i,r}_1 (\psi^{i,r}_2)^\dagger
\ee  
is invariant under global group transformations and belongs to the neutral theory ${\cal O}$. See figure \ref{int-twist}. It  commutes with operators in ${\cal O}_{\textrm{add}}(S)$ but cannot be generated additively by operators in ${\cal O}(R_1)$ and ${\cal O}(R_2)$ since the charged operators $\psi^{i,r}$ belong to the field algebra ${\cal F}$ but not to ${\cal O}$.  

In a dual way, there are {\sl twist} operators $\tau_g$ implementing the group operations in $R_1$ and acting trivially in $R_2$. These commute with  ${\cal O}(R_1)$ and ${\cal O}(R_2)$, that is, uncharged operators in $R_1$ or $R_2$, but do not commute with the intertwiners, which have charged operators in $R_1$. The twists can be chosen to satisfy the group operations
\be\
\tau_g \tau_h=\tau_{gh}\,,\hspace{1cm}U(g) \tau_h U(g)^{-1}=\tau_{g h g^{-1}}\,, \label{tt1}
\ee
where $U(g)$ is the global symmetry operation. 
For a non-Abelian group, the twists are not invariant. The combinations of twist operators invariant under the global group
\be
\tau_{c}=\sum_{h\in c} \tau_{h} 
\ee
are labelled by conjugacy classes of the group, $g c g^{-1}=c$. These operators belong to the neutral algebra ${\cal O}$.  Hence, if the full model ${\cal F}$ including charge creating operators satisfy duality and additivity, this is not the case of the neutral model ${\cal O}$. We have
\bea
({\cal O}_{\textrm{add}}(R_1 R_2))' &=& {\cal O}_{\textrm{add}}(S)\vee \{\tau_{c}\}\,,\label{copi1}\\  
({\cal O}_{\textrm{add}}(S))' &=& {\cal O}_{\textrm{add}}(R_1 R_2) \vee \{{\cal I}_r\}\,.\label{copi2}
\eea
 This shows explicitly that, retaining additivity, duality fails for the two-component region $R_1R_2$ and its complement $S$. The reason is the existence in the model of operators (twists and intertwiners) in these regions, which cannot be additively generated inside the same regions by operators localized in small balls. However, the intertwiners and twists can be generated additively in ${\cal O}$ but in bigger regions with trivial topology. 

For finite groups, the number of independent twists coincides with the number of intertwiners. This is because the number $n_C$ of conjugacy classes of the group is equal to the number of irreducible representations. For Lie groups, we have an infinite number of irreducible representations, but the same occurs for conjugacy classes. 

We can choose the intertwiners to satisfy a closed algebra. We get the fusion algebra 
\be
 {\cal I}_{r_1} {\cal I}_{r_2}=\sum_{r_3} n_{r_1 r_2}^{r_3} {\cal I}_{r_3}\,,\hspace{1cm}{\cal I}_{\bar r}=({\cal I}_r)^\dagger\,,\hspace{1cm} {\cal I}_1=1\,,\label{fusion}
\ee
where $\bar{r}$ is the representation conjugate to $r$, and  $n^{r_3}_{r_1 r_2}$ are the fusion matrices of the group representations,
\be
[r_1]\otimes [r_1]=\oplus_{r_3} \,n^{r_3}_{r_1\, r_2} \, [r_3]\,,
\ee
 giving the number of irreducible representations of type $r_3$ in the decomposition of the tensor product of $r_1$ and $r_2$. Because $n_{r_1 r_2}^{r_3}=n_{r_2 r_1}^{r_3}$ the algebra (\ref{fusion}) is Abelian. The same can be said of the algebra of the twists. From (\ref{tt1}) we get
\be\label{fusc}
\tau_{c_1} \tau_{c_2}= \sum_{c_3} m^{c_3}_{c_1 c_2}   \tau_{c_3}\,, 
\ee
with $ m^{c_3}_{c_1 c_2}$ the fusion coefficients of the conjugacy classes. 

The two Abelian algebras of twists and intertwiners do not commute with each other. For finite groups, it can be shown they can be embedded in the non-Abelian matrix algebra of $|G|\times |G|$ matrices. A similar embedding works for Lie groups but the embedding algebra needs to be infinite-dimensional. For Abelian symmetry groups, the commutation relations take a very simple form
\be\label{ccrel}
\tau_g \, {\cal I}_r = \chi_r(g) \, {\cal I}_r \, \tau_g\,,  
\ee 
where $\chi_r(g)$ is the group character.

\subsection{Generalized global symmetries for gauge theories}

Another candidate for giving us trouble in the algebra region relations are gauge symmetries. It will turn out that in this case the problems are for regions with non-trivial $\pi_1$ or $\pi_{d-3}$ homotopy groups, that is, regions with non-contractible loops or $d-3$ dimensional spheres. 

Let us consider first a free Maxwell field in $d=4$.  This is the Gaussian theory of the electric and magnetic fields, with equal time commutation relations
\be
[E^i(\vec{x}),B^j(\vec{y})]=i \varepsilon^{ijk}\, \partial_k \delta^3(\vec{x}-\vec{y})\,.
\ee
Equivalently, the theory can be described by the normal oriented electric and magnetic fluxes $\Phi_E$, $\Phi_B$ on two-dimensional surfaces with boundaries $\Gamma_E$ and $\Gamma_B$.  For such fluxes we have a commutator which is topological, it is proportional to the linking number of $\Gamma_E$ and $\Gamma_B$,
\be
[\Phi_E,\Phi_B]=\frac{i}{4\pi}\int_{\Gamma_E}\int_{\Gamma_B} \frac{\vec{x}_1-\vec{x}_2}{|\vec{x}_1-\vec{x}_2|^3}\, d\vec{x}_1\times d\vec{x}_2  \,.\label{deci}
\ee  
We will assume these fluxes to be smeared over positions of $\Gamma_E$ and $\Gamma_B$ such that the flux operators are well defined (and not operator-valued distributions). The smearing region for $\Gamma_E$ and $\Gamma_B$ lies inside a region with the topology of a ring $R$ and its complement $R'$ respectively. In $d=4$ the topology of $R'$ is the same as the topology of $R$. It is $S^1\times R^2$ and has non-trivial $\pi_1(R)$. 

\begin{figure}[t]
\begin{center}  
\includegraphics[width=0.35\textwidth]{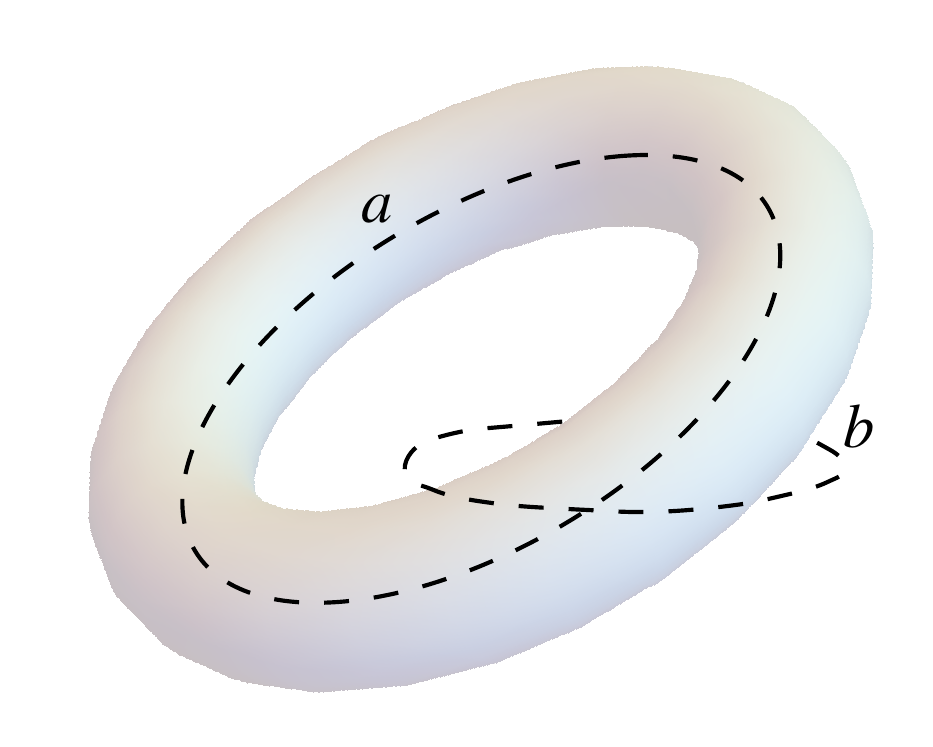}
\captionsetup{width=0.9\textwidth}
\caption{A ring like region with a non breakable Wilson loop $a$, and its complement with a linked  't Hooft loop $b$.}
\label{toro}
\end{center}  
\end{figure}  

Because $\nabla E=\nabla B=0$ the fluxes are conserved, and then the surface over which they are computed can be deformed keeping the boundary fixed. By deforming the surface of the flux we can take it away from some local operator lying in the original surface, and this implies that the fluxes will commute with the locally generated operators associated with the complementary ring.   

 We can write a bounded  electric flux operator ('t Hooft loop) $T^{g}=e^{i g \Phi_E}$, and a magnetic operator (Wilson loop) $W^{e}=e^{i q \Phi_B}$, for any $g,q$. See figure \ref{toro}.  The commutation relations for these operators in simple linked loops follows from (\ref{deci})
\be
T^{g}W^{q}= e^{i\, q\, g } \, W^q T^g\,.\label{cr}
\ee 
This non-commutativity implies these operators cannot be locally generated in the ring in which they are based. For example, if $T^g$ were locally generated in $R$ (where its boundary lies) this would imply, by the arguments given above, it necessarily commutes with $W^q$ based on the complementary ring. But this is not possible according to (\ref{cr}).

Therefore, the algebra of a ring $R$ and its complement $R'$ cannot be taken additive without violating duality because the commutant of the additive algebra of the ring contains both the electric and magnetic loops (for $d=4$) of any charge based on $R'$, and this is not additive. We have
\bea
{\cal A}_{\textrm{max}} (R')\equiv ({\cal A}_{\textrm{add}}(R))'=({\cal A}_{\textrm{add}}(R'))\vee \{W^q_{R'} T^g_{R'}\}_{q,g \in {\bf R}}\,,
\eea  
and analogously interchanging $R\leftrightarrow R'$. Here we have written $W^q_{R'}, T^g_{R'}$  for Wilson and 't Hooft loops based on $R'$.

One can repair duality at the expense of additivity by defining the algebras for rings to contain, on top of locally generated operators, some particular non locally generated ones that commute with other selected non locally generated operators in the complement. A natural condition is to select operators with electric and magnetic charges $(q,g)$, which are the same for any ring, such that our choice does not ruin translation and rotation invariance. Given two {\sl dyons} $(q,g)$, $(q',g')$ in the same ring, the one formed by their product, $(q+q', g+g')$, and the conjugate $(-q,-g)$, should also be present to close an algebra. Therefore the set of all dyons should be an additive subgroup of the plane, giving a lattice 
\be
(q,g)=n (q_1,g_1) + m (q_2,g_2)\,,\label{lat}
\ee
where $n,m\in Z$, and $(q_1,g_1)$ and $(q_2,g_2)$ are the generating vectors of the lattice.
Locality between what it would be a ``dyon'' $(q,g)$ in $R$ and another one $(\tilde{q},\tilde{g})$ in $R'$ (i.e. the vanishing of the phase in (\ref{cr})) gives the Dirac quantization condition.
If we want to construct a net that satisfies duality we need to take a maximal set of charges that satisfy the Dirac quantization condition. This forces us to choose
\be
q_1 \,g_2-g_1\,q_2=2 \pi\,.\label{sis}
\ee
The nets constructed in this way will satisfy duality but of course, they are not additive. 

Additivity can be recovered if we couple the theory to charged fields. 
For example, if we have a field $\psi$ of electric charge $e$  
 we can now consider Wilson line operators of the form
\be
\psi(x)e^{i q \int_x^y dx^\mu A_\mu } \psi^\dagger(y) \,.
\ee  
Taking products of consecutive Wilson lines, and allowing for fusion (or the OPE) of the fields with the opposite charge at the extremes of the lines we want to join, the Wilson loop $W^{q}$ in $R$ (with the specific charge $q$), becomes an operator in the additive algebra of $R$. 
In the same way, if we have magnetic charges $g$, $T^{g}$ corresponding to this charge should be additive in $R$, and with a dyon $(q,g)$ we can break the operators $T^{g}W^{q}$.
 For the theory to still satisfy locality the charges have to satisfy the Dirac quantization condition. This is now converted into the Dirac quantization condition for the charges. 
 As mentioned before, this condition is seen here as a consequence of causality in the net of algebras for the theory without charges. 
In this way, by adding a full set of charged fields with charges corresponding to a net with duality property, we can make the theory  ``complete'', in the sense of both satisfying duality and additivity. But of course, it is now a different model.  If we do not add charged operators for a full lattice, for example, if there are quantized electric charges and no monopoles, there will still be some problems of algebras and regions.

In presence of charged fields, the flux operators $T^g W^q$ continue to exist even if $(q,g)$ does not belong to the lattice. But they now depend on a surface rather than a closed curve, since $\nabla E=\nabla B=0$ is modified, and the fluxes on a surface cannot be deformed to another surface with the same boundary. Then, in this case, the operator belongs to a topologically trivial region and cannot be associated with a ring.

Consider now the case of non-Abelian Lie groups. We could have expected the algebras of non-local operators would be more complicated in this case. But they actually turn out to be simpler. 
Let us start with the Wilson loops. These are defined for each representation $r$ of the gauge group as
\be
W_{r}\equiv \textrm{tr}_{r} \mathcal{P} \,e^{i\oint_C dx^{\mu} A^r_{\mu}}\;,
\ee
where $C$ is a loop in space-time and $\mathcal{P}$ the path ordering.  

We now want to find whether Wilson loops are additive or not. A Wilson loop of representation $r$ could be certainly broken into pieces if there are charged fields $\phi_{r}$ transforming in representation $r$ because with these charged fields we can construct Wilson lines to break the loop. 
 Even if we are considering pure gauge theories without charges, we cannot escape the fact that, for non-Abelian gauge fields, the gluons are charged themselves. They are charged under the adjoint representation. Indeed, we can form the following Wilson line, terminated by curvatures,
\be
F_{\mu\nu}(x) \,P e^{i \int_x^y  dx^\sigma A_\sigma}\, F_{\alpha\beta}(y)\,,\label{indices}
\ee
where all fields are in the adjoint representation of the Lie algebra. Then, a loop in the adjoint representation can be generated locally by multiplying several of these lines along a loop. Since the adjoint Wilson loop is locally generated, the same can be said for all representations generated in the fusion of an arbitrary number of adjoint representations. 
Therefore, the truly non-local Wilson loops, those violating Haag duality, are labelled by the equivalence classes arising when we quotient the set of irreducible representations by the set of representations generated from the adjoint. This is equivalent to the representations of the centre $Z$ of the gauge group. The centre is of course an Abelian group. For example for $SU(2)$ it is just $Z_2$, formed by the identity and the $2\pi$ rotation which is $-{\bf 1}$. The adjoint representation has spin $1$ and generates all representations of integer spin. The fundamental representation is a representative of the unique class of unbreakable loops (composed by all representations with half-integer spin).  For $SU(n)$ the center is $Z_n$. 

In the same manner, there should be a dual operator lying in the complementary ring that does not commute with these non-local Wilson loops. These are called 't Hooft loops and are labelled by the elements of the centre $Z$ itself. They are singular gauge transformations corresponding to these elements of the centre across a surface bounded by the loop.

In general, it can be shown that the algebra of non-local operators for these topologies always form Abelian groups in $d\ge 4$. In a field theory language, these Abelian groups form what is called generalized symmetries. The commutation relations for non-local operators are 
\be
T^{z}W^{r}= \chi_r(z) \, W^r T^z\,,\label{gggrrr}
\ee
where $z$ is an element of the centre $Z$ of the gauge group,  $r$ a representation of $Z$, and  $\chi_r(z)$ is the representation character.  

As we see the groups of these symmetries are small. Being gauge redundancies that do not affect the states, all other non-Abelian symmetries do not give place to physical symmetries at all. These generalized symmetries disappear if we couple the gauge field to charges in fundamental representations.

\subsection{Entropic order parameters}

We have seen that some QFT can have more than one algebra choice for the same region, i.e. ${\cal A}_{\textrm{add}}$ and ${\cal A}_{\textrm{max}}$, with or without the non-local operators. It is then natural to look for an entropic quantity that takes into account this multiplicity. In particular, it will give a statistical measure on the non-local operators.   

The simplest idea would be to subtract the entropies of the two algebras for the same region $R$. However, this turns out to have some ambiguities because we are evaluating entropies in different algebras. To produce a well-defined quantity we need to introduce the concept of a conditional expectation. 

A conditional expectation is a linear map $\varepsilon : \mathcal{M}\rightarrow \mathcal{N}$,  from an algebra to a subalgebra, $\mathcal{N}\subset\mathcal{M}$, that is positive (maps positive operators into positive operators), and carries the unit into the unit, leaving the target algebra invariant. Another defining property is
\be
\hspace{-1mm} \varepsilon\left(n_{1}\,m\,n_{2}\right)=n_{1}\varepsilon\left(m\right)n_{2}\,,\hspace{3mm} \forall m\in\mathcal{M},\,\forall n_{1},n_{2}\in\mathcal{N}.\label{ce_def_prop}
\ee
These maps are the mathematical definition of what a coarse-graining  our observables means. Examples are tracing out part of the system
\be 
\mathcal{M}=\mathcal{N}\otimes\mathcal{O}\,\,\,\,\,\,\,\,\,\, \varepsilon(n\otimes O)\equiv \frac{\textrm{tr}(O)}{d_{\cal O}}\, n\otimes 1_{\mathcal{O}}\;,
\ee
or retaining the neutral part of a subalgebra under the action of a certain symmetry group
\be 
m\in \mathcal{M}\,,\hspace{.5cm} \varepsilon(m)\equiv \frac{1}{|G|}\sum\limits_{g}\tau_g \,m\,\tau_{g}^{-1}\in {\cal N}\;.\label{grupo}
\ee
Here the importance of these maps is that they allow lifting states from the subalgebra to the algebra: if $\omega$ is a state in ${\cal N}$, $\omega\circ \varepsilon$ is a state in ${\cal M}$. 

In general, there can be several conditional expectations between an algebra and a subalgebra. However, when ${\cal N}$ does not have a centre, and ${\cal N}'\cap {\cal M}$ is trivial, there is only one conditional expectation. This is expected to be the case in the QFT algebras  ${\cal A}_{\textrm{add}}$ and ${\cal A}_{\textrm{max}}$ we are interested in. Therefore, there is only one conditional expectation. This may be implemented by the dual non-local operators. For example, for the case of two balls and a global symmetry, the conditional expectation (\ref{grupo}), where the elements of the group are the twists, produces the unique conditional expectation.     
   
We have two algebras and the vacuum in both of them. We can use the conditional expectation to produce a state in the big algebra from the vacuum in the smaller one. In this way, we get two states for the big algebra and can compute a relative entropy.
 We define the  entropic parameters in the following manner. If the algebra of the non additive operators $a$ lives in a certain region $R$, this provides us with a natural inclusion of algebras
\be 
 {\cal A}_{\textrm{add}}(R)\subseteq  {\cal A}_{\textrm{add}}(R)\vee \{a\}={\cal A}_{\textrm{max}}(R)\;,
\ee
and the associated conditional expectation
$
\varepsilon : {\cal A}_{\textrm{max}}(R)\rightarrow {\cal A}_{\textrm{add}}(R)
$. 
This leads to the following entropic order parameter
\be 
S_{{\cal A}_{\textrm{max}}(R)}(\omega \mid \omega\circ\varepsilon)\,.
\ee

A parallel story works for the disorder parameters $b$. We remind they live in the complementary region $R'$, and they provide us with the following inclusion of algebras
\be 
{\cal A}_{\textrm{add}}(R')\subseteq  {\cal A}_{\textrm{add}}(R')\vee \{b\}={\cal A}_{\textrm{max}}(R')\;,
\ee
a  conditional expectation $\varepsilon'$, and the following dual entropic order parameter
\be 
S_{{\cal A}_{\textrm{max}}(R')}(\omega \mid \omega\circ\varepsilon')\,.
\ee

Notice that a conditional expectation $\varepsilon(a)=\delta_{a,1}$ eliminates all non local operators. Therefore, the state $\omega\circ \varepsilon$ automatically gives zero expectation values for these operators. The relative entropies above are therefore measures of distinguishability between the vacuum and another state that is similar to the vacuum except that it kills non-local operators. As the vacuum expectation values of the non-local operators grow larger, the larger the relative entropy will be. 

If we have an Abelian group, it can be shown that 
\be
S_{{\cal A}_{\textrm{max}}}(\omega|\omega \circ \varepsilon)=S_{{\cal A}_{\textrm{max}}}(\omega\circ E)-S_{{\cal A}_{\textrm{max}}}(\omega)\,.\label{652}
\ee
This particular difference of entropies is well defined in the continuum even though the entropies are not. Eq. (\ref{652}) shows that the difference of entropies is monotonous. Note that it is the difference of entropies between two states in the same algebra and not the entropies of two different subalgebras. This latter quantity depends on the regularization. 
 
\subsection{The certainty and uncertainty relations}

We have seen entropic order parameters can be defined both for operators such as intertwiners and Wilson loops and for the dual ones, such as twists and 't Hooft loops. It turns out that both are related through an entropic certainty relation we now introduce.  

We have the following diagram of algebras and conditional expectations
\bea\label{cdiaor}
{\cal A}_{\textrm{add}}(R)\vee \{a\} & \overset{\varepsilon}{\longrightarrow} &{\cal A}_{\textrm{add}}(R)\nonumber \\
\updownarrow\prime \!\! &  & \,\updownarrow\prime\\
{\cal A}_{\textrm{add}}(R')& \overset{\varepsilon'}{\longleftarrow} & {\cal A}_{\textrm{add}}(R')\vee \{b\}\,.\nonumber 
\eea
The vertical lines connect commutant algebras, while the horizontal lines are connected by a conditional expectation that maps an algebra to a subalgebra. In this situation, one can show the relevant relative entropies are related by the following identity called the certainty relation, valid for pure global states $\omega$,  
\be
S_{{\cal A}_{\textrm{add}}(R)\vee \{a\}}\left(\omega|\omega\circ\varepsilon\right)+S_{{\cal A}_{\textrm{add}}(R')\vee \{b\}}\left(\omega|\omega\circ\varepsilon'\right)=\log |G|\;.\label{ede}
\ee
In a general complementarity diagram like (\ref{cdiaor}) $|G|$ is called the index of the conditional expectations. In the case we are interested here where the underlying algebra of the non-local operators come from a group, this index is simply the number of elements in the group.  
The certainty relation is a cousin of the equality of entropies for commutant algebras and applies to pure states too. However, it applies to complementary algebras here where the two sets of non-local operators do not commute to each other.   Interestingly, each relative entropy measures the statistics of one set of non-local operators. This relation shows they are not independent when the whole algebra of local operators in the regions are also considered.   

From this expression and the positivity of relative entropy, we obtain the individual bounds
\be 
S_{{\cal A}_{\textrm{add}}(R)\vee \{a\}}\left(\omega|\omega\circ\varepsilon\right)\leq\log|G|\nonumber\,, \hspace{.6cm}
S_{{\cal A}_{\textrm{add}}(R')\vee \{b\}}\left(\omega|\omega\circ\varepsilon'\right)\leq\log|G|\;.
\ee
Other bounds can be obtained using the monotonicity of relative entropy. For example, if we know the expectation values on subalgebras of non-local operators that close in themselves, we can estimate the relative entropy using the upper and lower bounds
\be
S_{\{a\}}\left(\omega|\omega\circ\varepsilon\right)\le S_{{\cal A}_{\textrm{add}}(R)\vee \{a\}}\left(\omega|\omega\circ\varepsilon\right)\le \log |G|- S_{\{b\}}\left(\omega|\omega\circ\varepsilon'\right)\,.
\ee
Restricting to these algebras we have an entropic uncertainty relation 
\be
S_{\{a\}}\left(\omega|\omega\circ\varepsilon\right) + S_{\{b\}}\left(\omega|\omega\circ\varepsilon'\right)\le \log |G|\,.
\ee
This is an entropic version of the ordinary uncertainty relations satisfied by the expectation values of non-commuting operators like $a,b$. It tells that expectation values of $a$ and $b$ cannot saturate to the maximal values simultaneously. in that case, each of the relative entropies would be $\log G$, violating the bound.

The way the certainty relation is realized is easily guessed in certain limits. If the expectation values of the $a$ operators tend to zero, for example when the region $R$ is very thin, the state with and without the conditional expectation will not be easily distinguished, and the order parameter goes to zero,
\be
S_{{\cal A}_{\textrm{add}}(R)\vee \{a\}}\left(\omega|\omega\circ\varepsilon\right)\rightarrow 0\,,\hspace{.7cm}S_{{\cal A}_{\textrm{add}}(R')\vee \{b\}}\left(\omega|\omega\circ\varepsilon'\right)\rightarrow \log|G|\,.
\ee
Analogously, when the expectation values of $b$ tend to zero we have 
\be
S_{{\cal A}_{\textrm{add}}(R)\vee \{a\}}\left(\omega|\omega\circ\varepsilon\right)\rightarrow \log |G|\,,\hspace{.7cm}S_{{\cal A}_{\textrm{add}}(R')\vee \{b\}}\left(\omega|\omega\circ\varepsilon'\right)\rightarrow 0\,.
\ee

Summarizing, symmetries and generalized symmetries are associated with algebra-region problems. Order parameters for these symmetries are then naturally suggested from the fact that two different states can be produced out of the vacuum for the same algebra. The relative entropy between these states further satisfies a surprising relation that ties the statistics of complementary dual non-local operators.

\subsection{Topological contributions to the entropy and the density of charged states}

Let us now focus on the case of global symmetries. There is a breaking of Haag duality in a pair of disconnected regions due to the existence of certain intertwiners~(\ref{Ir}). These are neutral operators formed by a charged operator in one connected component and a compensatory anti-charge operator in the other. There is one intertwiner per irreducible representation. In the complementary region, which has the topology of a spherical shell, there is also a breaking of duality due to the existence of twists operators~(\ref{tt1}), representing the symmetry group locally. The twists do not commute with the intertwiners~(\ref{ccrel}).  

We take two disconnected regions $R_1$ and $R_2$ and their complement, the ``shell''  $S=(R_1 R_2)'$. As described above, there are  two choices for the algebra of $R_1 R_2$, namely, the additive algebra $\mathcal{O}_{R_1R_2}$ or, alternatively, the additive algebra plus the intertwiners $\mathcal{O}_{R_1R_1}\vee \mathcal{I}$. Similarly, we have two algebras for $S$, the additive one $\mathcal{O}_{S}$ and the additive one plus the intertwiners $\mathcal{O}_{S}\vee \tau$. 
We have two relevant relative entropies 
$
S_{\mathcal{O}_{R_1R_2}\vee \mathcal{I}}\left(\omega|\omega\circ E_\mathcal{I}\right)$ and $S_{\mathcal{O}_{S}\vee \tau}\left(\omega|\omega\circ E_\tau\right)$, which add up to $\log|G|$, if the state is pure.  

In this case, we have at our disposal the theory ${\cal F}$ containing charged operators too. Using this other theory we can write the relative entropy in another useful form. To do that we have to introduce another entropy property. This is called 
 the conditional expectation property. For a general conditional expectation $\varepsilon : \mathcal{M}\rightarrow \mathcal{N}$, and for a state $\omega$ of ${\cal M}$ and a state $\phi$ on ${\cal N}$, we have,
\be
S_{\cal M}(\omega|\phi\circ \varepsilon)=S_{\cal N}(\omega|\phi)+S_{\cal M}(\omega|\omega\circ \varepsilon)\,.
\ee
In particular, when $\omega$ is invariant under $\varepsilon$ we get
\be
S_{\cal M}(\omega|\phi\circ \varepsilon)=S_{\cal M}(\omega\circ \varepsilon|\phi\circ \varepsilon)=S_{\cal N}(\omega|\phi)\,.\label{660}
\ee

We apply these formulas to the additive algebras ${\cal F}_{12}$ and ${\cal O}_{12}$ corresponding to the region $R_1R_2$, and the  states $\omega$ and $\omega_1\otimes \omega_2$. Assuming the state $\omega$ is invariant under the global symmetries we get 
\be
 S_{\mathcal{O}_{R_1R_2}\vee \mathcal{I}}\left(\omega|\omega\circ E_\mathcal{I}\right)= S_{{\cal F}_{12}}\left(\omega|\omega\circ E_\mathcal{I}\right)=I_{{\cal F}}(R_1,R_2)-I_{\cal O}(R_1,R_2)\,.
\label{mutualdif}\ee 
The first equation follows from (\ref{660}), where the conditional expectation generated by the global symmetry group is used.  On the other hand, to obtain the second equation one has to use the conditional expectation property with the conditional expectation generated by the twists.   

On the right-hand side of (\ref{mutualdif}) we have the difference between the mutual information on the two models. This formula tells that this mutual information difference, which is naturally positive by the inclusion of algebras, is also monotonic, and is equal to a relative entropy. 

This formula can be used to understand how the regularized entropies of the two models differ (see section \ref{regent}). For that, we have to take the special limit in which $R_2$ is the complement of $R_1$, except for a thin regularization region $S$ in between. In this case $S_{\mathcal{O}_{R_1R_2}\vee \mathcal{I}}\left(\omega|\omega\circ E_\mathcal{I}\right)$ can be understood as (twice) the difference between regularized entropies between the model with charges ${\cal F}$ and the orbifold ${\cal O}$.  

What happens in that limit is not difficult to understand. Whenever $R_1$ and $R_2$ touch each other (even in a small region of the boundary) we can place intertwiners very near the boundary such that their expectation values are maximal. The twist, on the other hand, has to live in a very thin region, and its expectation value goes to zero. In consequence, in this limit, we have 
\be
\Delta I= \log |G|\,.\label{qq}
\ee

This is rather surprising in a certain way. Let us consider the case in which the fields are very massive. In that case, the mutual information will be almost zero when the regions are separated by a distance larger than the gap scale $m^{-1}$. However, when the distance between them crosses this scale, mutual information (both of them) rapidly acquires non zero value and the difference saturate to $\log |G|$. In odd spacetime dimensions, the difference of mutual information represents the difference of the universal constant terms  $F$ in the regularized entropy. Then, this sudden change of value with the cutoff contradicts the idea that deep in the infrared the contributions from high energy physics should be local on the boundary of the region. As we have discussed in chapter \ref{EEQFT}, short-distance contributions which are local on the boundary cannot change the constant term. Then, we conclude there must be non-local contributions along the boundary when the cutoff reaches the gap scale.         

 This new feature comes from the orbifold. This contains a non-local operator, the twist, seated in the shell (the entropy of the two balls in the orbifold is equal to the one of the shell with the twist by $S({\cal A})=S({\cal A}')$). The expectation value of a smeared twist changes from $1$ to $0$ as we cross the gap scale, independently of the size of the sphere. This changes non locally the high energy contribution to the entropy. The non-locality comes from the constraint of zero charge in the orbifold model. A simple example is a massive free fermion in $d=3$, where the symmetry is the fermionic symmetry $\psi\rightarrow -\psi$. This model has $F=0$. But the orbifold has $F=\log(2)/2$ (computed as half of the mutual information regularization). This is a negative contribution to the entropy, which is subtracted from the area term.  To get this term, the cutoff scale $\epsilon$ in the mutual information should be such $\epsilon m\ll 1$. 
 
 \begin{figure}[t]
\begin{center}  
\includegraphics[width=0.35\textwidth]{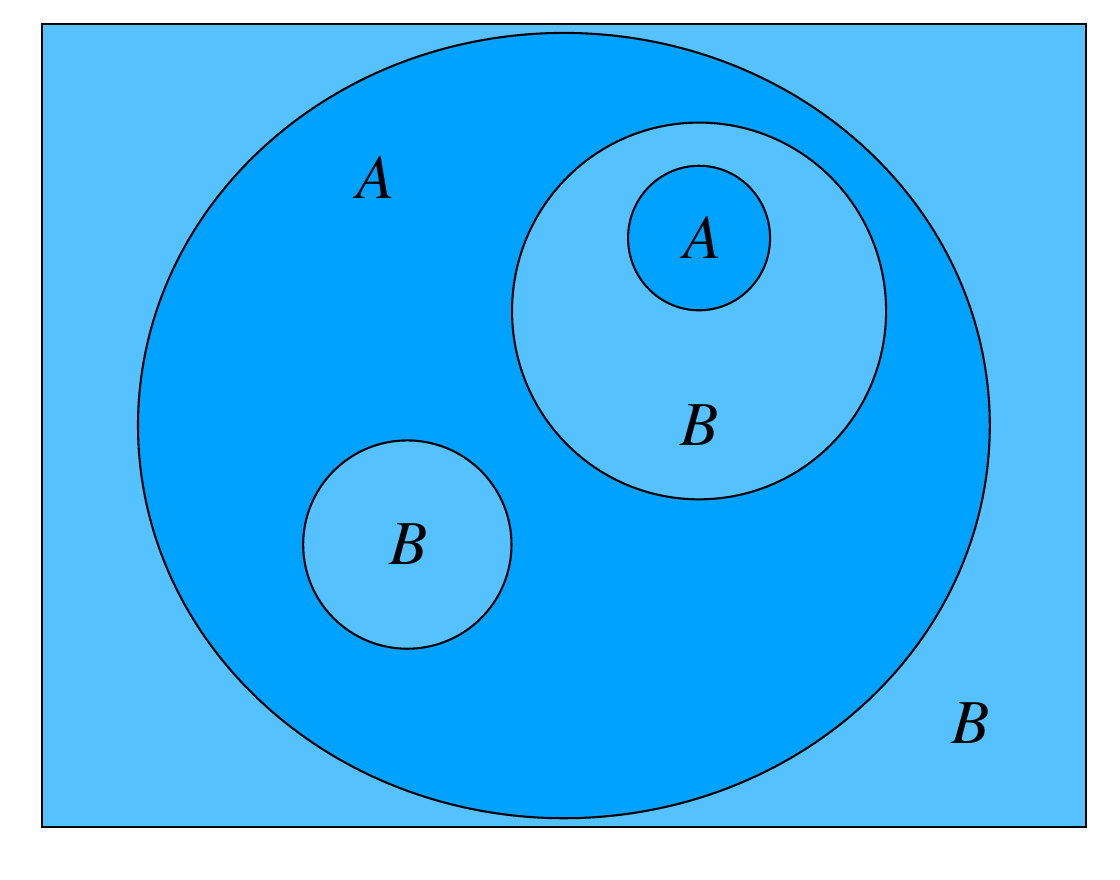}
\captionsetup{width=0.9\textwidth}
\caption{Complementary regions $A$ and $B$ with many boundaries in common (here $n_\partial=4$). There is an independent twist seated on each of the boundaries.}
\label{topo}
\end{center}  
\end{figure}
 
 These types of contributions are called topological. In fact, we see they are independent of the shape of the region since (\ref{qq}) always saturates to the same number. When we consider regions with more boundaries (see figure \ref{topo}) we get a contribution proportional to the number of boundaries. This is because the number of independent twists coincides with the number of boundaries $n_\partial$. We get    
\be
\Delta I= n_\partial \log |G|\,.\label{qq1}
\ee
This has to be understood as a negative term in the orbifold entropy. A topological contribution cannot exist without the area term (which reflects the existence of non-topological degrees of freedom), because if it were alone it would make the mutual information $I_{\cal O}$ negative. 

Topological contributions may question our assertion that the RG charges at the IR measure quantities of the IR theory itself. In fact, to obtain them the cutoff $\epsilon$ has to be set smaller than the UV scales, which may be seen as a susceptibility to the UV physics. However, the details of the UV physics are irrelevant for the final result of the RG charge at the IR.  This depends only on the constraints of the IR model itself.  

In this section and the preceding one we have assumed $|G|$ was a finite number. If the group is a Lie group the certainty relation is divergent. The relative entropy that diverges is $S_{\mathcal{O}_{S}\vee \tau}\left(\omega|\omega\circ E_\tau\right)$ because of the continuous parameters of the twists. The intertwiner relative entropy is finite, and we have for nearly touching regions 
\be 
\Delta I\sim \frac{|{\cal G}|}{2} \, \log(A/\epsilon^{d-2})\,,
\label{ghgh} \ee
where $A$ is the area of the boundary, and $|{\cal G}|$ is the dimension of the Lie algebra. This can be obtained by evaluating upper and lower bounds with relative entropies in subalgebras of twists and intertwiners. 
A heuristic interpretation for this formula is that 
there is a large number of independent intertwiners (or charge fluctuations) along the boundary that acts as random variables. The distribution in the space of charges is then an effective Gaussian distribution with variance $A/\epsilon^{d-2}$. The relative entropy on this Gaussian distribution gives the logarithm in (\ref{ghgh}).

\bigskip

Instead of talking about two disjoint regions in the vacuum of a QFT with a global symmetry we could also consider the case of a thermal state in the QFT and consider the thermofield double purifying this thermal state. Then, the same reasoning applies to the mutual information between the two sides of the thermofield double. The temperature plays now a role analogous to the inverse of the separating distance between the two disjoint regions in vacuum. We have $\Delta I\rightarrow 0$ in the zero-temperature limit and $\Delta I\rightarrow \log|G|$ in the large temperature limit. This can be proved rigorously using the certainty relation.  This calculation has been recently used to prove an asymptotic limit for the density of charged states at high energies and finite volume in a QFT with a global symmetry. This is 
\be
\lim_{E \rightarrow \infty} \rho_r(E)= \frac{d_r^2}{|G|}\,  \rho(E)\,,
\ee
where $d_r$ is the dimension of the representation and $\rho_r(E)$ is the density of states corresponding to representation $r$. The same holds for the large volume limit instead of the high energy limit.

\subsection{The logarithmic term of the Maxwell field}

The entanglement entropy of a Maxwell field has raised controversies over the years.  
On one hand, the logarithmic term for a sphere is supposed to be given by the anomaly, with a coefficient $-31/45$. On the other hand, it is not difficult to decompose the  Maxwell field into vector spherical harmonics and see that for each angular mode $l\ge 1$ there are two modes (the two helicities), and each of them has the same radial Hamiltonian and commutators as the spherical $l$ mode of a scalar. There is no mode $l=0$ for the Maxwell field.  Therefore, this counting says that the logarithmic coefficients are related as 
\be
s^{\textrm{Maxwell}}_{\textrm{log}}=2\,(s^{\textrm{scalar}, d=4}_{\textrm{log}}- s^{\textrm{scalar}, d=2}_{\textrm{log}})=2(-1/90-1/6)=-16/45\,,   
\ee
where the $d=2$ scalar corresponds to the $l=0$ mode of the scalar in $d=4$. 
The coefficient $-16/45$ is indeed what one gets from the mutual information for a free Maxwell field.

The explanation of this mismatch is in origin quite similar to the topological model discussed above. The Maxwell field has constraints $\nabla E=\nabla B=0$ that could be lifted by very massive electric and magnetic charges. If the $\epsilon$ of our regularization scale is smaller than this mass scales $\epsilon m\ll 1$, mutual information across the shell will be able to sense the charge fluctuations and get a new contribution. As in the topological model, this new contribution will change the non-local part of the entropy, because it is affecting non-local correlations. These non-local correlations in the present case are related to the expectation values of Wilson and 't Hooft loops in the shell (that is, magnetic and electric fluxes). The expectation values of these loops have a perimeter law
\be\label{perr}
\langle T \rangle,\langle W \rangle \sim e^{-\frac{\textrm{perimeter}}{\epsilon}}
\ee  
for large $\epsilon m \gg 1$ and an area law 
\be \label{arr}
\langle T \rangle,\langle W \rangle \sim e^{-\alpha \frac{\textrm{area}}{\epsilon^2}}\,,
\ee 
for $\epsilon m \ll 1$, with $\alpha$ the coupling constant for charged particles. This transition never happens if there are no charged particles. 

The change in the logarithmic term in the entropy can be computed and exactly fits the mismatch above. The coefficient $-16/45$ is the correct one for a purely free Maxwell field while $-31/45$ corresponds to a Maxwell field interacting with electric charges and magnetic monopoles. If electric charges are quantized and there are no magnetic monopoles, for example, we do not get an agreement with the anomaly, and the coefficient is only corrected halfway to $-47/90$. This is because magnetic constraints have not been lifted in this case.  

In the topological model, we need to cross the mass scale in order to sense the non-local correlations the infrared model has. Here, for the interacting field, there is an effective generalized $U(1)\times U(1)$ symmetry at the infrared, given by the non-local loop operators. Once we cross the scale of the charges, these generalized symmetries are shown to be fictitious, since loop operators become flux surface operators when there are charges. This is the reason for the sudden parametric change in the expectation values from (\ref{perr}) to (\ref{arr}) for large size loop operators as a function of their small width. This prompts the parametric change on the logarithmic term in the mutual information. As we see, this is neither a purely UV nor a purely IR phenomenon.     

In the calculation of the logarithmic term of the sphere by mapping to de Sitter space in section \ref{tow}, we did not compute the mutual information, but rather the entropy. The present discussion reveals that in models with generalized symmetries this bare entropy calculation by mapping to de Sitter space should be supplemented with adequate boundary conditions on the horizon.  The boundary conditions provide effective additional information on the algebra-region ambiguities.

We emphasize that the Maxwell field interacting with heavy electric and magnetic charges does not have algebra-regions problems and is a complete theory. In this sense, the entropy contribution discussed in this section differs from the orbifold ones discussed before. Here, they are rather the result of effective algebra region problems (for rings like regions rather than spherical shells) that appear in the infrared limit theory.

\subsection{Operator order parameters and breaking of additivity-duality}

According to the previous discussion, associated with the failure of additivity and duality, there is a set of non locally generated operators that enables multiple choices of algebras for a given region. We have seen these features of the net of algebras implies the existence of generalized symmetries. The non-local operators are in fact the ones implementing these generalized symmetries.

In a somewhat opposite direction, here we argue that given the existence of operators with certain properties that are adequate to sense symmetry breaking there must be necessarily some algebra region problems. 

To be concrete, let us focus on the case of loops in gauge theories.  The Wilson loop was proposed as an order parameter for confinement. If we have an area law for the Wilson loop expectation value there is confinement.  In that case of spontaneous symmetry breaking (SSB), we have a perimeter law for the Wilson loop and an area law for the complementary operator, the 't Hooft loop. In fact, both instances can be understood on the same footing as SSB of one of the two dual generalized symmetries.  
    
We will argue that for any QFT it is not possible to construct a loop order parameter displaying an area law by employing only operators that are locally generated on the ring. For example, this is the case of QCD with fundamental quarks, where the fundamental Wilson loop is additive (can be broken by Wilson lines), and the formation of mesons forbids an area law. We can have only a sub-perimeter law behaviour (perimeter law, or even a constant law) in such additive loop cases. This implies that the existence of such (confinement) order parameter requires a non locally generated operator and the failure of the additivity property for ring-like regions. So algebra-region problems can also be detected through expectation values without computing commutants or talking about entropies. 

\begin{figure}[t]
\begin{center}  
\includegraphics[width=0.4\textwidth]{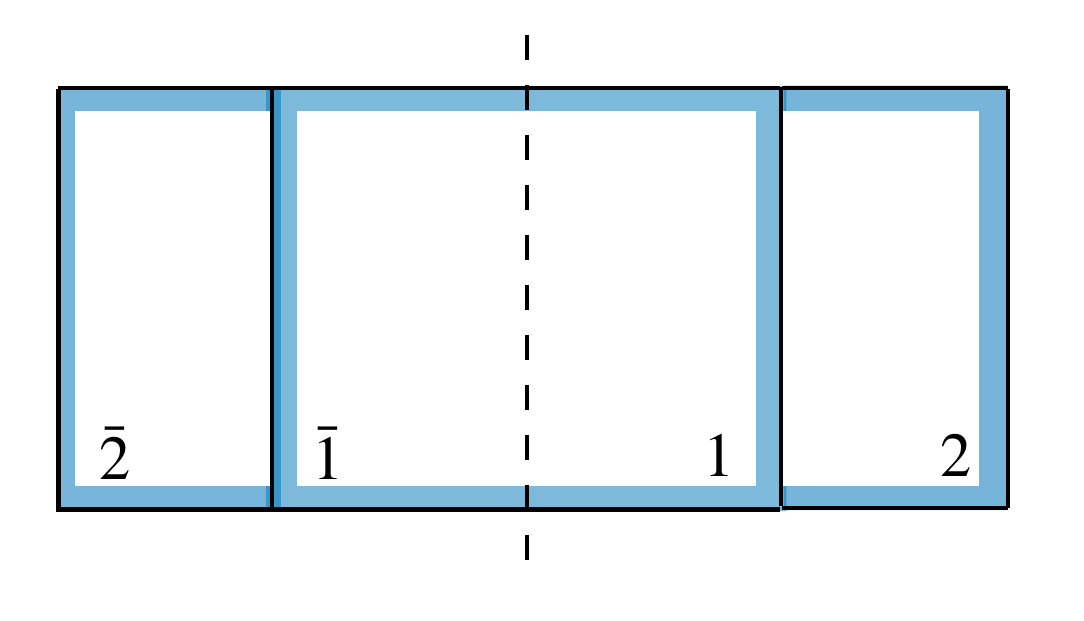}
\captionsetup{width=0.9\textwidth}
\caption{The construction  that shows the convexity of the quar-antiquark  potential.}
\label{fig-bachas}
\end{center}  
\end{figure}

Let us first recall that the exponential decay of the expectation value of a (appropriately smeared) line operator with size is always bounded from below by an area law. To explain this, we refer to figure \ref{fig-bachas}, which shows four rectangular loop type operators. These are formed by products of two half-loops (labelled $1$ and $2$) reaching just to a plane of reflection,  and their reflected CRT images (labelled $\bar{1}$ and $\bar{2}$ respectively). The application of reflection positivity in the Euclidean version, or CRT positivity in real time (Tomita Takesaki positivity of the Rindler wedge, see section \ref{tt} and \ref{rw}) leads to 
\be
\langle W(1,\bar{2})\rangle \langle W(2,\bar{1})\rangle \le   \langle W(1,\bar{1})\rangle \langle W(2,\bar{2})\rangle \,. \label{calcu}
\ee
Writing $\langle W\rangle = e^{-V(x,y)}$, with $x,y$ the two sides of the rectangle, it follows from this relation,  and the analogous one in the $y$ axis, that the potential $V(x,y)$ must be concave
\be\label{conc}
\partial_x^2 V(x,y)\le 0\,,\hspace{.8cm} \partial_y^2 V(x,y)\le 0\,. 
\ee
Then, the slopes $\partial_x V(x,y)$, $\partial_y V(x,y)$ never increase. They will converge to a fixed value in the limit of large size. If these values are non zero we have an area law. If they are zero, we have a sub-area law behaviour instead. No loop operator expectation value can go to zero faster than an area law $\sim e^{-c\,A}$ as the size tends to infinity. This calculation holds for any loop, whether locally or non locally generated in the ring, provided they are locally generated in the plane. The derivation can be justified more rigorously in a lattice model.  

Now we think only in loop operators formed additively in a ring. It is more convenient to use circular loops for our present purposes. As the loops are locally generated, we can imagine forming a partial operator $W(l_1,l_2)$ in an arc $(l_1,l_2)$ of the ring of longitudinal size $l=l_2-l_1$. The idea is that we construct now a loop of a certain size not by increasing the size of a smaller loop as above, but by increasing the size of an operator in an arc until the arc closes into a ring. 

\begin{figure}[t]
\begin{center}  
\includegraphics[width=0.25\textwidth]{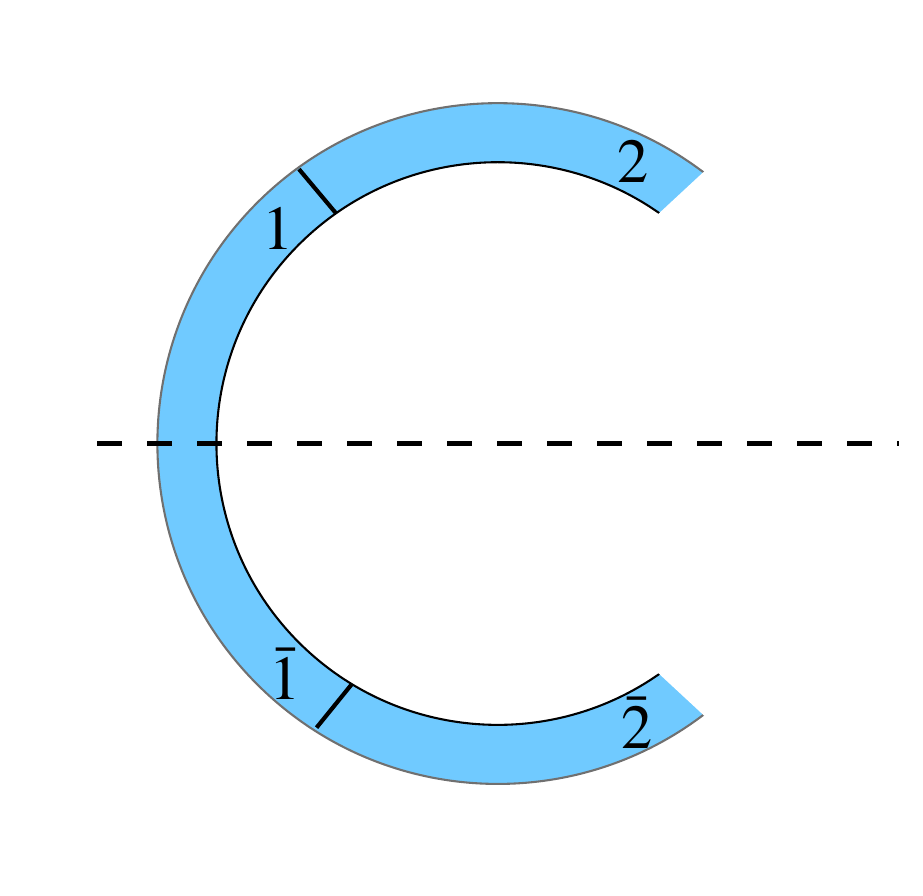}
\captionsetup{width=0.9\textwidth}
\caption{Reflection positivity applied to line operators along angular sectors of a ring. Operators $1$ and $2$ reach to the reflection plane and $\bar{1}$ and $\bar{2}$ are their CRT reflections.}
\label{fig-sectors}
\end{center}  
\end{figure}

Assume rotational invariance and define the potential
\be
\langle W(l_1,l_2)\rangle =e^{-V(l)}\,.\label{hh}
\ee
 In this case, we can use again CRT positivity, as shown in figure \ref{fig-sectors}. This results in
\be 
 V''(l)\le 0\,.
\ee   
Therefore the slope of $V'(l)$ is non increasing and
\be
\langle W \rangle \ge e^{-2 \pi R \,V'(0)-V(0)} \,. \label{boundito}
\ee
 Eq. (\ref{boundito}) gives a perimeter law, or more precisely a sub-perimeter law behaviour. In particular, this excludes the possibility of an area law or any law increasing more than linearly in the perimeter.   

The application of the same idea to the case of non locally generated loop operators in the ring fails. The reason is that we cannot define the partial (non-closed) line operators. For example, we can try using a non-gauge invariant Wilson line.  This introduces several problems when some gauge fixing is chosen to compute the expectation value. On the other hand, if we do not gauge fix, the expectation value of this line operator is zero, and the potential infinity. This prevents the calculation to give any useful bound.   

In conclusion, order parameters with area law require non-local operators in the ring. For all types of generalized symmetries, we expect by the same reason that order parameters for symmetry breaking are associated with algebra-region problems.

\subsection{Entropic order parameters and phases}

We briefly describe now how the entropic order parameters that capture the physics of generalized symmetries are able to distinguish different phases of QFT's such as spontaneous symmetry breaking of a global symmetry, Higgs and confinement phases of gauge theories, and conformal phases as well.

Let us start with a global symmetry. Let us consider two balls separated by a distance $r$ and let $r$ go to infinity. If the symmetry is unbroken the intertwiner expectation value tends to zero and so does the intertwiner entropic order parameter for the balls. This is a power-law decay for conformal fix points and an exponential one for a gapped one. 
In the case of spontaneous symmetry breaking, the charged operator get non zero expectation values, and so does the intertwiner.  As a consequence, the entropic order parameter tends to be a constant at large distances.

For the twist order parameter, we can consider a thin shell as the shell radius increases. The certainty relation connecting with the intertwiner physics can be used to understand the different phases. In the case of SSB, taking many separated intertwiners crossing the shell, with a number that grows with the volume $V$ of the shell, it can be proved that the twist order parameter decays to zero with the volume of the ball enclosed by the shell
\be
S_\tau\sim e^{-c\, V}\,. 
\ee
This is in contrast with the area law that holds for the case of unbroken symmetry.

\begin{figure}[t]
\begin{center}  
\includegraphics[width=0.45\textwidth]{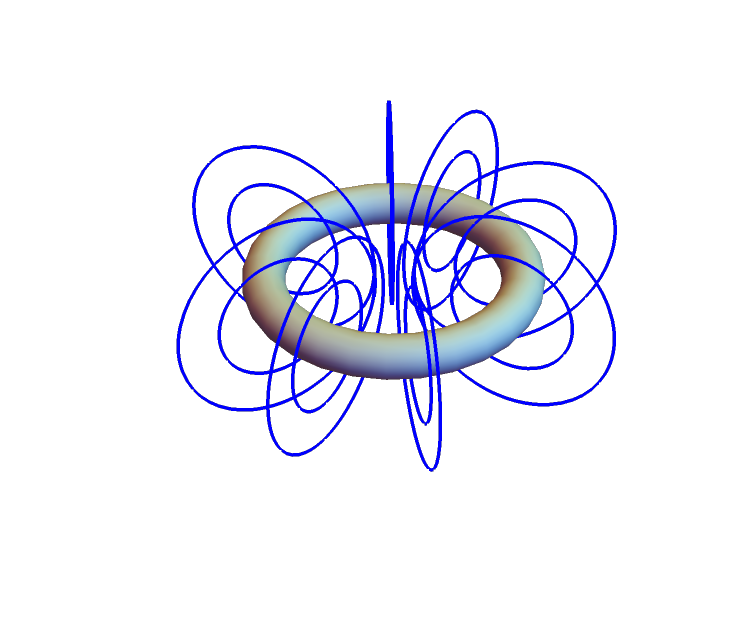}
\captionsetup{width=0.9\textwidth}
\caption{An area worth of 't Hooft loops crossing a Wilson loop.}
\label{madeja}
\end{center}  
\end{figure}

For gauge theories, let us consider, for example, the confinement phase. It is well known that the Wilson loop of a fundamental representation was initially devised as an order parameter for it. The expectation value of such fundamental Wilson loop can decay exponentially fast with the area of the loop. This behaviour is indicative of confinement since this implies a linear quark-antiquark potential. On the other hand, a perimeter law scaling of the Wilson loop excludes the possibility of confinement. 
The entropic order parameter for thin, large rings, will also display an area law 
\be
S_W\sim e^{-c \, A}\,.
\ee
The area law for the Wilson loop order parameter is tied to a constant law for the 't Hooft loop order parameter according to the certainty relation.  To see how these are related, we can place an area worth of 't Hooft loops crossing the area $A$ enclosed by the Wilson loop, as seen in figure \ref{madeja}.  In a gapped phase, these 't Hooft loops are decoupled to each other. If they have approximately constant expectation values, the 't Hooft loop order parameter of the whole set of thin loops together will approach saturation exponentially fast in the number of decoupled loops. In consequence, they will induce an area law for the Wilson loop, through the certainty relation. The case of spontaneous symmetry breaking of gauge theories is dual to the confinement case: we have area law for the 't Hoof loop parameter and constant law for the Wilson one.

The usual terminology in terms of area/perimeter laws for line operators is a bit cumbersome because we can enlarge line operators in one direction only. For example, there is an improvement from perimeter to constant law for a loop as we increase the width to become wider than the SSB scale.  In contrast, in the conformal case, we have a perimeter law for thin loops that cannot be improved to a constant law.  Further, this constant law in the SSB scenario does not persist without a perimeter term for exponentially large loops of constant width. The reason for these nuisances is that line operators are UV and IR operators at the same time, according to the two widely different scales involved in their geometry.

Entropic order parameters have the advantage that they depend only on the geometry of the region. A natural route as we move from the UV to the IR is to consider scaling regions, rather than long loops. For scaling regions $R\rightarrow \lambda \, R$, $\lambda\in (0,\infty)$, the relative entropy corresponding to a region $R$ of certain fix topology could approach zero or tend to the maximal value ($\log|G|$ or the logarithm of the order of a subgroup) as we move $\lambda$ to zero or infinity. In this case, we say there is saturation, and this happens independently of the precise shape of $R$. The two possible saturation values, $0$ or $\log |G|$, are achieved in a dual way for dual order parameters. Depending on the rate of approach to saturation we have confinement/SSB (exponentially fast in $\lambda$) or a Coulomb phase (power law in $\lambda$). 
  In contrast, the conformal case is characterized by limit values of the entropic order parameters that are not saturated and depend on the conformal geometry of $R$. The following table summarizes the behaviour of the entropic order parameters for the phases of gauge theories.  

\bigskip

\bigskip

\hspace{-1.3cm}
\begin{tikzpicture}[level 1/.style={sibling distance=9cm},level 2/.style={sibling distance=5cm},
level distance=3.9cm,
  every node/.style = {shape=rectangle, rounded corners,
    draw, align=left,
    top color=white, bottom color=black!20}]]

  \node {Wilson loops and `t Hooft loops relative entropies 
  \\$S_W(\lambda R)$ and $S_T(\lambda 
  R)$  for scaled regions, as $\lambda\rightarrow 0$ or $\lambda\rightarrow \infty$}
    child { node {Saturation: $S_W(\lambda R)$ and $S_T(\lambda R)$ \\ dualy converge to $\log \vert G\vert$ or 0 \\ independently of shape}
     child { node { Exponentially fast in $\lambda$: \\ {\bf Higgs/confinent phases  }}} child { node { Power law in $\lambda$ (power \\ depends on shape):\\ Electric and magnetic \\ {\bf Coulomb phases}}}}
    child { node {$S_W(\lambda R)$ and $S_T(\lambda R)$ converge to intermediate \\ values between $\log \vert G\vert$ and $0$  depending on \\ the shape of $R$}
      child { node {{\bf Conformal phase}} }
      };
      
\end{tikzpicture}

\bigskip

\subsection{Exercises}

\begin{itemize}

\item[1.-] Consider the algebra ${\cal M}$ of a qubit ($2\times 2$ matrices), the Abelian subalgebra ${\cal N}$ generated by the Pauli matrix $\sigma_1$, and the conditional expectation $\varepsilon(m)=(m+\sigma_1 m\sigma_1)/2$ mapping $\varepsilon:{\cal M}\rightarrow {\cal N}$. The conditional expectation $\varepsilon'(a +b \sigma_1)=a$ maps ${\cal N}'={\cal N}$ into ${\cal M}'={{\bf 1}}$. Show that the certainty relation holds with a $\log(2)$ on the right hand side for all pure states, and the sum of relative entropies is less that $\log(2)$ for impure global states.

\item[2.-] Suppose you have a group $G$ of twists and you know the expectation values $\langle \tau_g \rangle$ due to a state $\omega$. Compute the relative entropy $S(\omega|\omega\circ \varepsilon)$ in the group algebra in terms of these expectation values, where $\varepsilon(\sum_g a_g \tau_g)=a_1$ eliminates the non trivial twists. 

\end{itemize}

\subsection{Notes and references}
The content and presentation of this chapter are mainly based on the two articles \cite{casini2020entanglement,Casini:2020rgj}. A short review is \cite{Casini:2021zgr}. Generalized symmetries are introduced in a traditional field theory language in \cite{aharony2013reading}, generalizing an older paper by 't Hooft \cite{tHooft:1977nqb}, which also introduces 't Hooft loops.  There is a large body of related literature about symmetries in the algebraic approach to QFT that focuses on superselection sectors rather than on duality. For a guide see for example \cite{Haag:1992hx}. In the mathematical literature algebra-region problems for the case of CFT in $d=2$ is described in \cite{brunetti1993modular}.

The explicit computations for the chiral free scalar are from \cite{Arias:2018tmw}.

Entropic order parameters are introduced and studied in \cite{casini2020entanglement,Casini:2020rgj}. See some previous analogous computations in $d=2$ by R. Longo and F. Xu \cite{Longo:2017mbg}.  The certainty relation was introduced in \cite{casini2020entanglement}. A proof of this relation in a  general context for matrix algebras by J. M. Magan and D. Pontello is in \cite{magan2021quantum}. In the context of type III algebras see the papers by S. Hollands \cite{hollands2020variational}, and F. Xu \cite{xu2020relative}. The index ($|G|$ in the examples discussed here) can be defined algebraically and is called the Jones index \cite{Jones1983,KOSAKI1986123,cmp/1104179850}. It heuristically measures how many times one algebra includes another. In the present case, this inclusion is between the algebras with and without non-local operators.
    
 Topological terms for the orbifold are computed in \cite{casini2020entanglement}. Topological terms in the entropy for more general topological theories are considered in \cite{Kitaev:2005dm,Levin:2006zz}.  The asymptotic density of charged states was conjectured by D. Harlow and H. Ooguri in \cite{Harlow:2021trr}. The proof using the certainty relation is by J. M. Magan in \cite{Magan:2021myk}. This last paper also contains a discussion of the relation of this topic with the so-called ``entanglement equipartition'' in the symmetry-resolved entanglement entropy \cite{PhysRevB.98.041106,Murciano:2020vgh}. 
 
 The correction to the logarithmic term of the entropy of a Maxwell field is in \cite{casini2020logarithmic}. The argument for the concavity of the quark potential in \cite{bachas1986concavity}. The argument for sub-perimeter law for additive operators is from \cite{Casini:2020rgj}. 
Computations of entropic order parameters for gauge theories are in \cite{Casini:2020rgj}. The case of weakly coupled gauge theories is treated in \cite{Casini:2021tax}.
\newpage

\bibliography{trieste}{}
\bibliographystyle{utphys}

\end{document}